\documentclass[useAMS,usenatbib]{mn2e}
\usepackage{rotating, graphicx}
\usepackage{subfig}
\usepackage{pdflscape}
\usepackage{natbib}
\usepackage{amsmath}
\usepackage{courier}
\usepackage{color}
\usepackage{float}
\restylefloat{table}
\usepackage{footnote}

\title[A nebular analysis of the central Orion Nebula with MUSE]{A nebular analysis of the central Orion Nebula with MUSE}
\author[A. F. Mc Leod]{A. F. Mc Leod$^{1}$\thanks{E-mail amcleod@eso.org}, P. M. Weilbacher$^{2}$, A. Ginsburg$^{1}$, J. E. Dale$^{3}$$^{,}$$^{4}$, S. Ramsay$^{1}$ \newauthor and L. Testi$^{1}$$^{,}$$^{4}$$^{,}$$^{5}$ \\
$^{1}$European Southern Observatory, Karl-Schwarzschild-Str. 2, D-85748 Garching bei M{\"u}nchen, Germany\\
$^{2}$Leibniz-Institut f{\"u}r Astrophysik Potsdam (AIP), An der Sternwarte 16, D-14482 Potsdam, Germany\\
$^{3}$Universit{\"a}ts-Sternwarte M{\"u}nchen, Scheinerstr. 1, D-81679 M{\"u}nchen, Germany\\
$^{4}$Excellence Cluster 'Universe', Boltzmannstr. 2, D-85748 Garching bei M{\"u}nchen, Germany\\
$^{5}$INAF/Osservatorio Astrofisico of Arcetri, Largo E. Fermi, 5, 50125 Firenze, Italy}

\begin{document}

\date{Accepted yyyy month dd. Received yyyy month dd; in original form yyyy month dd}

\pagerange{\pageref{firstpage}--\pageref{lastpage}} \pubyear{2002}

\maketitle

\label{firstpage}

\begin{abstract}
A nebular analysis of the central Orion Nebula and its main structures is presented. We exploit MUSE integral field observations in the wavelength range 4595-9366 \AA\ to produce the first O, S and N ionic and total abundance maps of a region spanning 6' x 5' with a spatial resolution of 0.2". We use the S$_{23}$ ( = ([SII]$\lambda$6717,31+[SIII]$\lambda$9068)/H$\beta$) parameter, together with [OII]/[OIII] as an indicator of the degree of ionisation, to distinguish between the various small-scale structures. The only Orion Bullet covered by MUSE is HH 201, which shows a double component in the [FeII]$\lambda$8617 line throughout indicating an expansion, and we discuss a scenario in which this object is undergoing a disruptive event. We separate the proplyds located south of the Bright Bar into four categories depending on their S$_{23}$ values, propose the utility of the S$_{23}$ parameter as an indicator of the shock-contribution to the excitation of line-emitting atoms, and show that the MUSE data is able to identify the proplyds associated with disks and microjets. We compute the second order structure function for the H$\alpha$, [OIII]$\lambda$5007, [SII]$\lambda$6731 and [OI]$\lambda$6300 emission lines to analyse the turbulent velocity field of the region covered with MUSE. We find that the spectral and spatial resolution of MUSE is not able to faithfully reproduce the structure functions of previous works.

\end{abstract}


\section{Introduction}
The Orion Nebula (M 42) is the closest Galactic HII region and corresponds to one of the most observed objects in the sky. It therefore not only serves as a template for the comparison with observations of other (Galactic and extragalactic) HII regions, but also as a very good testing ground for new instruments. The number of papers written about M 42 in the last six decades is in the three digit regime, demonstrating how important and how well studied this region is. A thorough review of the main features, the geometry, the population of stars, the outflows and the main physical aspects of the HII region conditions is given in \cite{2001ARA&A..39...99O}.

\begin{figure*}
\centering
\includegraphics[scale=0.6]{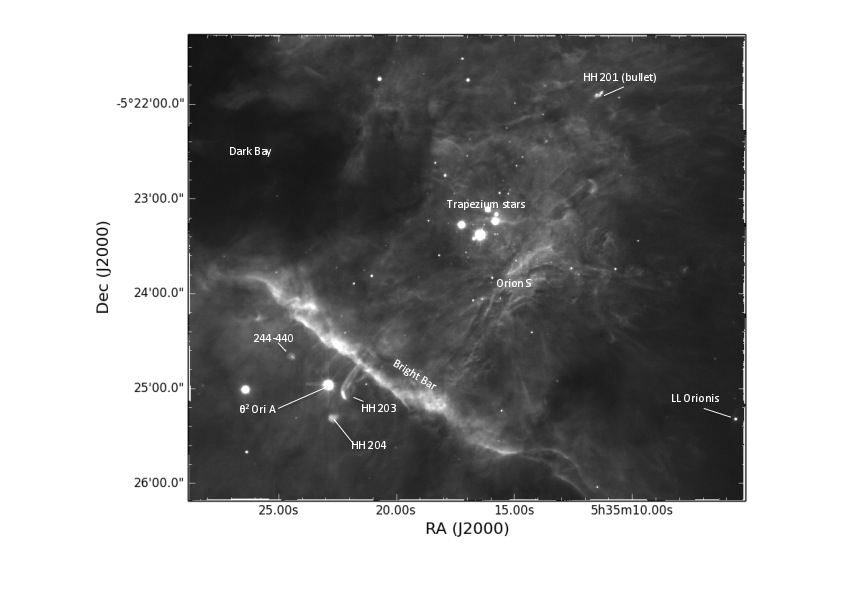}
\caption{The most commonly discussed features of the central Orion Nebula on a MUSE non-continuum subtracted, integrated intensity map of [SII]$\lambda$6717.}
\label{orienta}
\end{figure*}

The central part of M 42 is a treasure chest for star formation feedback studies: it is home to four massive stars (one O- and three B-type stars) that make up the so called Trapezium cluster, of which $\theta^{1}$ Ori C (of spectral type O7 \citealt{1954TrSht..25....1P}, \citealt{1988AJ.....95.1744V}) is the most luminous. These stars are ionising the surrounding material and giving rise to the vast number of detected nebular emission lines which trace ionisation fronts such as the Bright Bar and the Orion S cloud. Furthermore, because of the ongoing star formation in the region, it hosts many optical Herbig-Haro (HH) and molecular outflows, as well as a large population of young stellar objects called \textit{proplyds} (short for protoplanetary disk, \citealt{1994ApJ...436..194O}). The three-dimensional structure of this region is that of a blister-like HII region in front of the Orion Molecular Cloud (OMC-1), where the emitting region is $\sim$ 0.1 pc thick and 1 pc in lateral dimension \citep{2001ARA&A..39...99O}. The most common features of this region are shown in Fig. \ref{orienta}, in an integrated [SII]$\lambda$6717 map.

Because of its vicinity ($\sim$ 420 pc, \citealt{2014ApJ...786...29S}), in combination with the fact that it is associated with recent star formation and a comparatively high surface brightness, it is the perfect object to study elemental abundances and therefore help understand not only the chemical evolution of the interstellar medium (ISM), but also the process of nucleosynthesis. Many studies have been dedicated to the elemental abundances in the Orion Nebula using slit spectroscopy (e.g. \citealt{1992ApJ...389..305O}, \citealt{1998MNRAS.295..401E}, \citealt{2004MNRAS.355..229E}, \citealt{2008ApJ...675..389M}), finding that the abundances of heavy elements in Orion are only somewhat higher than solar ones. In the era of integral field spectroscopy, new studies have emerged that exploit the combination of simultaneous imaging and spectroscopy on large spatial and spectral scales (e. g. \citealt{2007A&A...465..207S}, \citealt{2011MNRAS.417..420M}). A common analysis in the study of HII regions is given by line ratios of nebular emission lines which are used as abundance tracers (e.g. [NII]/H$\alpha$, [SII]/H$\alpha$ tracing the nitrogen and sulphur abundances respectively) or as tracers of the degree of ionisation (e.g. [OII]/[OIII]). Another sulphur abundance parameter computed for both Galactic and extragalactic HII regions is given by S$_{23}$ = ([SII] + [SIII])/H$\beta$ \citep{1996MNRAS.280..720V}, which is commonly used to determine star formation histories and evolutionary scenarios, as this parameter is found to vary as a function of position within a galaxy. Together with [OII]/[OIII], S$_{23}$ can also be used to analyse the ionisation structure of HII regions \citep{2010MNRAS.408.2234G}. Together with a spatially resolved velocity map, the combination of the S$_{23}$ parameter and [OII]/[OIII] has been used by \cite{2015MNRAS.450.1057M} (MC15 henceforth) to detect a previously unknown outflow in the famous Pillars of Creation in the Eagle Nebula (M 16) in MUSE science verification observations, as the location where the outflow is currently emerging from the pillar material has distinctively high S$_{23}$ and low [OII]/[OIII] values. Future MUSE observations will determine whether this new method can be used to detect outflows in molecular cloud structures. In this work, we exploit this method to characterise the outflow and proplyd population in M 42.

The same nebular emission lines used to determine ionic and total abundances have also been used to study the turbulent motions in M 42 with a statistical approach by computing the second order velocity structure function, generally defined as $S_{2}(\textbf{r})=\langle|\textbf{v}(\textbf{r'})-\textbf{v}(\textbf{r''})|^{2}\rangle$ (where $\textbf{r}=\textbf{r'}-\textbf{r''}$ is the separation between any given pair of points), and comparing the shape of the structure function to that of theoretical models of turbulence (e. g. \citealt{1941DoSSR..30..301K}, \citealt{1951ZA.....30...17V}). The main studies about structure functions in Orion used the [OIII] \citep{1988ApJS...67...93C}, the [OI] \citep{1992ApJ...387..229O} and later the [SIII] lines \citep{1993ApJ...409..262W}, which show that the structure function has a steep slope at small values of $r$ and a transition scale after which the slope is shallower for larger values of $r$, except for [OI] where the slope is seen to remain constant over almost all measured scales. In general, the cited studies indicate similarities between the observations and the \citeauthor{1951ZA.....30...17V} predictions (see Section 4 and\cite{2001ARA&A..39...99O}). 

The new optical integral field unit (IFU) MUSE at the VLT offers, for the first time, a very powerful combination of sub-arcsecond spatial resolution and medium spectral resolution over a field of view of 1' $\times$ 1' and a large spectral range. With this instrument it is now possible to compute not only ionic and elemental abundance maps, but also compute velocity information from the same observations in an unbiased manner for a very large field. In this work we derive the abundance maps of oxygen, nitrogen and sulphur, analyse how the S$_{23}$ and [OII]/[OIII] emission line ratios vary across the central part of the Orion Nebula and how one can use them to distinguish between the different types of outflows. Furthermore, we attempt a kinematical analysis by computing the second order structure function. This work demonstrates the potential of IFU spectroscopy to probe different types of feedback mechanisms (ionisation, jets/outflows) with a combination of line ratios, kinematics and physical properties. The paper is organised as follows: we briefly present the observations in Section 2, compute and discuss the abundance and line ratio maps of the entire mosaic in Section 3; the structure functions are discussed in Section 4, while in Section 5 we present detailed studies of several selected regions. Finally, the conclusions are presented in Section 6.

\section[]{IFU observations}
\begin{figure}
\hspace{-15pt}
\includegraphics[scale=0.34]{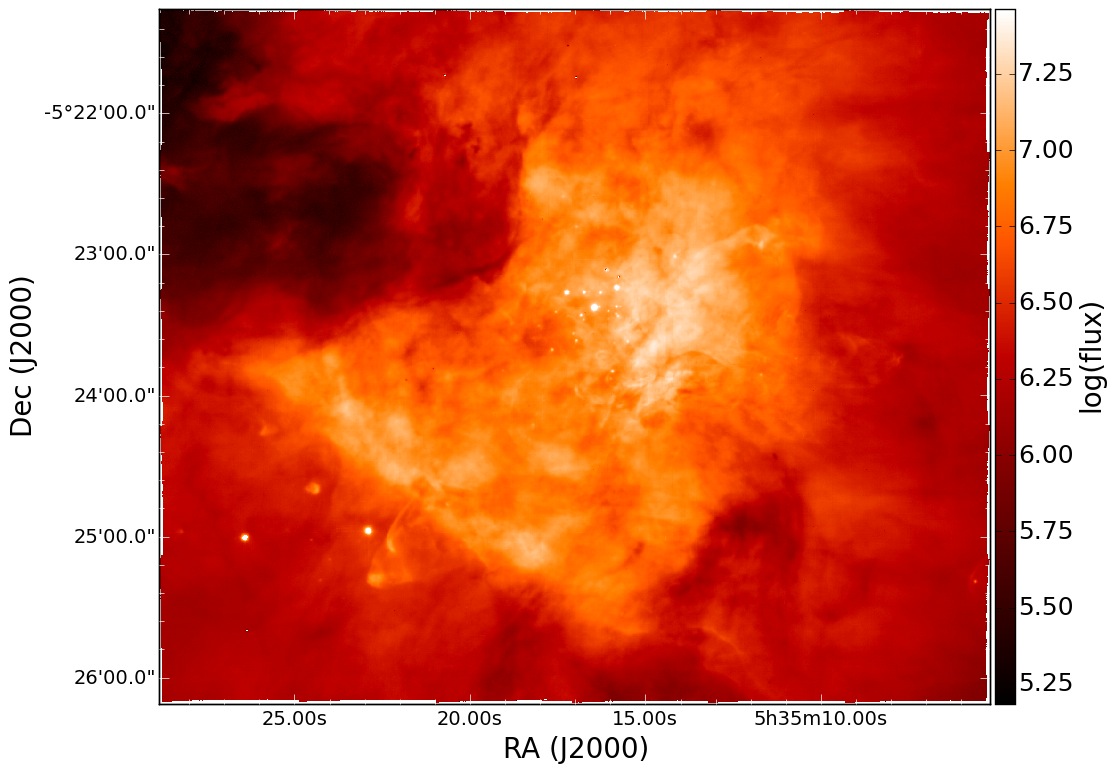}
  \caption{H$\alpha$ continuum-subtracted integrated intensity map, linearly auto-scaled (the flux is measured in 10$^{-20}$ erg s$^{-1}$ cm$^{-2}$ pixel$^{-1}$).}
  \label{ha}
\end{figure}

The MUSE integral field observations of the Orion Nebula were taken during the instrument's commissioning run \citep{2014Msngr.157...13B} on February 16th, 2014. The 6'$\times$5' mosaic consists of 60$\times$5 seconds exposures, where each of the 30 mosaic pointings was observed twice with a 90 degree rotation dither pattern. The data reduction was carried out in the E\textsc{so}R\textsc{ex} environment with the MUSE pipeline \citep{2012SPIE.8451E..0BW}. For a detailed description of the data reduction we refer to \citealt{2015A&A...582A.114W} (W15 henceforth). The observations were carried out in the Wide Field Mode with a field of view of 1'$\times$1', in the wavelength range 4595 - 9366 \AA\, and a sampling of 0.2"$\times$0.2"$\times$0.85 \AA.
An integrate intensity map of H$\alpha$ is shown in Fig. \ref{ha}. Maps of other emission lines, as well as maps of the most relevant physical parameters (e. g. electron density, electron temperature, extinction) are shown and discussed in W15. 

\section{Abundance maps}
\subsection{Ionic and total abundances}
Abundance determination in the Orion Nebula is supported by a long record of publications dating back to the last century (e. g. \citealt{1977MNRAS.179..217P}, \citealt{1993ApJ...413..242R}, \citealt{1991ApJ...374..580B}, \citealt{1998MNRAS.295..401E}, \citealt{2007A&A...465..207S}, \citealt{2011MNRAS.417..420M}), and it serves as the comparison ground for abundances in other extra- and galactic HII regions. One of the major problems in determining abundances is the dependance of line emissivity on the electron temperature T$_{e}$: the emissivity of recombination lines (RLs) decreases with increasing T$_{e}$, while the emissivity of collisionally excited lines (CELs) increases with T$_{e}$ in an exponential manner \citep{2001ARA&A..39...99O}. This means that in the case of temperature inhomogeneity along lines of sight (as is the case for Orion, \citealt{1967ApJ...150..825P}) different regions will be more or less sensitive to RLs or CELs, and ratios of RLs are therefore preferentially used to determine abundances because of their weaker dependance on temperature. As the RLs covered by MUSE (OII$_{\lambda4650}$ and OII$_{\lambda4661}$) tend to be very weak and noise-dominated, in this work we only make use of CEL line ratios to determine abundances.  

The large spatial and spectral coverage of MUSE offers the first unbiased possibility of computing ionic abundance maps for many atoms simultaneously. For this, we used integrated intensity maps, corrected for extinction as described in MC15 with R$_{V}\approx$ 5.5, using the python package \textsc{pyneb} \citep{2015A&A...573A..42L} together with the electron density and temperature maps derived in W15. The atomic data used for the computations are shown in Table \ref{atomdata}. The oxygen abundance ratios O$^{+}$/H$^{+}$ and O$^{++}$/H$^{+}$ were computed from the [OII]$\lambda$7320,30 and [OIII]$\lambda$4959,5007 lines respectively and assuming T([OII])$\sim$ T([NII]) and T([OIII]) $\simeq$ T([SIII]), as MUSE does not cover the [OII]$\lambda$3727,29 and [OIII]$\lambda$4363 lines to compute T([OII]) and T([OIII]). The [SII] electron density map was used for the abundance determination. The [OII]$\lambda$7320,30 lines are generally low in intensity and can suffer from the contamination of OH rotational line emission at 7330 \AA, and, as stated in W15, the MUSE Orion data cube was not corrected for sky background. However, the nebular emission in the central Orion Nebula is very bright compared to the sky, and the maximal contribution of sky emission to the measurements is of 5\% to the [OII] lines. Because in the outer regions (e.g. the Dark Bay) the [OII] lines, as well as the [SII] and the [NII]$\lambda$5755 lines are significantly weaker and the level of noise very high, we masked the emission line maps used to compute the abundance based on the flux of the weakest lines of the three species ([OII]$\lambda$7330$>$4$\times$10$^{-16}$ erg s$^{-1}$ cm$^{-2}$ pixel$^{-1}$, [SII]$\lambda$6717$>$4$\times$10$^{-16}$ erg s$^{-1}$ cm$^{-2}$ pixel$^{-1}$, [NII]$\lambda$5755$>$4$\times$10$^{-17}$ erg s$^{-1}$ cm$^{-2}$ pixel$^{-1}$). The sulphur abundance was computed with [SII]$\lambda$6717, [SII]$\lambda$6731 and [SIII]$\lambda$9068, while for nitrogen we made use of the [NII]$\lambda$5755, [NII]$\lambda$6548 and [NII]$\lambda$6584 lines. To account for undetected lines corresponding to higher ionisation states we adjust the abundance determination with the appropriate ionisation correction factors (ICF), according to \cite{2008MNRAS.383..209H} for sulphur and \cite{1998MNRAS.295..401E} for nitrogen.

\begin{figure*}
\mbox{
\subfloat[]{\includegraphics[scale=0.35]{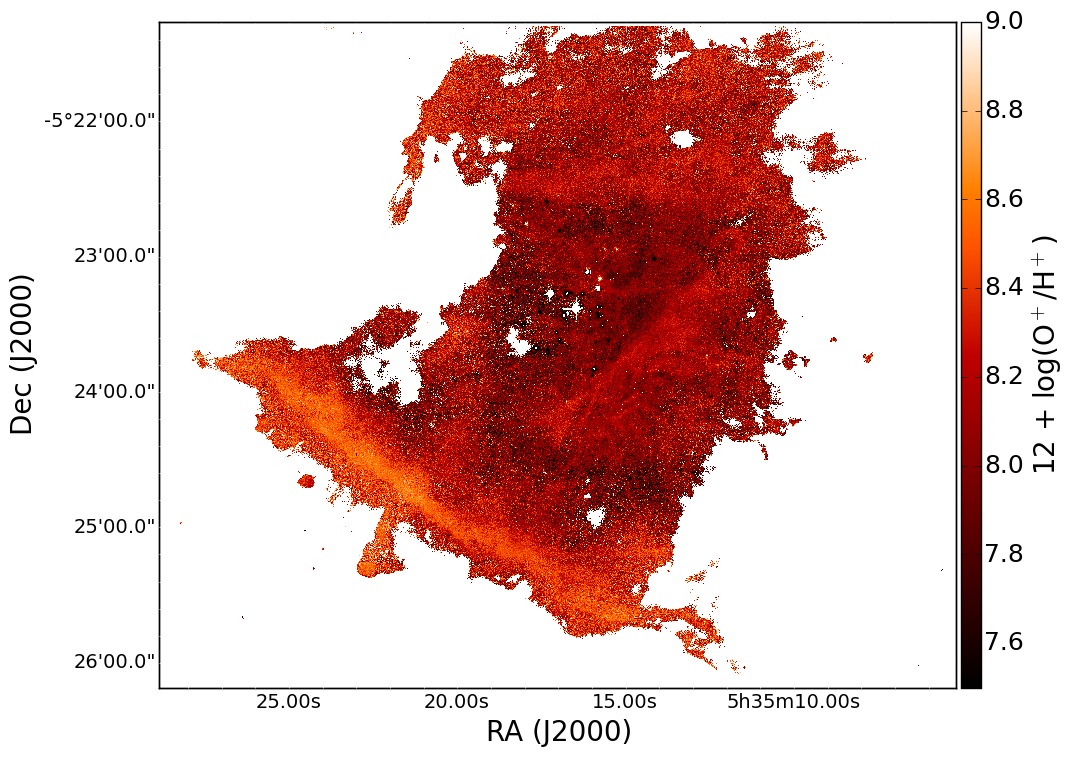}}
\subfloat[]{\includegraphics[scale=0.35]{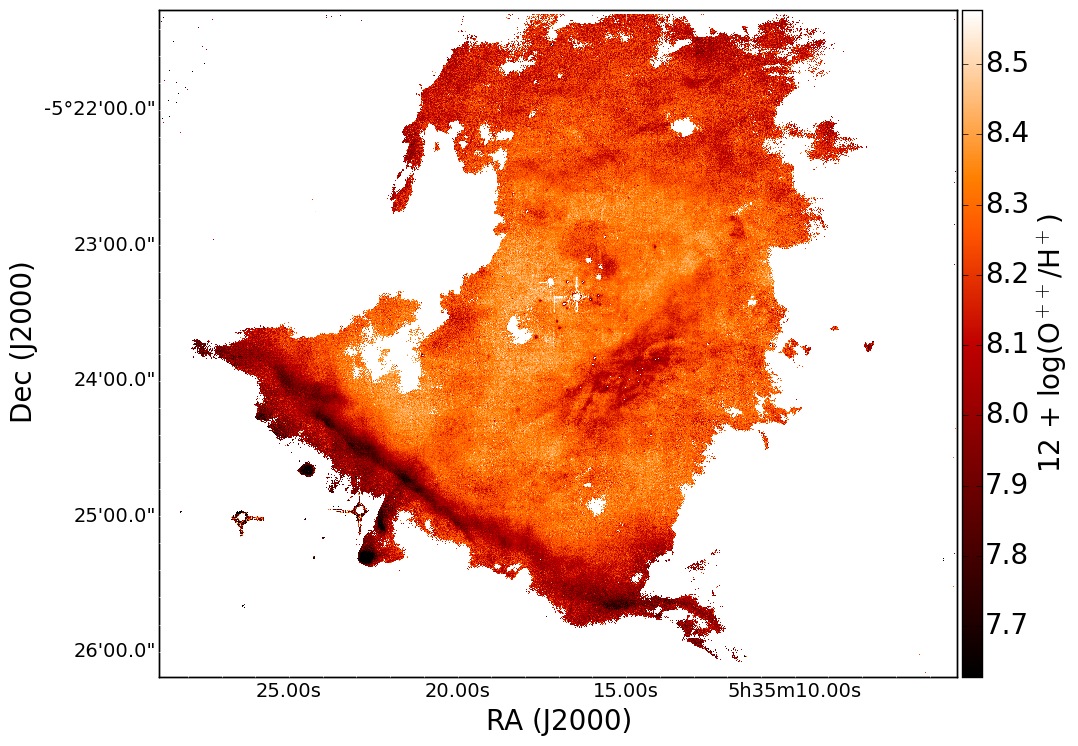}}}
\mbox{
\subfloat[]{\includegraphics[scale=0.35]{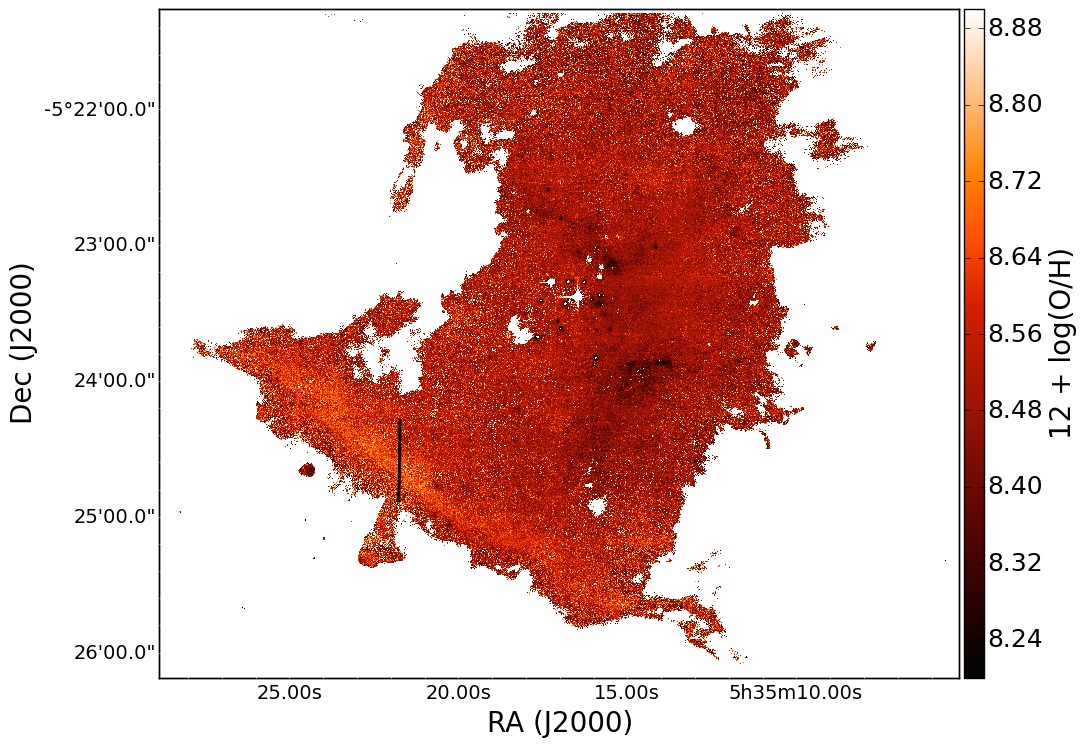}}}
  \caption{Maps of the O$^{+}$ (panel a), O$^{++}$ (panel b) and total O (panel c) abundances, see text Section 3.1. The black line in panel (c) marks the position of the slit used to compute the profiles shown in Fig. \ref{prof}.}
  \label{O_abun}
\end{figure*}

\begin{figure*}
\mbox{
\subfloat[]{\includegraphics[scale=0.35]{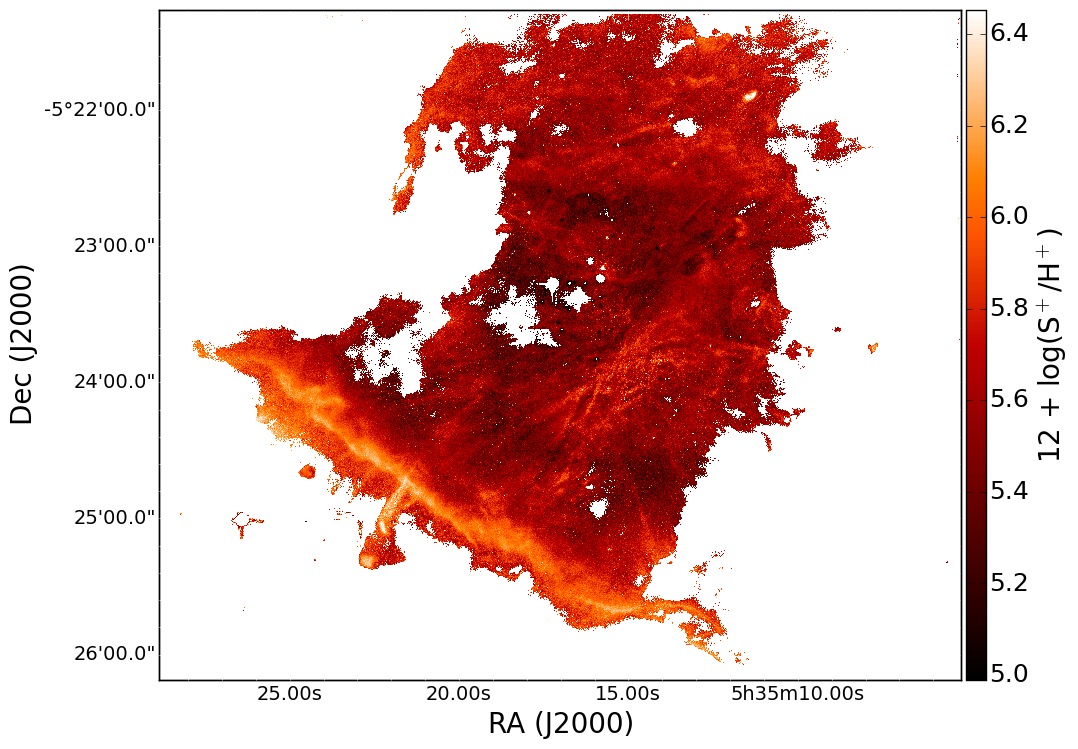}}
\subfloat[]{\includegraphics[scale=0.35]{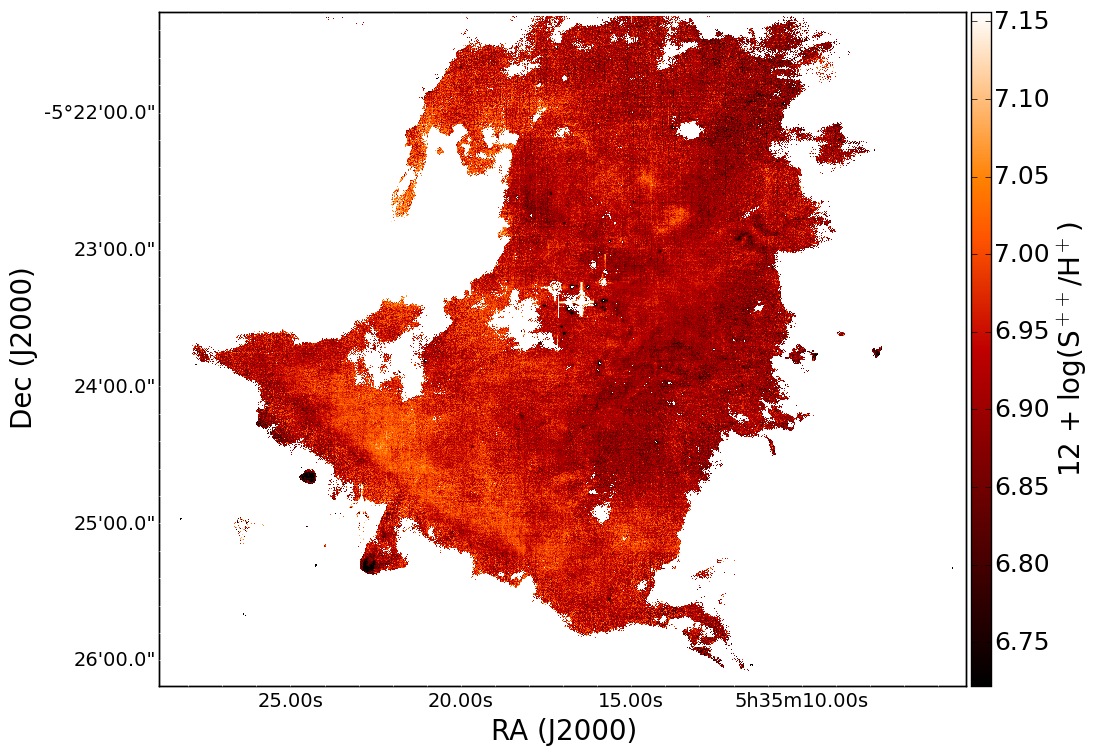}}}
\mbox{
\subfloat[]{\includegraphics[scale=0.35]{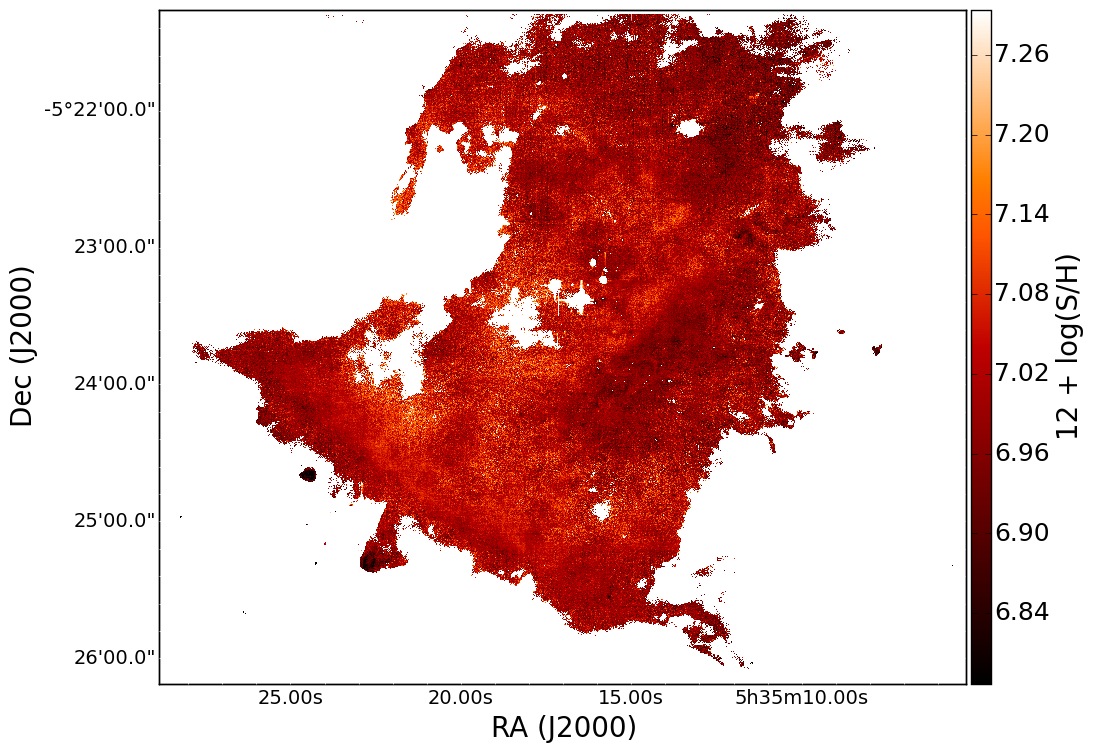}}}
  \caption{Maps of the S$^{+}$ (panel a), S$^{++}$ (panel b) and total S (panel c) abundances, see text Section 3.1.}
  \label{S_abun}
\end{figure*}

\begin{figure*}
\mbox{
\subfloat[]{\includegraphics[scale=0.35]{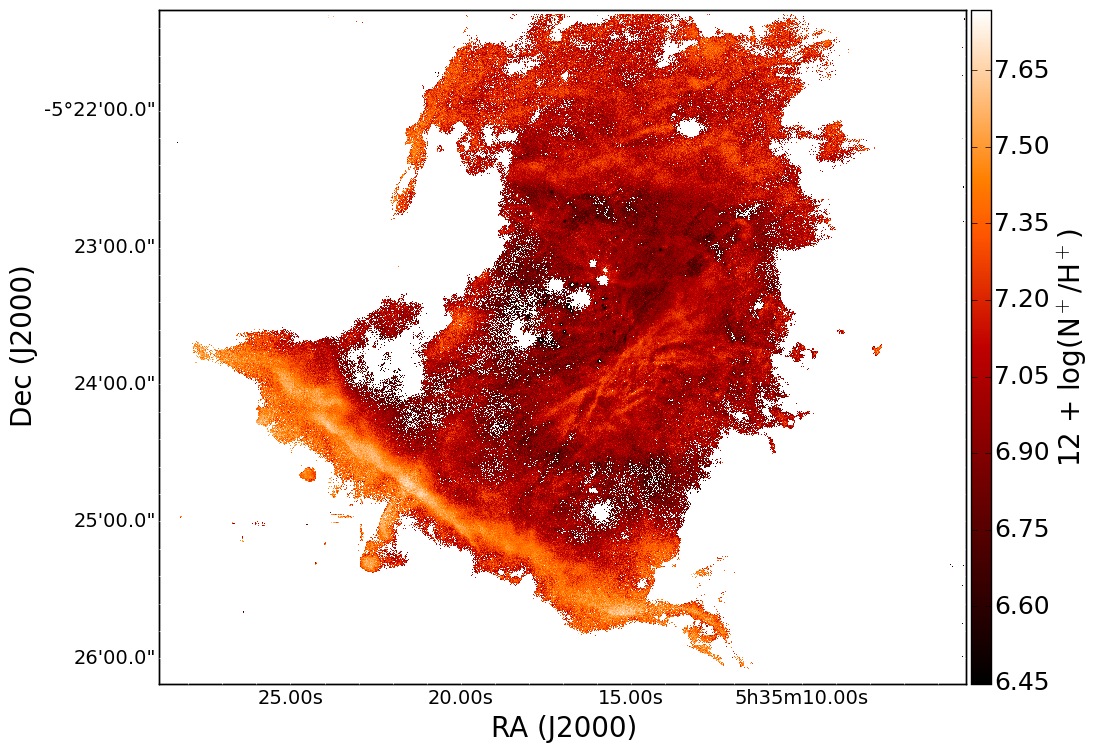}}
\subfloat[]{\includegraphics[scale=0.35]{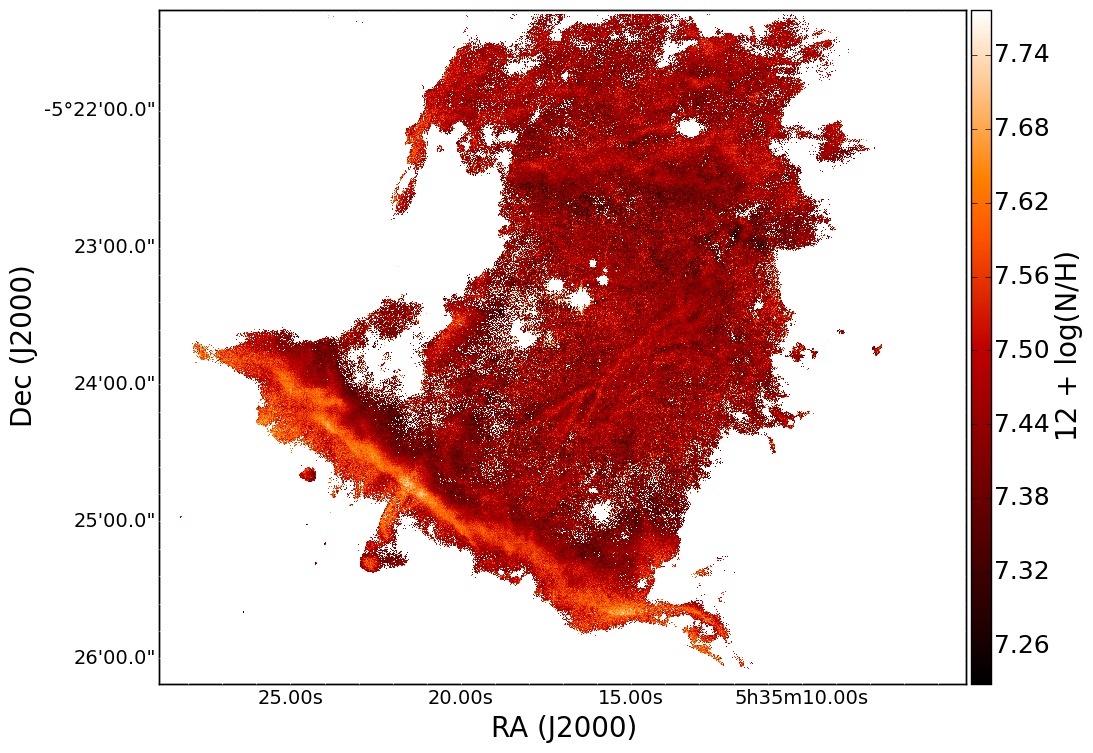}}}
  \caption{Maps of the N$^{+}$ (panel a) and total N (panel b) abundances, see text Section 3.1.}
  \label{N_abun}
\end{figure*}

The abundance maps are shown in Fig. \ref{O_abun} through Fig. \ref{N_abun} (the stellar emission has been removed from these maps by fitting and subtracting the continuum on a  pixel-by-pixel basis, stars that were saturated appear as white in the images because they are masked out). Mean values for circular regions (with a radius of 2.5") of the Trapezium cluster and the Bright Bar, as well as of a 8.5"x3" box as to match the slit Position 2 in \cite{1998MNRAS.295..401E} are shown in Table \ref{abun} (to distinguish between our extraction and the actual slit used in \citealt{1998MNRAS.295..401E}, we label the latter as P2E and the region used for this work as P2). Also shown in the same table are ionic and total abundances obtained by \cite{2004MNRAS.355..229E} (henceforth referred to as E04) for P2.

\begin{table}
\scriptsize
\begin{tabular}{lcc}
\hline
\hline
Ion & Transition & Collision \\
 & probability & strength\\
 \hline
 O$^{+}$ & \cite{1982MNRAS.198..111Z} & \cite{2006MNRAS.366L...6P}\\
  & \cite{1996atpc.book.....W} & \cite{2007ApJS..171..331T} \\
O$^{++}$ & \cite{1996atpc.book.....W} & \cite{1999ApJS..123..311A} \\
 & \cite{2000MNRAS.312..813S} & \\
S$^{+}$ & \cite{podo} & \cite{2010ApJS..188...32T}\\
S$^{++}$ & \cite{podo} & \cite{1999ApJ...526..544T}\\
N$^{+}$ & \cite{1997AAS..123..159G} & \cite{2011ApJS..195...12T}\\
\hline
\end{tabular}
\caption{Atomic data used in \textsc{pyneb} for the ionic abundance determination.}
\label{atomdata}
\end{table}

The total abundance maps computed from the MUSE data show a certain degree of structure: the Bright Bar, the Orion S region, the HH 203 and 204 objects, as well as some of the proplyds are clearly seen in the maps. The fact that the abundances are not constant across the nebula but show a certain degree of structure has already been discussed in \cite{2011MNRAS.417..420M} (henceforth referred to as MD11) and \cite{2012MNRAS.421.3399N}. Specifically, MD11 find that for the Bright Bar (BB) their mean value of 8.49$\pm$0.03 for the O/H ratio agrees with previous estimates of 8.50 (e.g. \citealt{1998MNRAS.295..401E}, \citealt{2006ApJ...644.1006B}), that the the spatial variation of O/H in the BB varies within the computed typical error, while the range of O/H values is slightly higher for the Orion S region. They also find a coincidence of the lowest O$^{++}$/H$^{+}$ values with the largest electron temperature uncertainties, the highest total oxygen abundance values in spatial agreement with the higher ionic O$^{+}$/H$^{+}$ abundance values, and a structural similarity of O$^{+}$/H$^{+}$ and O/H to the electron density (N$_{e}$) map, indicating a strong dependance of the oxygen abundance on the physical parameters. They suggest two possible explanations for this: (i) possibly, the N$_{e}$ derived from the [SII] lines is not appropriate for the determination of O$^{+}$/H$^{+}$, as it does not reflect the true density of the O$^{+}$ zone; (ii) although reflecting the physical conditions of both S$^{+}$ and O$^{+}$, the computed N$_{e}$ values do not correspond to the true values in regions like the BB or Orion S, as in these regions the density approaches the critical density of the [SII] and [OII] lines, and higher N$_{e}$ values are computed. In fact, these authors discuss how, by correcting the N$_{e}$ to lower values, they recover the mean value for the O abundance. Without any additional information about specifically where to apply such correction, we show uncorrected maps.
 
To test the reliability of the abundances obtained in this work, we extract a 16"x16" sub-region of the BB, matching the observations of MD11. For the entire sub-region, we find a mean O/H ratio of 8.60$\pm$0.09, which is higher than the typical Orion O/H ratio of 8.50. We also compare our temperature, density, O$^{+}$/H$^{+}$ and O$^{++}$/H$^{+}$ of the sub-region with MD11, and find that our N$_{e}$ map (derived from the [SII] lines) shows: (i) lower values, typically of about 1000-2000 cm$^{-3}$; (ii) a noise-affected T$_{e}$ ([NII]) map; (iii) a comparable O$^{++}$/H$^{+}$ map; (iv) and higher O$^{+}$/H$^{+}$ as well as O/H values. We speculate that the higher O$^{+}$/H$^{+}$ values found for the BB in this work are a result of the noisy T$_{e}$ and [OII] maps. This could be tested with data with higher S/N.

\begin{figure}
\mbox{
\hspace{-30pt}
\subfloat[]{\includegraphics[scale=0.45]{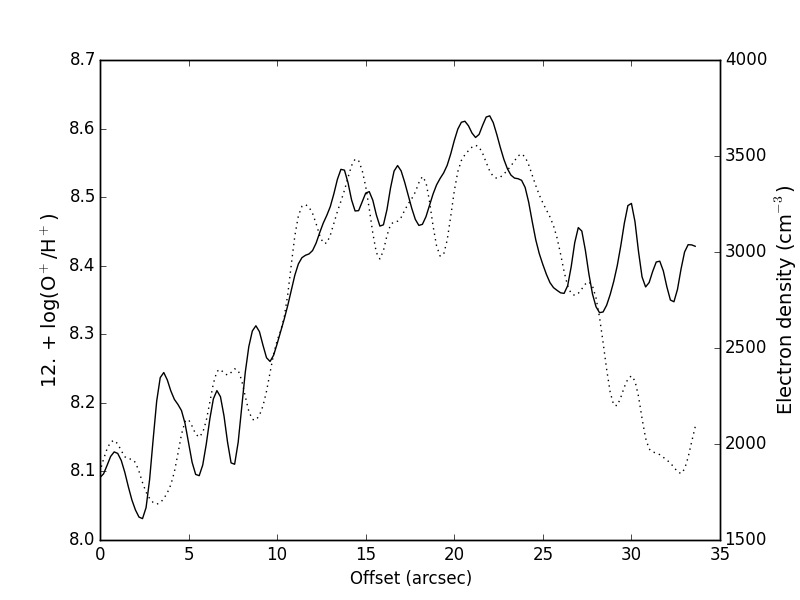}}}
\mbox{
\hspace{-30pt}
\subfloat[]{\includegraphics[scale=0.45]{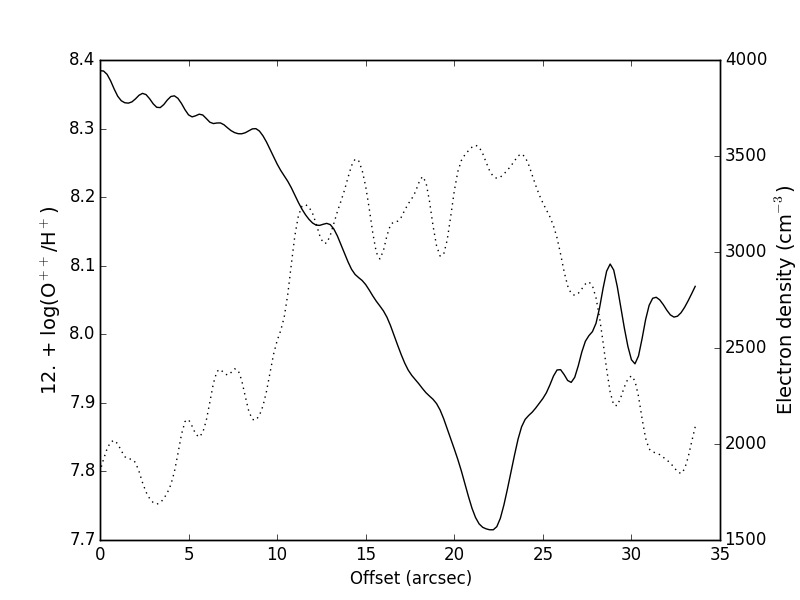}}}
\mbox{
\hspace{-30pt}
\subfloat[]{\includegraphics[scale=0.45]{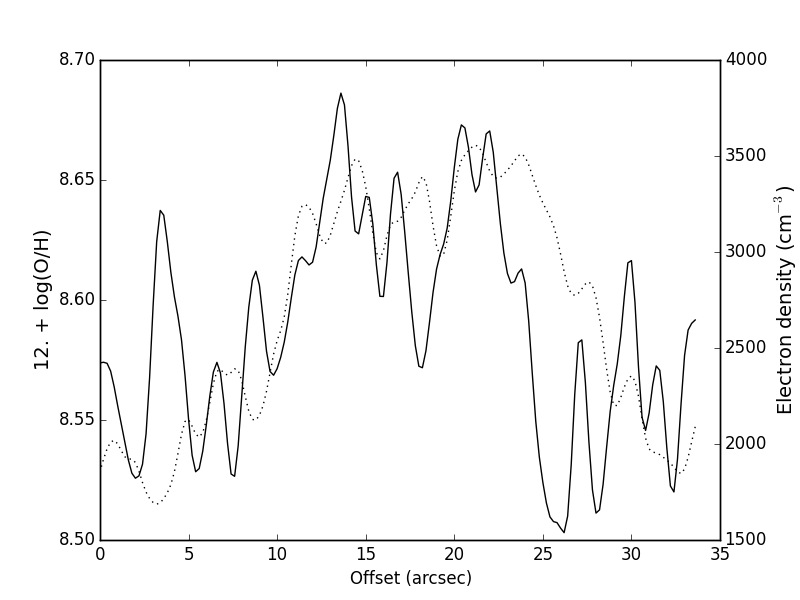}}}
  \caption{Profiles of the O$^{+}$ (top), O$^{++}$ (middle) and O abundances along the slit (along the north-south direction) shown in Fig. \ref{O_abun} (solid lines), together with the profile of the electron density derived from the [SII] lines (dotted lines) along the same slit (marked in Fig. \ref{O_abun}c). The originally very noisy profiles were smoothed with a Gaussian kernel. See text Section 3.1.}
  \label{prof}
\end{figure}

To demonstrate the strong dependence of the ionic and total abundances on the electron density, in Fig. \ref{prof} we plot profiles (smoothed with a Gaussian kernel) of these along a slit positioned on the Bright Bar (the slit is shown in Fig. \ref{O_abun}c, the profile is computed in direction from north to south): the O$^{+}$ and the O abundances show higher values in correlation with the higher N$_{e}$ values, while the opposite trend is the case for O$^{++}$. However, we cannot resolve this issue by artificially lowering N$_{e}$. Furthermore, the structuring is also seen in S$^{+}$/H$^{+}$, N$^{+}$/H$^{+}$ and N/H: we therefore suggest that the dominant mechanism that leads to the structures seen in the abundance maps is (as already discussed in MD11) that in regions like the Bright Bar and the Orion S cloud, densities approach the critical densities of [SII], [NII] and [OII] lead to untrustworthy artefacts.

Table \ref{abun} shows that the abundances found in this work do not agree with E04 (Table \ref{temden} specifies the electron density and temperatures extracted from the same regions listed in Table \ref{abun}: for P2 we find a lower N$_{e}$ as the assumed density of E04, but higher T$_{e}$ values): in this work, O${+}$ is overestimated by about 0.22 dex, while O$^{++}$ and O are underestimated by about 0.15 and 0.04 dex respectively; S$^{+}$ is overestimated by 0.16 dex, S$^{++}$ and S are underestimated by 0.09 and 0.04 dex respectively; for nitrogen, N$^{+}$ is overestimated by 0.10 and N underestimated by 0.17 dex. Summarising this, we see how the abundances computed via \textsc{pyneb} from the MUSE data overestimate the low ionisation potential ions (O$^{+}$, S$^{+}$, N$^{+}$), and they underestimate the ions with high ionisation potential as well as the total abundances (O$^{++}$, O, S$^{++}$, S, N). However, within the margin of errors, the MUSE \emph{total} abundance maps are able to recover literature total O and S abundance measurements. This is not the case for N, as only one ionisation state is observed and the total abundance computation relies on the ICF.

\begin{table*}
\begin{center}
\scriptsize
\caption{Ionic and total abundances. Mean values for circular regions with 2.5'' radius centred on selected regions as well as slit Position 2 from Esteban et al., 2004, obtained in this work (P2) and abundances computed in Esteban et al. (P2E) for t$^{2}=$0.000. The last line indicates the approximate ranges of abundance values found in MD11 from integral field observations of a sub-region of the Bright Bar. See text Section 3.1.}
\begin{tabular}{lccccccccc}
\hline
\hline
Region & Coordinates & O$^{+}$ & O$^{++}$ & O & S$^{+}$ & S$^{++}$ & S & N$^{+}$ & N \\
 & (J2000) & & & & & & & &  \\
\hline
Bright Bar & 5 35 21.91 -5 24 38.20 & 8.56$\pm$0.08 & 7.93$\pm$0.05 & 8.65$\pm$0.06 & 6.11$\pm$0.06 & 6.98$\pm$0.03 & 7.04$\pm$0.03 & 7.51$\pm$0.04 & 7.60$\pm$0.03 \\
Trapezium & 5 35 16.4 -5 23 11.9 & 8.09$\pm$0.14 & 8.31$\pm$0.04 & 8.52$\pm$0.06 & 5.53$\pm$0.09 & 6.92$\pm$0.03 & 7.01$\pm$0.04 & 7.01$\pm$0.07 & 7.44$\pm$0.05  \\
P2 & 5 35 14.54 -5 23 33.86 & 7.98$\pm$0.12 & 8.28$\pm$0.05 & 8.47$\pm$0.04 & 5.56$\pm$0.15 & 6.92$\pm$0.02 & 7.02$\pm$0.03 & 7.00$\pm$0.10 & 7.48$\pm$0.05  \\
P2E, t$^{2}=$0.000 & 5 35 14.54 -5 23 33.86 & 7.76$\pm$0.15 & 8.43$\pm$0.01 & 8.51$\pm$0.03 & 5.40$\pm$0.06 & 7.01$\pm$0.04 & 7.06$\pm$0.04 & 6.90$\pm$0.09 & 7.65$\pm$0.09 \\
P2-P2E & & 0.22 & -0.15 & -0.04 & 0.16 & -0.09 & -0.04 & 0.10 & -0.17 \\
\hline
MD11 & Bright Bar & 8.00-8.47 & 7.86-8.24 & 8.41-8.58 & & & & &  \\
\hline
\end{tabular}
\label{abun}
\end{center}
\end{table*}

\begin{table}
\begin{center}
\caption{Electron densities and temperatures extracted from the same regions listed in Table \ref{abun} for this work and slit position 2 in Esteban et. al 2004 (P2E).}
\begin{tabular}{lccc}
\hline
\hline
Region & N$_{e}$ ([SII]) & T$_{e}$ ([NII]) & T$_{e}$ ([SIII])\\
 & (cm$^{-3}$) & (K) & (K)   \\
\hline
Bright Bar & 3430$\pm$430 & 9410$\pm$260 & 8870$\pm$190\\
Trapezium & 4530$\pm$1080& 10090$\pm$525 & 8780$\pm$210\\
P2 & 7310$\pm$3320 & 10535$\pm$520 & 8950$\pm$180\\
(this work) & & &  \\
P2E, t$^{2}=$0.000 & 8900$\pm$200& 10000$\pm$400 & 8320$\pm$40 \\
(Esteban et al. & & &\\
2004) & & & \\
\hline
\end{tabular}
\label{temden}
\end{center}
\end{table}

\subsection{Line ratios}
\begin{figure}
\hspace{-20pt}
\includegraphics[scale=0.45]{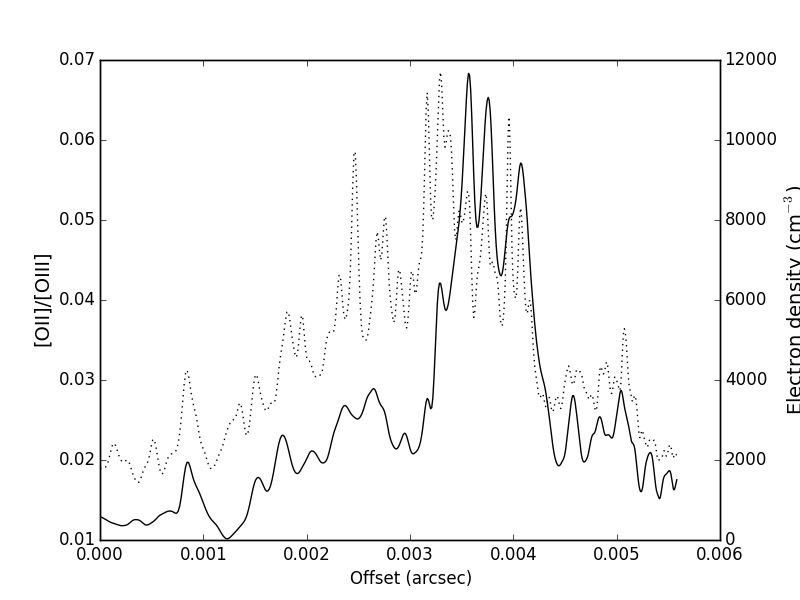}
\caption{[OII]/[OIII] and N$_{e}$ along the slit shown in Fig. \ref{SO23} (see text Section 3.2).}
\label{O23}
\end{figure}

\begin{figure*}
\mbox{
\hspace{-15pt}
\subfloat[]{\includegraphics[scale=0.34]{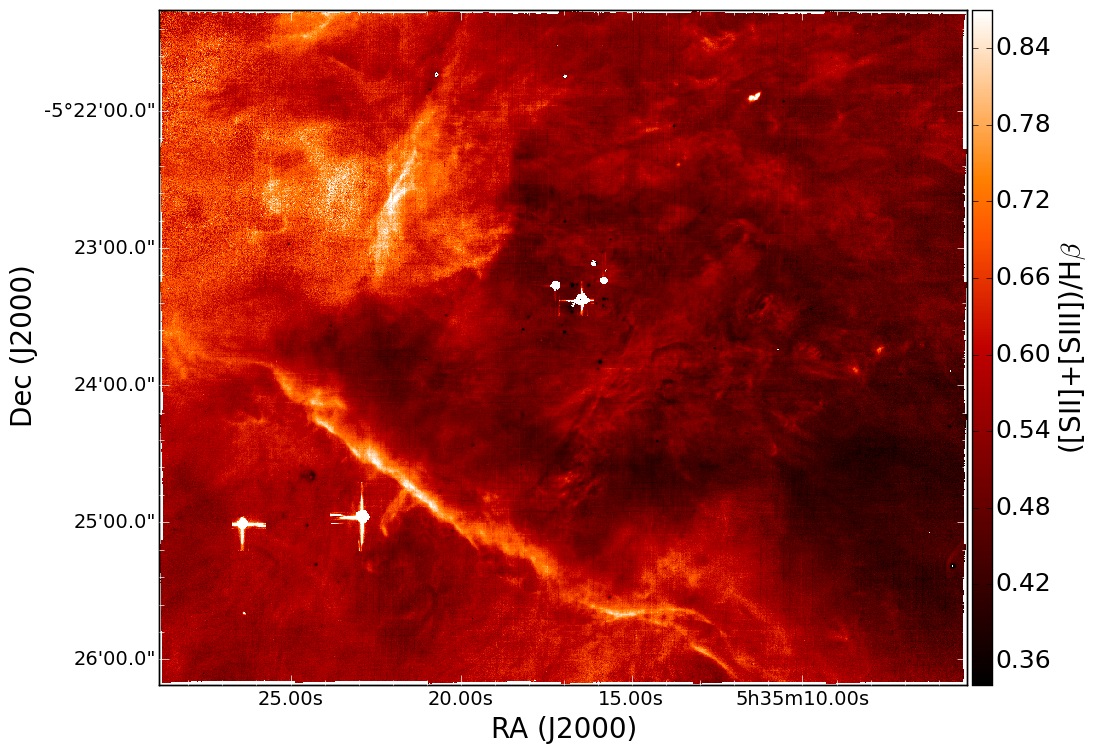}}
\subfloat[]{\includegraphics[scale=0.34]{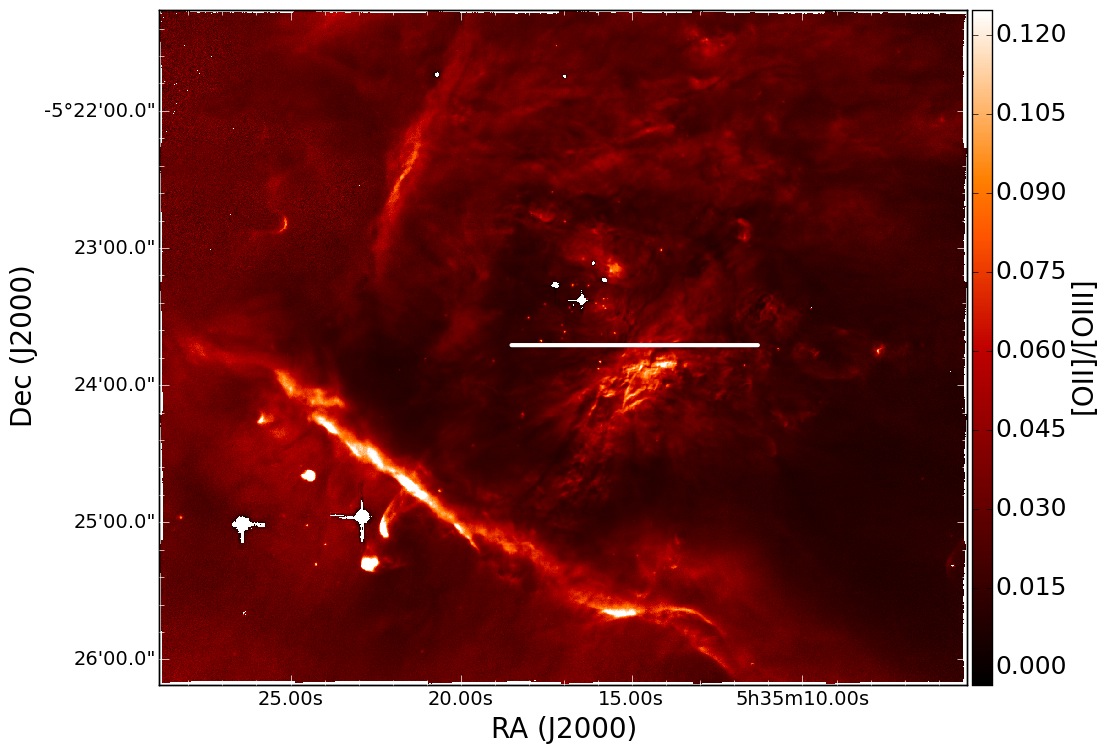}}}
\mbox{
\subfloat[]{\includegraphics[scale=0.5]{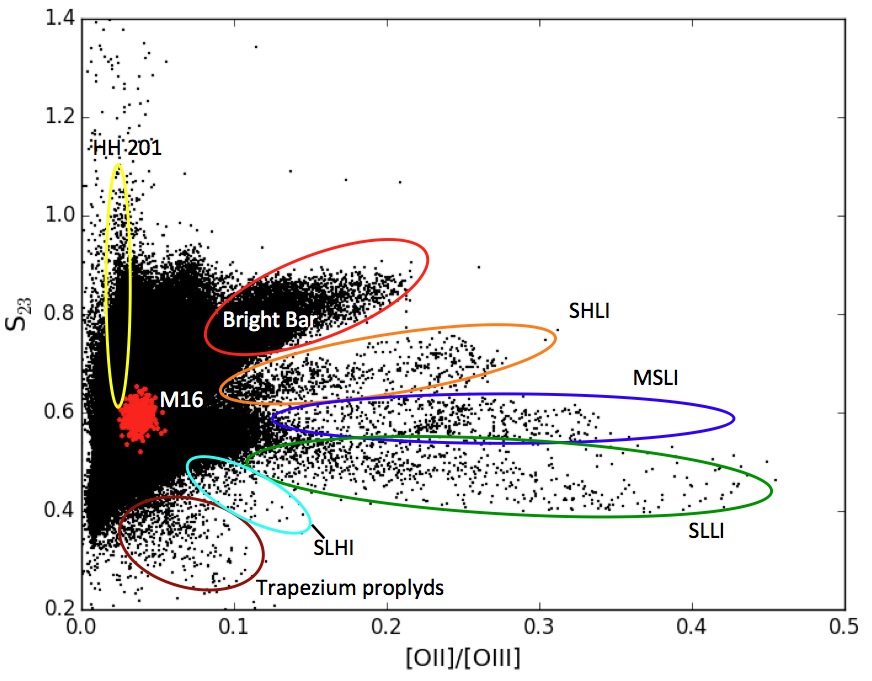}}}
  \caption{Continuum-subtracted maps of the S$_{23}$ ( = ([SII] + [SIII])/H$\beta$, panel a) and [OII]/[OIII] (panel b) parameters, linearly scaled. The white line in panel (b) indicates the slit used for Fig. \ref{O23}. Scatter plot of the two (black points, panel c) and the indicative positions of the following structures: the M 16 outflow (from MC15, red dots), the Bright Bar (red), HH 201 (yellow), the proplyds in the Trapezium cluster region (dark red), and the four classes of proplyds and outflows south of the Bright Bar (SHLI = S$_{23}$-high and low ionisation, MSLI = medium S$_{23}$ values and low ionisation, SLLI = S$_{23}$-low and low ionisation, SLHI = S$_{23}$-low and high ionisation, see text Section 5.2).}
  \label{SO23}
\end{figure*}

In our MUSE observations of pillar-like structures in the the Eagle Nebula (see MC15), the S$_{23}$ ( = ([SII]$\lambda$6717,31+[SIII]$\lambda$9068)/H$\beta$) parameter was used in combination with [OII]$\lambda$7320,30/[OIII]$\lambda$5007 (as an indicator of the degree of ionisation) and a velocity map obtained from the same dataset to identify a previously unknown outflow in one of the pillars. However, in MC15 we did not attempt a physical explanation of the empirical fact that S$_{23}$ can be used to trace outflows. Here, we repeated the S$_{23}$ vs. [OII]/[OIII] analysis, motivated by the fact that the central Orion Nebula hosts a very peculiar combination of massive ionising stars and a wealth of outflows, forming stars in the form of proplyds, and HH objects. We therefore refine the empirical S$_{23}$ vs. [OII]/[OIII] analysis of MC15 by exploring the rich environment of the Orion Nebula. Because the relative importance of photoionisation and shocks to the production of sulphur emission lines is different for different kind of objects (the Bar, HH outflows and proplyds), it can be used, together with [OII]/[OIII], to distinguish between them (as will be discussed in Section 5.2).ÊThe S$_{23}$ map is shown in panel (a) of Fig. \ref{SO23}, panel (b) shows the [OII]/[OIII] map, and panel (c) corresponds to a scatter plot of the two parameters. The maps and scatter plot shown in Fig. \ref{SO23} are continuum-subtracted: the strong stellar residuals of the saturated stars appear as white in the images because it is masked. Furthermore, the white line in Fig. \ref{SO23}b shows the slit used to compute the profile shown in Fig. \ref{O23} and discussed below.

We applied a technique called \textit{brushing}\footnote{This method was also applied in MC15. It is also known as \textit{graphical exploratory data analysis}, and it allows the user to manually select specific data points from an image or a plot by interactively drawing regions on the latter of the two.} to analyse the spatial correspondence of the data points in the scatter plot of S$_{23}$ vs. [OII]/[OIII] and trace them back to their position in the maps. Both in the [OII]/[OIII] and the S$_{23}$ maps the structures of the nebulosity are clearly distinguishable, e.g. the Bright Bar shows high values of both parameters, while the central part of the HII region is marked with lower values of both parameters, meaning that in the vicinity of the Trapezium cluster we find lower S$_{23}$ values and higher degrees of ionisation, as is expected. 

In Fig. \ref{SO23}c S$_{23}$ vs. [OII]/[OIII] is shown, and several of the main features are (indicatively) highlighted\footnote{Because of the large number of data points, we refrained from properly colour-coding single points in the scatter plot and trace them back to their spatial origin in the S$_{23}$ map as was done for HH 201 in Fig. \ref{bulletSO23}.}: the Orion bullet HH 201 (yellow ellipse) covered by the MUSE field shows high S$_{23}$ values as well as a high degree of ionisation, while the proplyds in the vicinity of the Trapezium stars (dark red ellipse) are found in a region with very low S$_{23}$ values but a rather high degree of ionisation; the HH objects and proplyds south of the Bright Bar clearly stand out, displaying a wide range in S$_{23}$ and ionisation values; furthermore, they can be divided into four distinct classes (described in Section 5.2), marked by the orange, blue, green and cyan ellipses; the Bright Bar is marked by the red ellipse. 

\begin{figure*}
\mbox{
\subfloat[]{\includegraphics[scale=0.5]{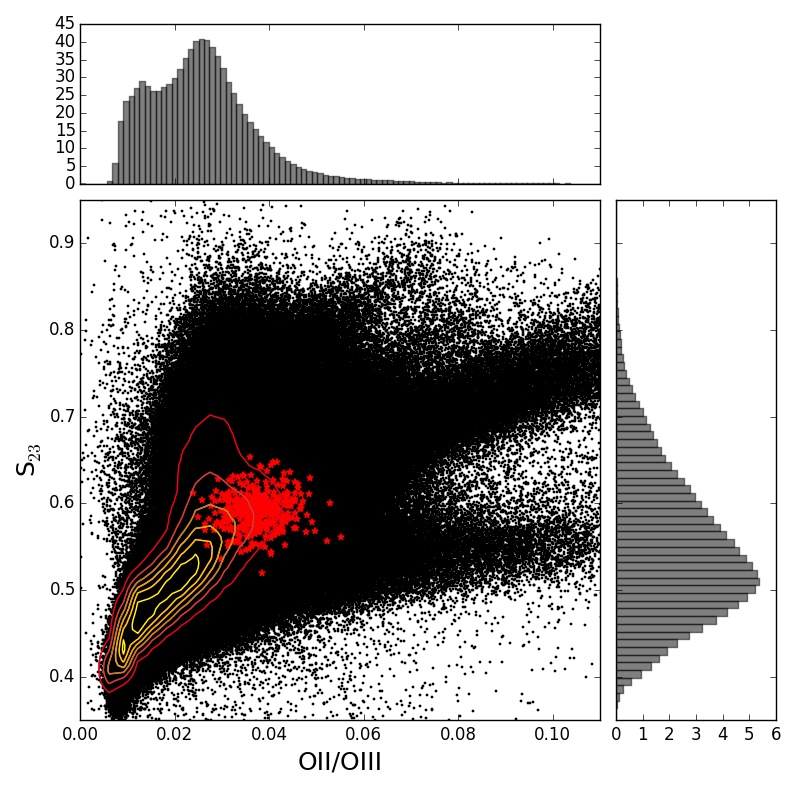}}}
\mbox{
\subfloat[]{\includegraphics[width=9.4cm, height=6.3cm]{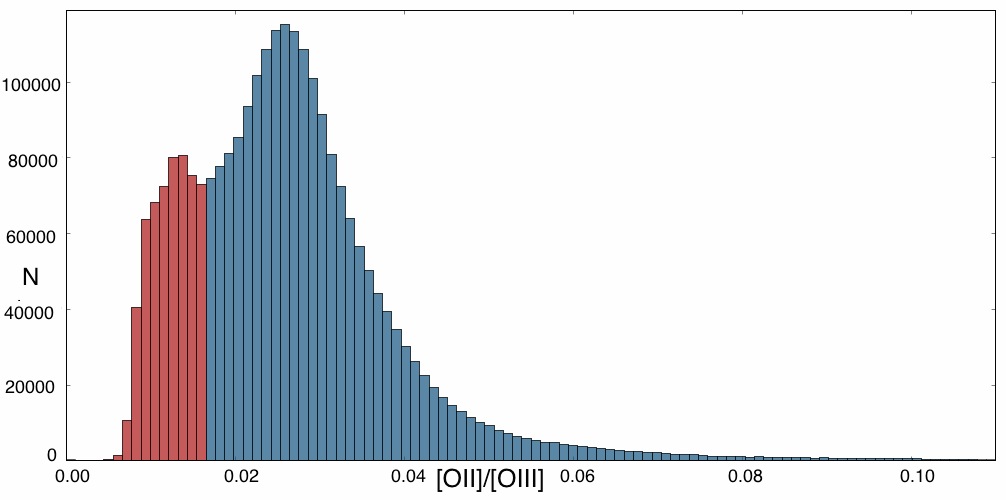}}
\subfloat[]{\includegraphics[width=9.4cm, height=6.3cm]{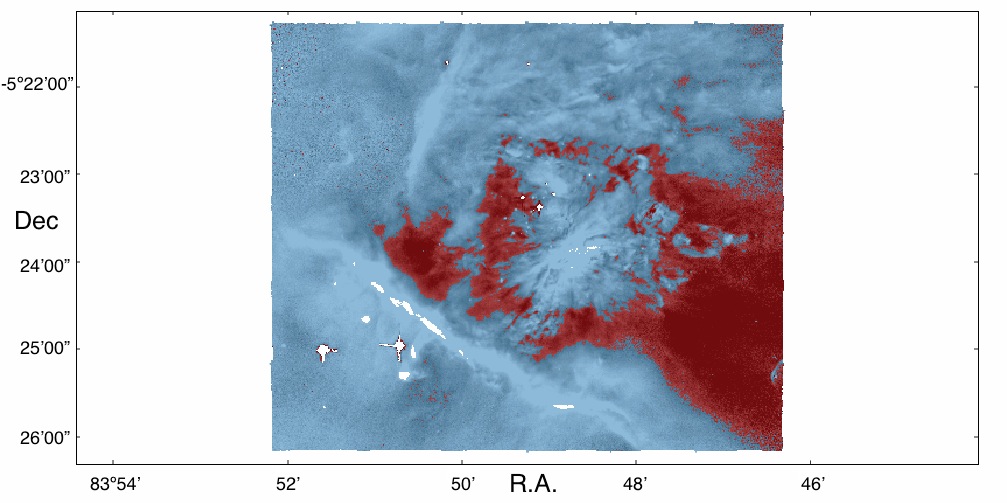}}}
  \caption{Panel (a)  zoom-in on the dense region of Fig. \ref{SO23}c  the contours correspond to the number of data points, the first contour level is $N=$ 3000 and the increment is 3000, and the red stars correspond to the M 16 outflow (MC15). The histograms are normalised counts. Panels (b) and (c) illustrate the spatial origin of the bimodal distribution of the [OII]/[OIII] histogram  highest (red) and lower (blue) degree of ionisation (the background [OII]/[OIII] histogram in panel b is the same as at the top of panel a, the red and the blue are the normalised histograms corresponding to the same-shaded areas in panel c). See text Section 3.2. }
  \label{fancy}
\end{figure*}

Fig. \ref{fancy}a is a zoom-in on the region of the S$_{23}$ vs [OII]/[OIII] parameter space where most of the data points lie. The [OII]/[OIII] histogram is bimodal, the spatial correspondence of the two distributions is shown in Fig. \ref{fancy}, panels (b) and (c). From this Figure, it is clear that the region immediately around the Trapezium stars, as well as the Orion S region, do not show the highest degree of ionisation, while the highly ionised matter surrounds these inner regions like a ring. This goes against the naive picture, where matter with the lowest [OII]/[OIII] values (i.e. the highest degree of ionisation) should be found in the immediate vicinity of the ionising stars. The reason for the higher [OII]/[OIII] values is not immediately clear. Because the strong dependence of [OII]/[OIII] on the electron density (as can be seen in Fig. \ref{O23}, where [OII]/[OIII] and N$_{e}$ are plotted along the slit marked in Fig. \ref{SO23}b), the following two scenarios can be a possible explanation for this empirical fact:
\begin{itemize}
\item density variations: in regions of higher density the ionisation parameter is quenched and the emission from low-ionisation lines (e.g. [OII], [SII]) enhanced
\item shocks: the above reasoning for density variations holds if shocks are present  as well, as shocks locally compress matter, producing density enhancements
\end{itemize}

A further possibility is a combination of these two scenarios, as the central Orion Nebula, where the Orion S cloud is located, hosts a wealth of HH outflows and shocks such as HH 529, HH 269, HH 202 \citep{2001ARA&A..39...99O}.

We will discuss selected subregions of the mosaic and the scatter plot in Section 5 in order to disentangle the very large number of data points and distinguish between the different populations.

\section{Kinematics}

Turbulence generates a velocity field characterised by stochastic hierarchical fluctuations, and because of this, a statistical approach such as the second order structure function has been widely used to analyse turbulent motions in both extragalactic (e.g. \citealt{1997ApJ...487..163M}, \citealt{2011MNRAS.413..721L}) and galactic (e.g. \citealt{1986ApJ...304..767O}, \citealt{1993ApJ...409..262W}, \citealt{1995ApJ...454..316M}, \citealt{1999A&A...346..947C}) HII regions. For Orion, this has been done, e.g. with the [OIII]$\lambda$5007 line in \cite{1988ApJS...67...93C}, with the [OI]$\lambda$6300 line in \cite{1992ApJ...387..229O} and with the [SIII]$\lambda$6312 line in \cite{1993ApJ...409..262W}.

\begin{figure*}
\mbox{
\hspace{-20pt}
\subfloat[]{\includegraphics[scale=0.38]{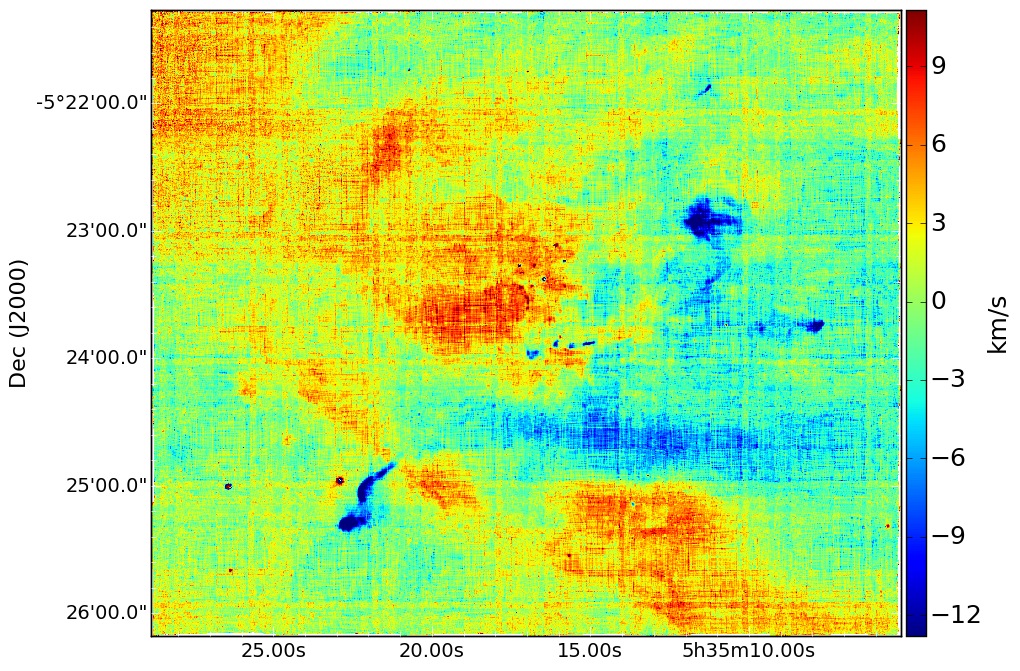}}
\subfloat[]{\includegraphics[scale=0.38]{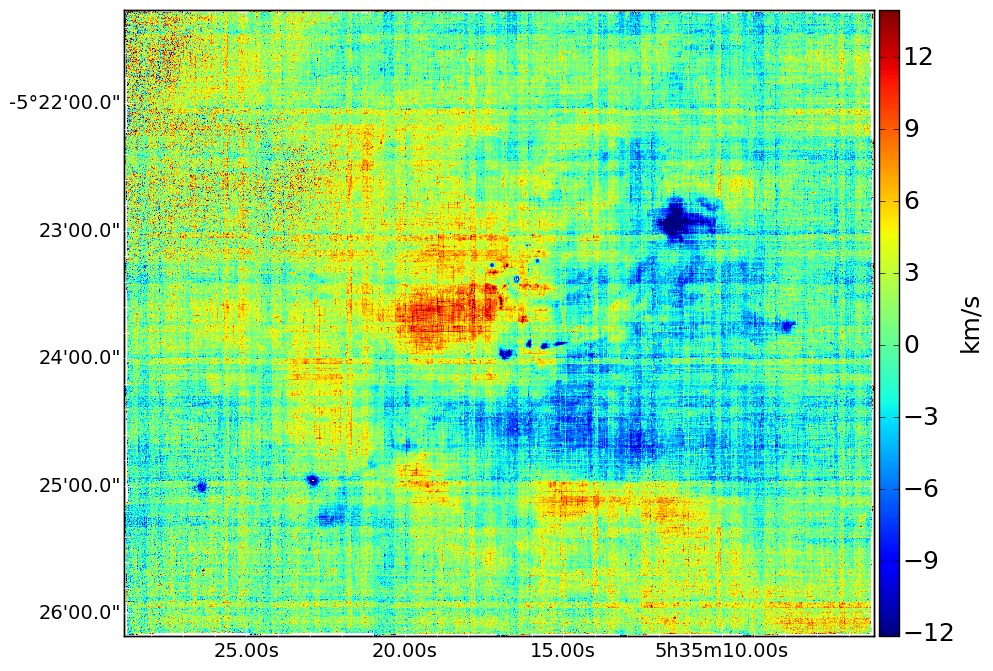}}}
\mbox{
\hspace{-20pt}
\subfloat[]{\includegraphics[scale=0.38]{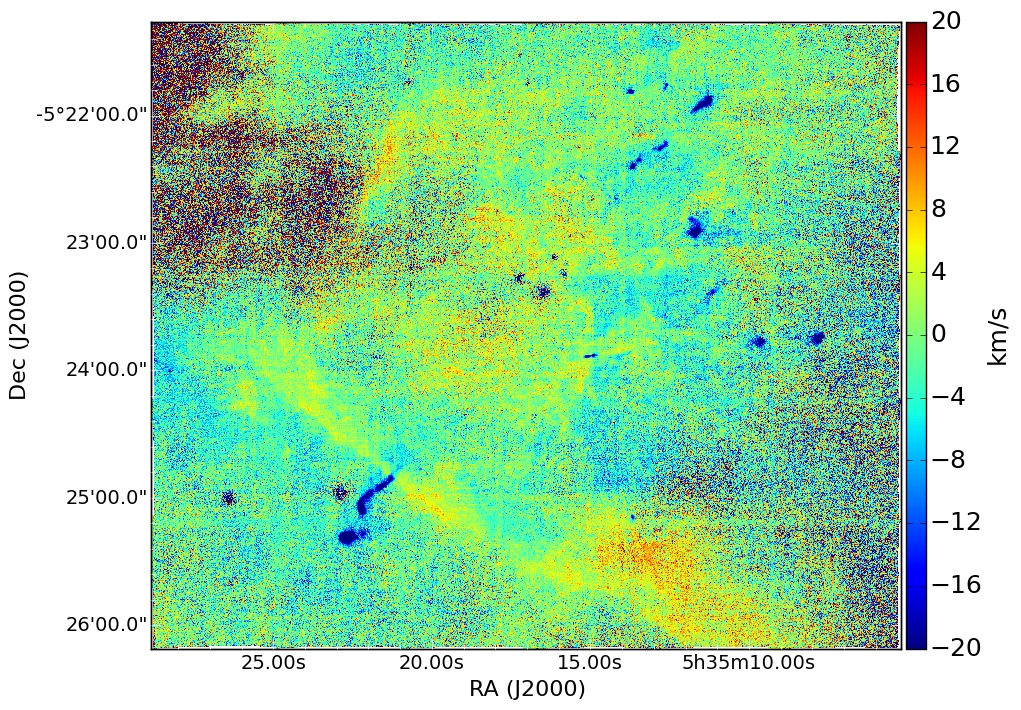}}
\subfloat[]{\includegraphics[scale=0.38]{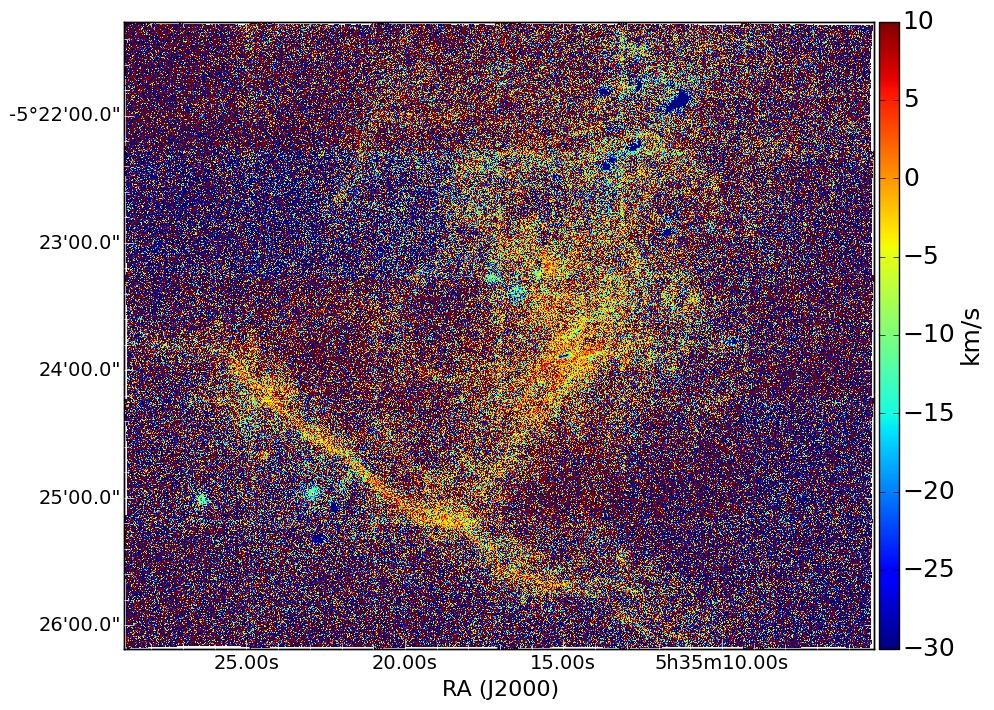}}}
  \caption{Velocity maps of H$\alpha$ (a), [OIII]$\lambda$5007 (b), [SII]$\lambda$6731 (c) and [OI]$\lambda$6300 (d). The indicated velocities correspond to velocities relative to the mean velocity (see text Section 5.1).}
  \label{vmaps}
\end{figure*}

Following the method of \cite{2015MNRAS.447.1341B}, we compute the second order structure function $S_{2}(dr)$ as the squared velocity difference between each pair of pixels. As the total number of pixels in the MUSE dataset is $>$ 2 x 10$^{6}$, we compute the structure function for a randomly selected sample of 10$^{3}$ pixels $j$, around which we radially bin all other pixels $i$. The size of the random pixel sample was chosen such that the optimum combination between computing time and smoothness was obtained. The structure function then corresponds to the mean of each radial bin, 

\begin{equation}
S_{2}(dr) = \langle (v_{i} - v_{j})^{2} \rangle_{bin}
\end{equation}
According to the predictions of \cite{1951ZA.....30...17V}, for a homogenous slab of emitting material of thickness $s$ the structure function should behave as 

\begin{equation}
S(r) \propto \begin{cases} r^{n+1} & \quad \text{if } r<s \\ r^{n} & \quad \text{if } r>s\\ \end{cases} 
\end{equation}
where $n=2/3$ in the case of Kolmogorov turbulence.
We computed $S_{2}$ for the velocity maps of H$\alpha$, [OI]$\lambda$6300, [OIII]$\lambda$5007 and [SII]$\lambda$6731. When comparing the shape of $S_{2}$ with the velocity map from which each was derived (Fig. \ref{vmaps} and red curves in Fig. \ref{stru} respectively), it is clear that for [OI]$\lambda$6300 and [SII]$\lambda$6731 the slope of $S_{2}$ is dominated by noise. 

\begin{figure}
\hspace{-25pt}
\includegraphics[scale=0.5]{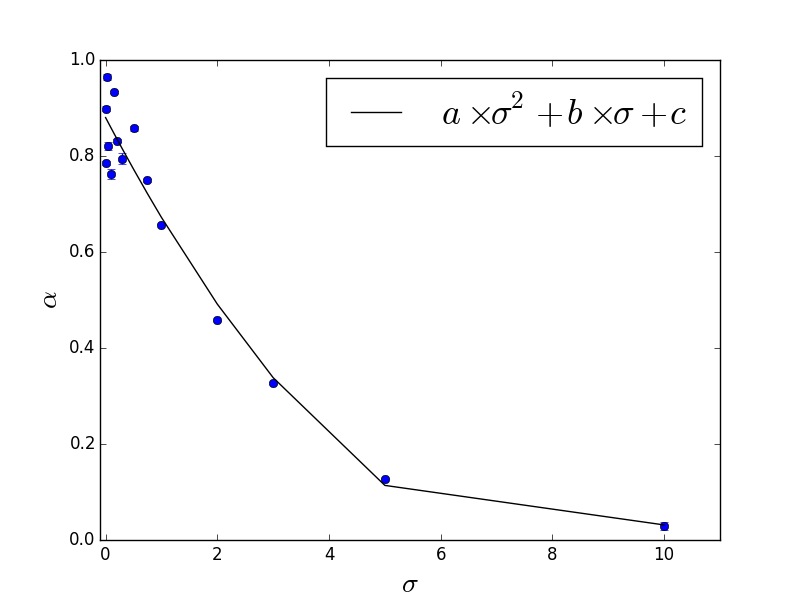}
  \caption{The slopes $\alpha$ resulting from a least-square fit to the of the structure functions computed from the [SII] maps of our simulated HII regions as a function of Gaussian noise with gradually increasing $\sigma$. The error bars correspond to the errors from the least-square fit (see text Section 4). A quadratic function was fitted with coefficients $a=0.0143\pm0.0020$, $b=-0.2285\pm0.0193$, $c=0.8922\pm0.0208$.}
  \label{s2_sims}
\end{figure}

\begin{figure*}
\mbox{
\hspace{-25pt}
\subfloat[]{\includegraphics[scale=0.48]{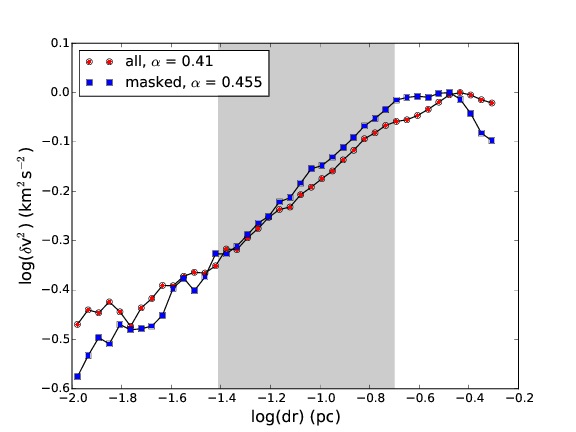}}
\subfloat[]{\includegraphics[scale=0.48]{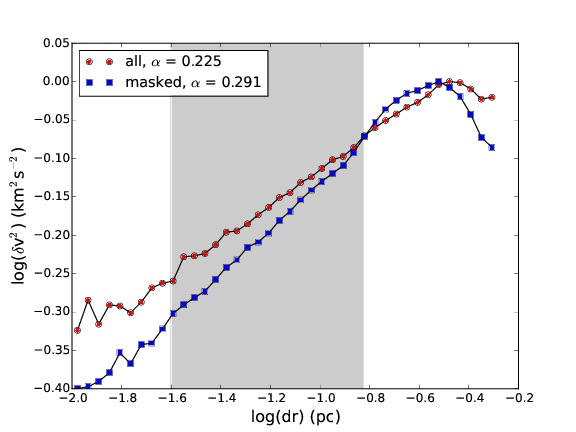}}}
\mbox{
\hspace{-25pt}
\subfloat[]{\includegraphics[scale=0.48]{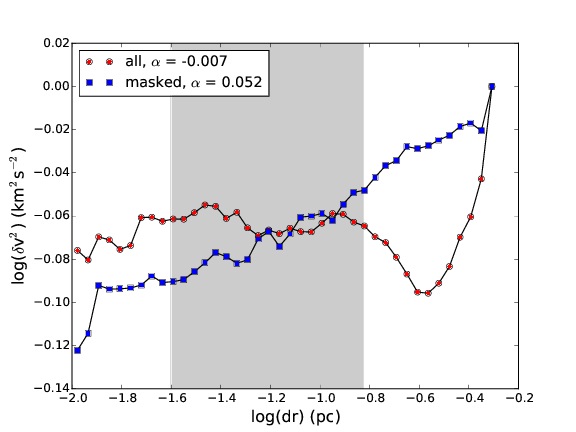}}
\subfloat[]{\includegraphics[scale=0.48]{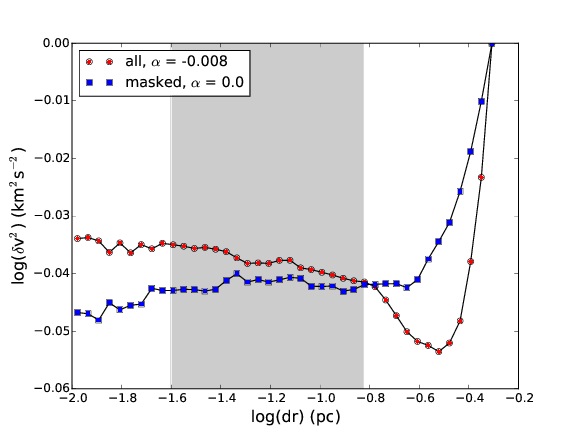}}}
  \caption{Structure functions (normalised to peak) of H$\alpha$ (a ), [OIII]$\lambda$5007 (b) and [SII]$\lambda$6731 (c) and [OI]$\lambda$6300 (d). The red circles are computed from the full velocity maps, the blue squares from the masked ones. The shaded areas correspond to the least-square fitting range used to compute the slope $\alpha$.  See text section 4.}
  \label{stru}
\end{figure*}

In order to quantify the effect of noise on the $S_{2}$ slope, we used the \textsc{mocassin} code (\citealt{2003MNRAS.340.1136E}, \citeyear{2005MNRAS.362.1038E}, \citeyear{2008ApJS..175..534E}) to create synthetic emission line maps from a snapshot of the Run UP simulation of an expanding HII region in a turbulent cloud from \cite{2013MNRAS.430..234D}. We selected a 5x5 pc subregion of the cloud containing a relatively simple isolated bubble, 0.38 Myr after ionisation was enabled (see Fig. A6). We then created first-moment maps by convolving the emission line maps with the smoothed-particle hydrodynamics (SPH) velocity field, as shown in Fig. A2a. This map has low intrinsic noise, thus it can be used as a baseline. We then added Gaussian noise with increasing values for the standard deviation $\sigma$ ($\sigma$ = 0.01, 0.05, 0.1, 0.5. 1, 5, 10 in panels b to h, not shown are $\sigma$ = 0.03, 0.15, 0.2, 0.3, 0.75, 2, 3). For each of these maps we then computed S$_{2}$, fitted power laws to the resulting structure functions and analysed the dependence of the slope $\alpha$ on $\sigma$. The result can be seen in Fig. \ref{s2_sims}, where a quadratic dependence of the slope on the level of noise is shown.

Because of this, we computed a mask based on the intensity of the mean value of the [OI]$\lambda$6300 line ($\sim$10$^{-16}$ erg s$^{-1}$ cm$^{-2}$ pixel$^{-1}$), discarding all pixels below this threshold, and applied this to all the above mentioned velocity maps (see Fig. A1)\footnote{We did not set a higher threshold, as this would have drastically reduced and limited the available length scales.}. The resulting structure functions are shown in Fig. \ref{stru}, where S$_{2}$ for both the unmasked (red circles) and the masked (blue squares) velocity maps for the four lines are plotted (to visualise the two on the same scale, the ordinate is always normalised to peak). By masking out some of the noise, the slopes of the structure functions are clearly steeper, but because we have no way of effectively getting rid of the noise and therefore computing unbiased structure function slopes both for the [SII] and in the [OI] lines, we concentrate the discussion on the structure functions derived from the [OIII] and the H$\alpha$ lines. These show power law slopes over all scales, indicating that turbulence is being driven on scales larger than the measured ones, and an energy cascade from the largest scales to the smallest. The slopes however do not correspond to the 2/3 Kolmogorov law. A comparison of the power law slopes with previous works shows that the MUSE data, because of a combination of lower spectral resolution, short exposure times and a low signal to noise ratio (especially for the weaker [OI] and [SII] lines), yields structure function slopes that are too shallow with respect to past papers. With high-resolution slit spectroscopy, \cite{1988ApJS...67...93C} find a slope of $\alpha \sim$ 0.86 for the [OIII]$\lambda$5007 line, which is a factor of about 2.6 steeper than $\alpha \sim$ 0.29 found in this work. For the [OI]$\lambda$6300 line, \cite{1992ApJ...387..229O} find an almost exact Kolmogorov slope of 2/3, while we are completely noise dominated and find $\alpha \sim$ 1. What we also do not recover from the MUSE observations is a break in the power law slopes as was previously found in the above mentioned studies. We will discuss the influence of noise on simulated structure functions in further detail in Mc Leod et al., in preparation. It is however of importance to state the implications of these results, should they be correct: a structure function slope that does not show a break hints at a uniform injection of turbulence on all observed scales (and that turbulence is injected at larger scales than the observed), while the fact that it is shallower than Kolmogorov indicates that the injected turbulence is not sufficient to maintain a Kolmogorov-type velocity field.

\section{Discussion of selected regions}

\subsection{The Orion bullet HH 201}
\begin{figure*}
\mbox{
\hspace{-10pt}
\subfloat[]{\includegraphics[scale=0.42]{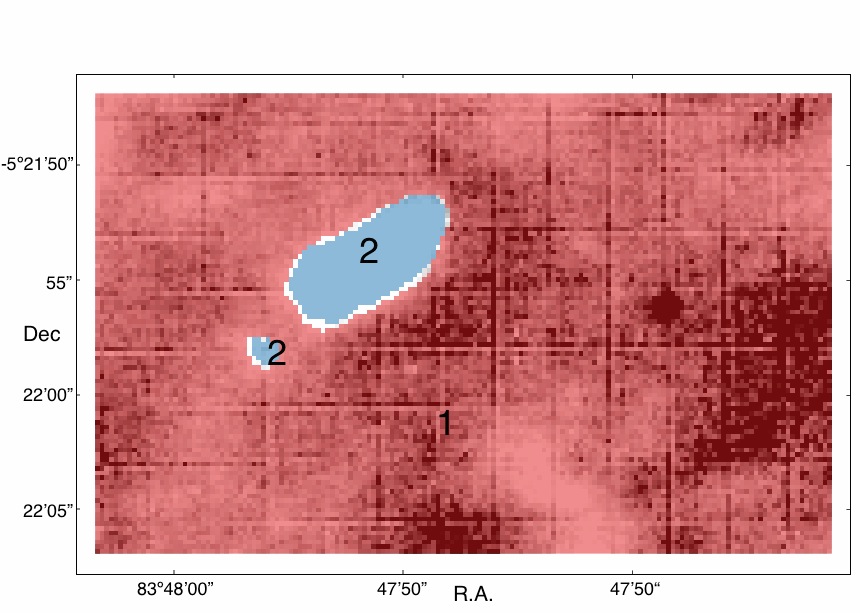}}}
\mbox{
\subfloat[]{\includegraphics[scale=0.4]{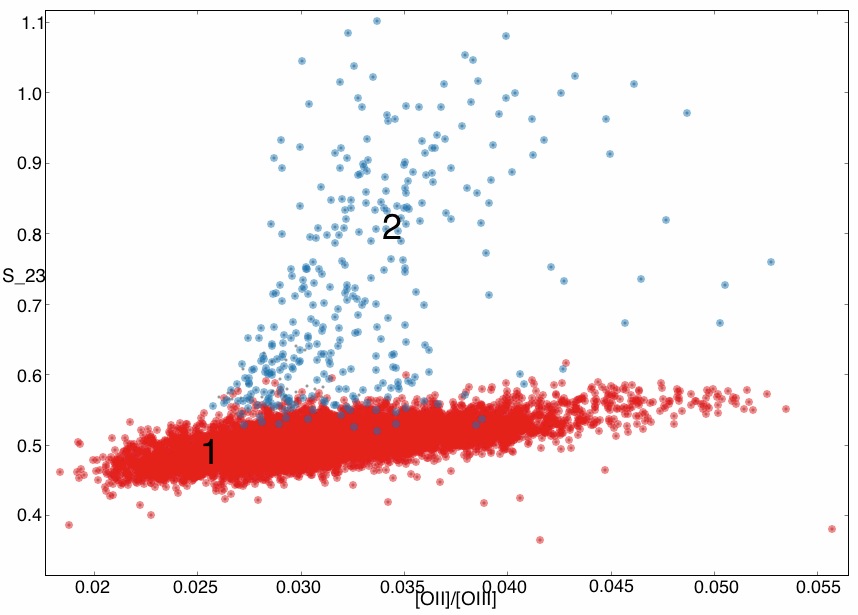}}}
  \caption{S$_{23}$ map of the Orion bullet covered by MUSE (panel a), colour coded according to the S$_{23}$ vs $[OII]/[OIII]$ plot in panel (b)  HII region(red circles, number 1)  and the bullet (blue circles, number 2).}
  \label{bulletSO23}
\end{figure*}

The only feature that, because of its very high S$_{23}$ values, stands out in the S$_{23}$ map shown in Fig. \ref{SO23}a is one of the so called Orion bullets in the upper right corner of the image, in fact the only bullet covered by the MUSE field. This object corresponds to HH 201 \citep{2003MNRAS.343..419G}, a bright shock from an outflow driven by the high-mass star forming region OMC-1 \citep{2000AJ....119.2919B}. To better analyse this region, we cropped the original mosaic to a 32" x 20" rectangle (centred at R.A. J2000 = 5:35:11.179, Dec J2000 = -5:21:57.22), as to cover both the bullet and a representative portion of the surrounding material. Just like the outflow detected in M 16 (see MC15), the bullet corresponds to a population in the S$_{23}$ versus [OII]/[OIII] parameter space (Fig. \ref{bulletSO23}b) that clearly deviates from the rest. This population, just as the M 16 outflow (red stars in Fig. \ref{SO23}c), displays very high S$_{23}$ values, as well as a relatively high degree of ionisation, but compared to the M 16 outflow, HH 201 has S$_{23}$ values higher by a factor of $\sim$ 1.5. Not only does HH 201 show higher S$_{23}$ values than the M 16 outflow, it also shows a higher degree of ionisation. This is not surprising, as the low-velocity M 16 outflow most probably originates from a deeply embedded protostar, and the outflow itself is just now emerging from the pillar material where the driving source is embedded; HH 201 on the other hand is a high-velocity outflow travelling in a less dense medium, it originates from a highly energetic explosive event \citep{2015A&A...579A.130B} rather than a protostar and it is closer (in projection) to the ionising O-star. 

\begin{figure*}
\mbox{
\hspace{-25pt}
\subfloat[]{\includegraphics[scale=0.35]{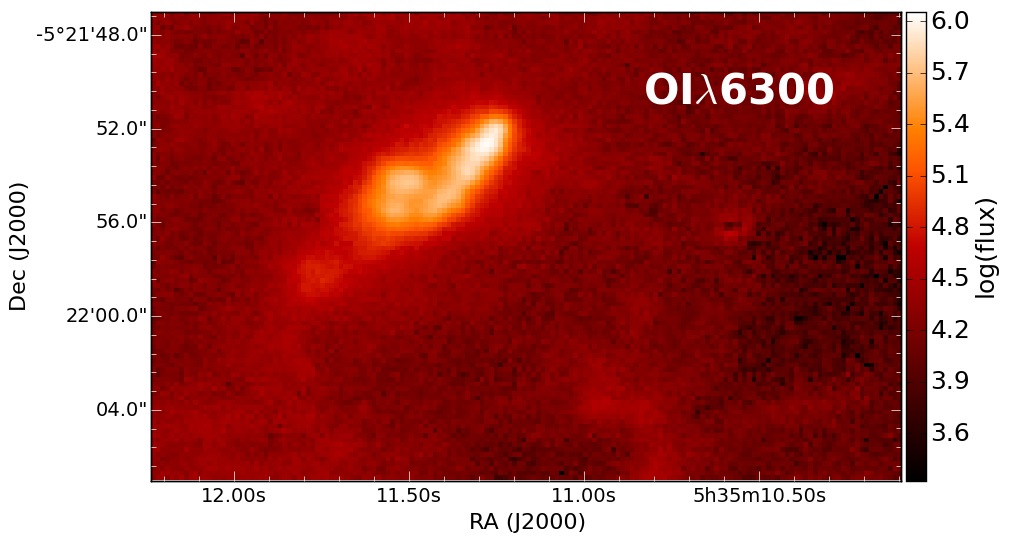}}
\subfloat[]{\includegraphics[scale=0.35]{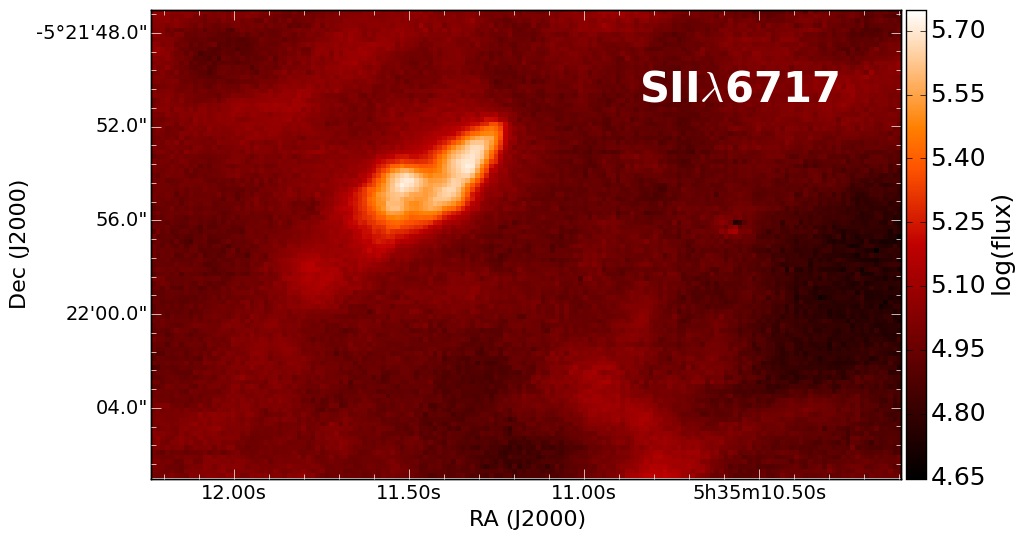}}}
\mbox{
\hspace{-25pt}
\subfloat[]{\includegraphics[scale=0.35]{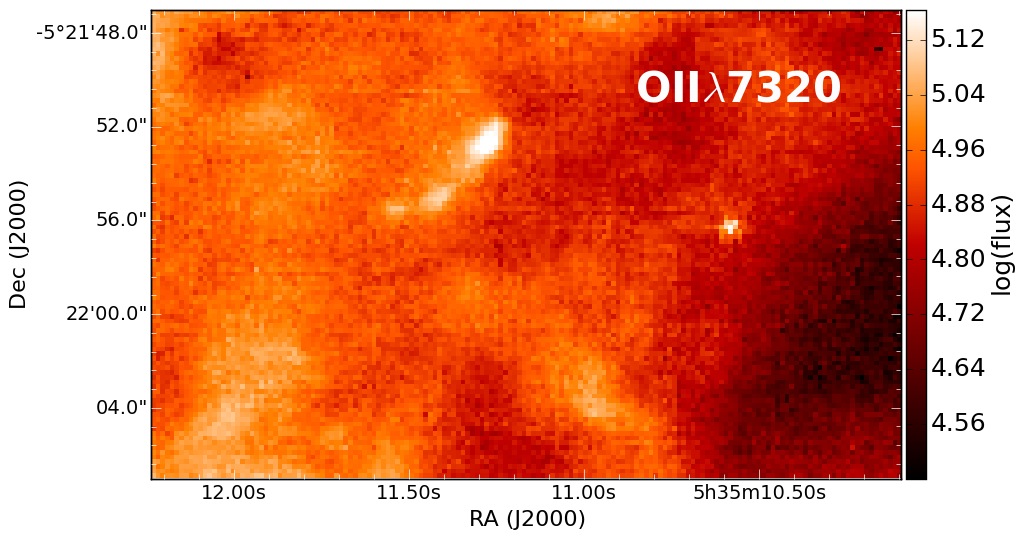}}
\subfloat[]{\includegraphics[scale=0.35]{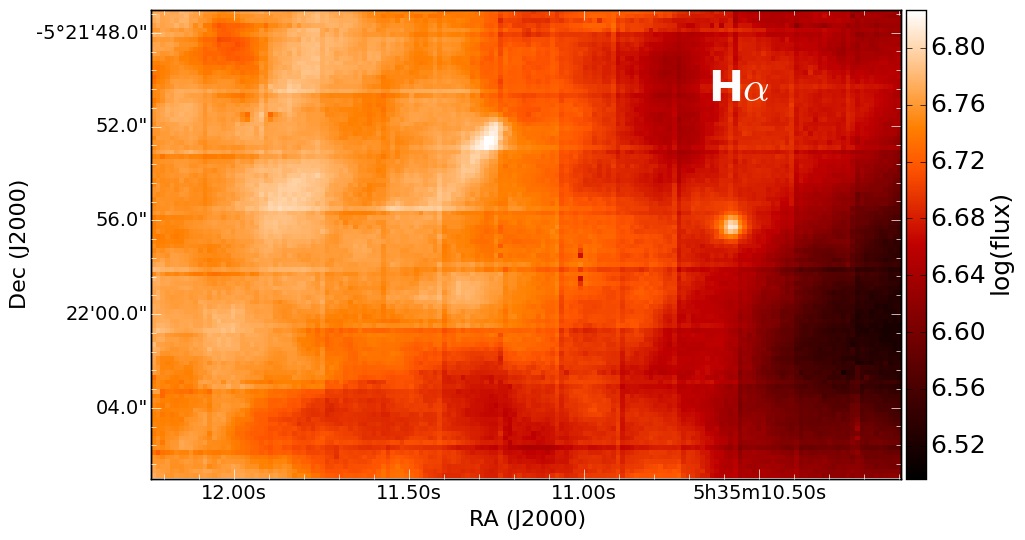}}}
\mbox{
\hspace{-25pt}
\subfloat[]{\includegraphics[scale=0.35]{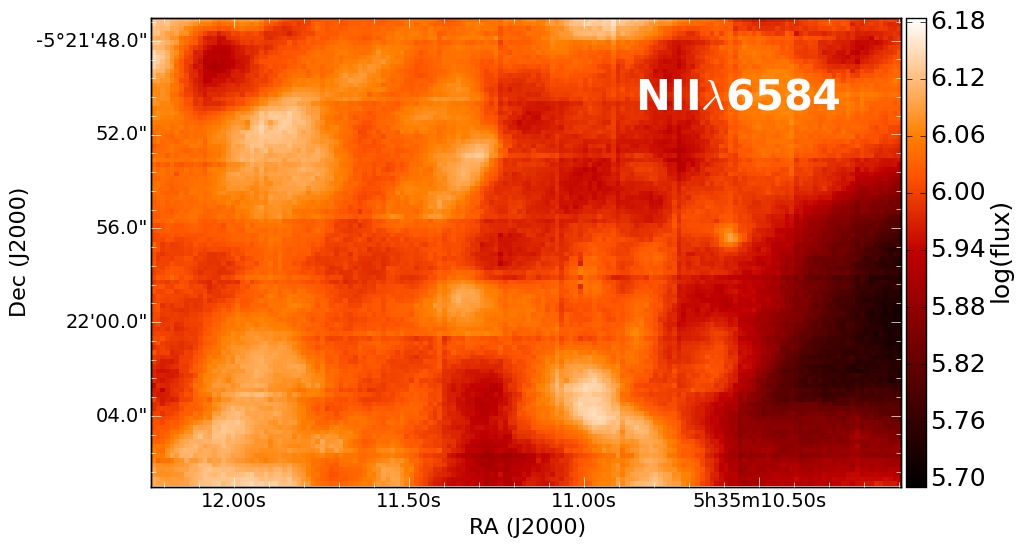}}
\subfloat[]{\includegraphics[scale=0.35]{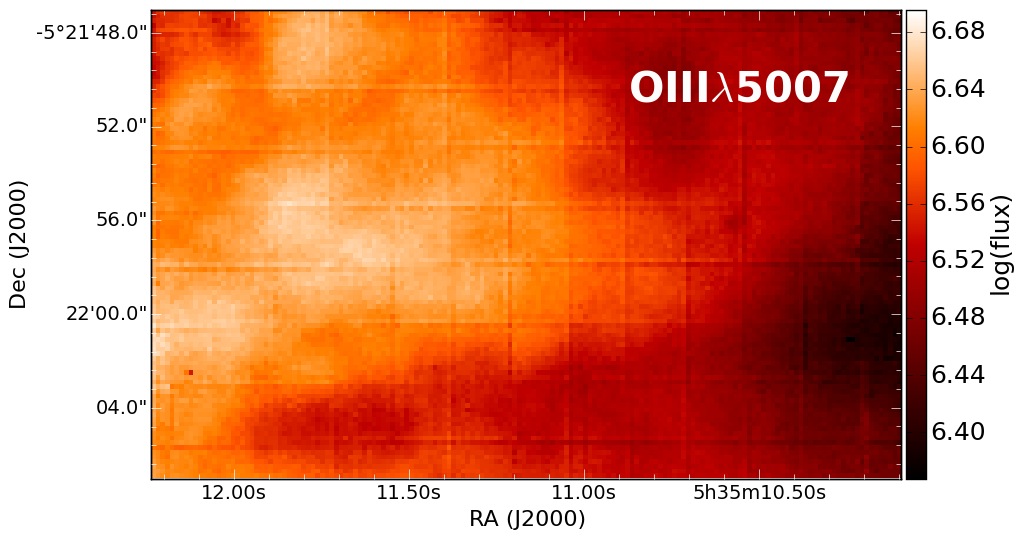}}}
  \caption{Continuum-subtracted and extinction corrected, integrated intensity maps of the Orion bullet covered by MUSE  [OI]$\lambda$6300 (a), [SII]$\lambda$6717 (b), [OII]$\lambda$7320, H$\alpha$ (d), [NII]$\lambda$6584 (e) and [OIII]$\lambda$5007 (f). The flux is measured in 10$^{-20}$ erg s$^{-1}$ cm$^{-2}$ pixel$^{-1}$, all maps are linearly scaled to minimum/maximum. Residual stellar emission from the continuum subtraction can be seen at R.A. 5:35:10.583, dec -5:21:56.50 (J2000).}
  \label{bullet_int}
\end{figure*}

\begin{figure*}
\mbox{
\subfloat[]{\includegraphics[scale=0.35]{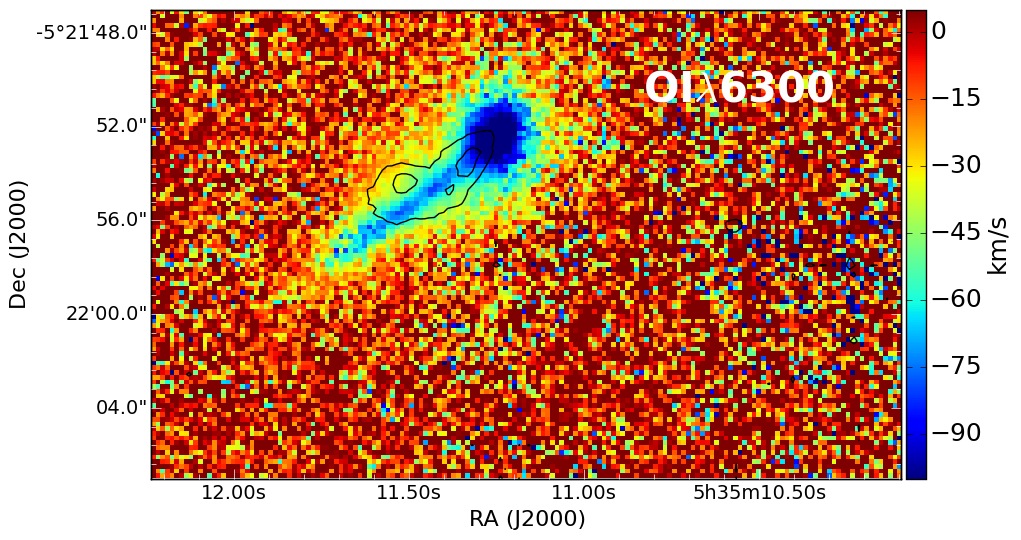}}
\subfloat[]{\includegraphics[scale=0.35]{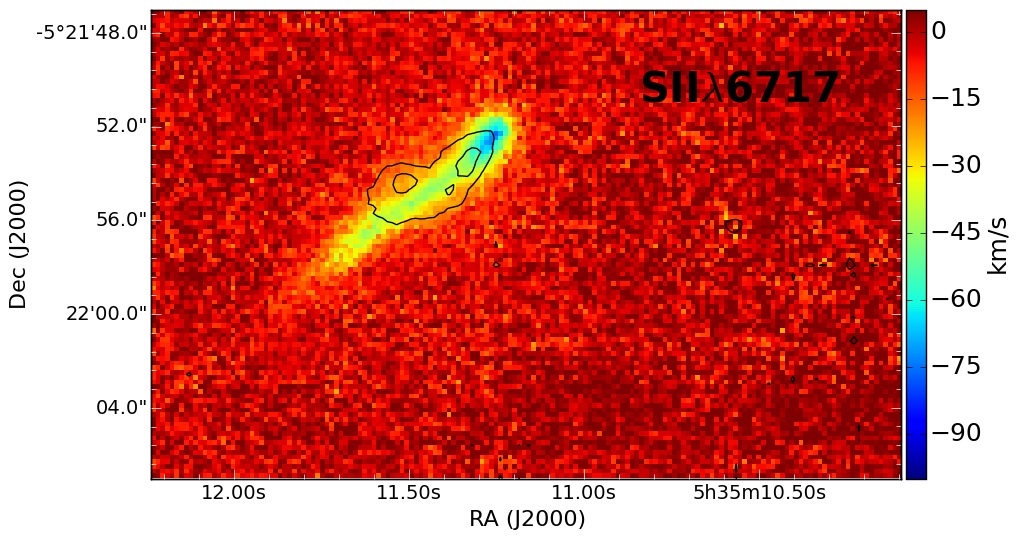}}}
\mbox{
\subfloat[]{\includegraphics[scale=0.35]{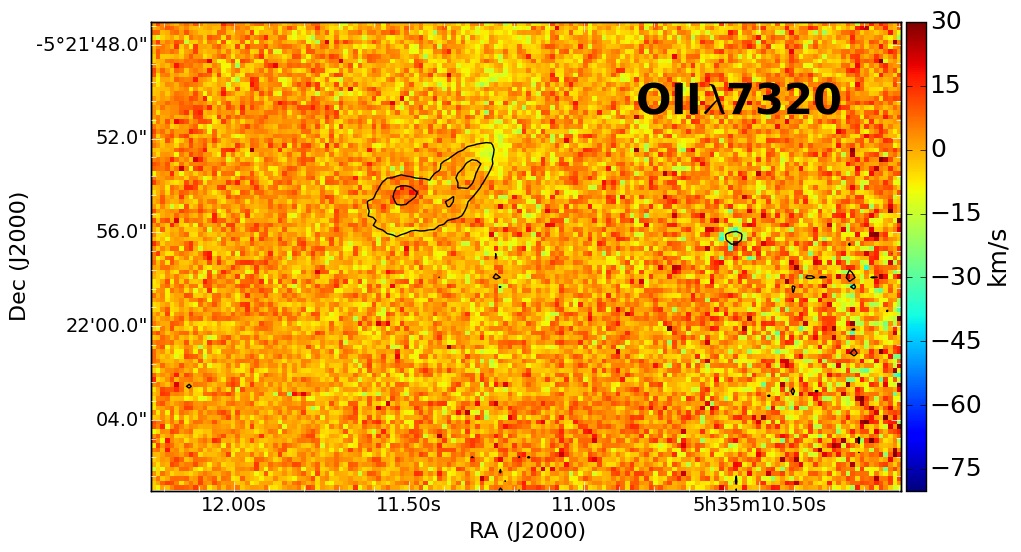}}
\subfloat[]{\includegraphics[scale=0.35]{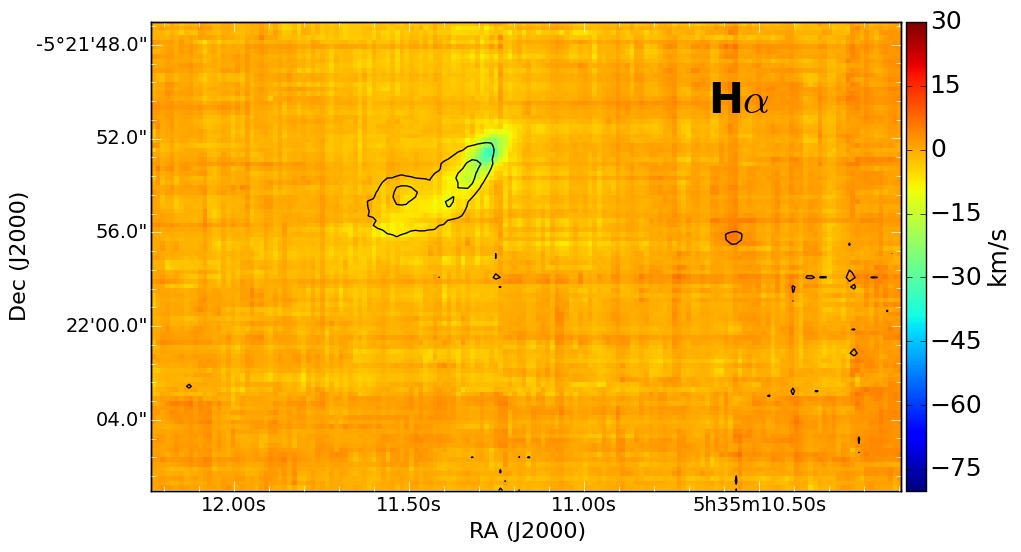}}}
\mbox{
\subfloat[]{\includegraphics[scale=0.35]{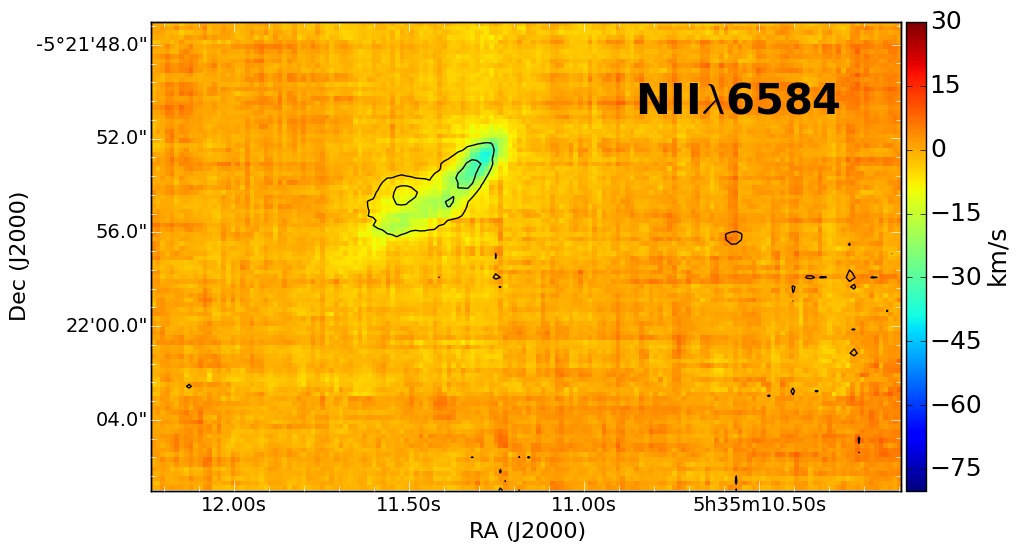}}
\subfloat[]{\includegraphics[scale=0.35]{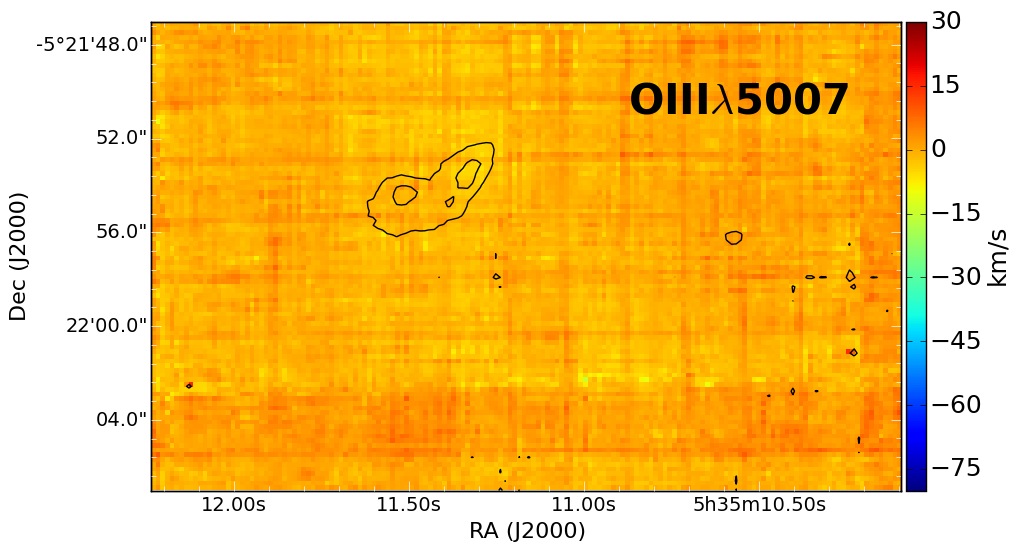}}}
  \caption{Velocity maps of the Orion bullet covered by MUSE  [OI]$\lambda$6300 (a), [SII]$\lambda$6717 (b), [OII]$\lambda$7320, H$\alpha$ (d), [NII]$\lambda$6584 (e) and [OIII]$\lambda$5007 (f). Black contours correspond to S$_{23}$ values (at R.A. 5:35:10.583, dec -5:21:56.50 a residual from the continuum-subtraction is seen). The indicated velocities correspond to velocities relative to the mean velocity of the surrounding medium (see text Section 5.1).}
  \label{bullet_vel}
\end{figure*}

The integrated intensity maps of [OI]$\lambda$6300, [SII]$\lambda$6717, [OII]$\lambda$7320, [NII]$\lambda$6584, H$\alpha$ and [OIII]$\lambda$5007 (Fig. \ref{bullet_int}a to f, due to an imperfect continuum-subtraction, the residual of a star in the right side of the images is seen) show that the contrast between the bullet and the surrounding medium gets weaker as a function of the ionisation state. The sharpest contrast is seen in the neutral [OI]$\lambda$6300 line, whereas the [OIII]$\lambda$5007 map only shows diffuse emission. The velocity maps of the same emission lines (Fig. \ref{bullet_vel}) show a similar behaviour: the bullet can be clearly identified in the [OI] and [SII] maps, where it assumes a blueshifted cometary shape with a head pointing away from the Trapezium cluster and a tail pointing back towards it. In the [OII] velocity map only the head can be identified as being slightly blueshifted, while the bullet cannot be seen at all in the [OIII] map. Furthermore, the [OI] and [SII] maps reveal a clear velocity difference between the head and the tail of about 50 km s$^{-1}$, the head showing the more negative velocities. The head-tail geometry and the velocity difference between the two could be the result of the high-velocity bullet ramming into a high-density region and being slowed down by the impact. The velocities reported here correspond to relative velocities, meaning that for each line we subtracted the mean velocity of the surrounding medium from the entire velocity map.

\begin{landscape}
\begin{figure}
\centering
\includegraphics[scale=0.85]{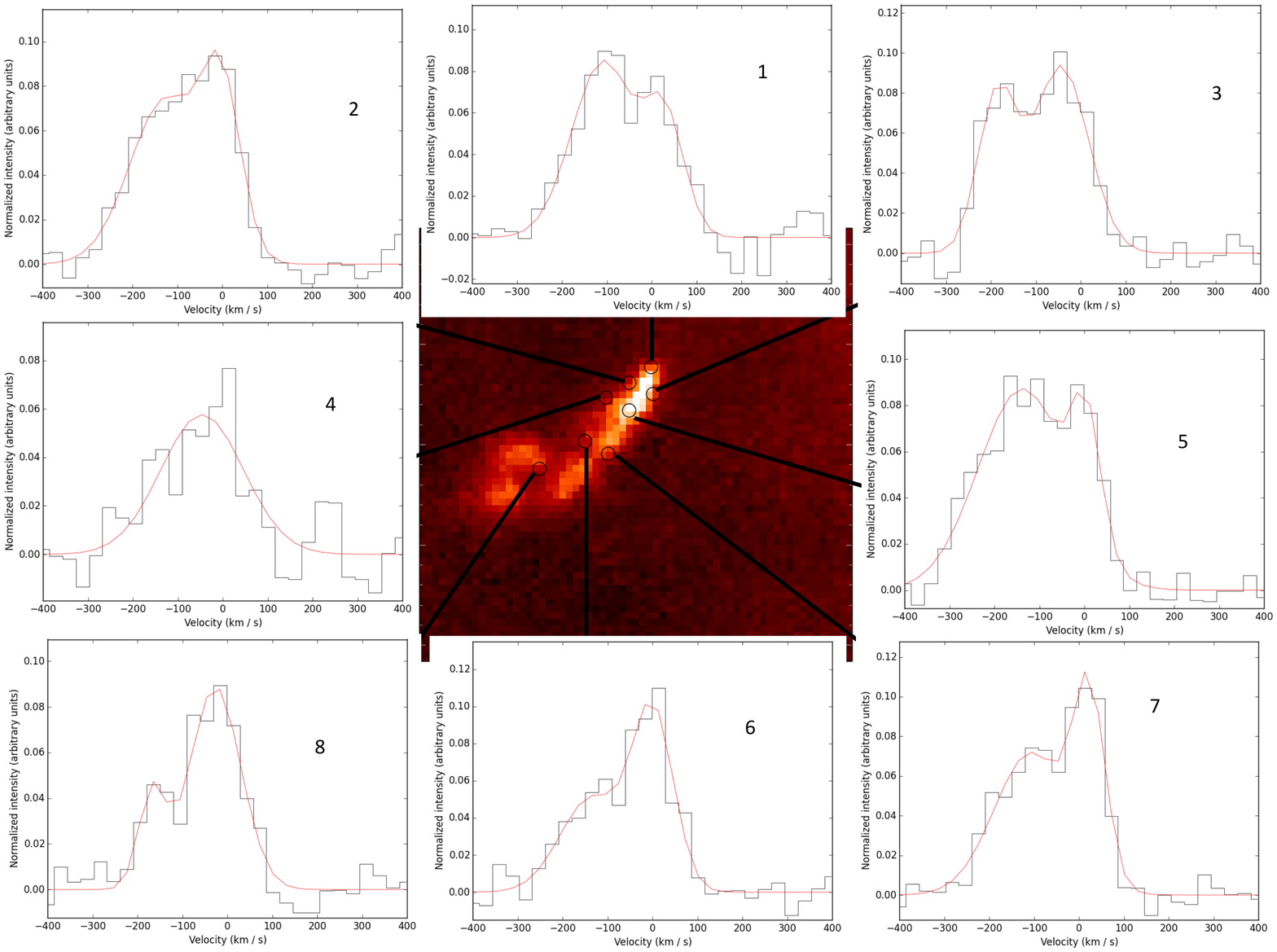}
  \caption{Integrated intensity map of the [FeII]$\lambda$8617 line (middle), and the spectra corresponding to the pixels marked by the centroid of the black circles. The spectra are continuum subtracted, and fitted with a two component Gaussian (the numbers correspond to the entries in Table \ref{specfit}, see also text Section 5.1 for details).}
  \label{specs}
\end{figure}
\end{landscape}

The Orion bullets are particularly bright in emission lines of ionised iron (\citealt{2015arXiv150204711B}, Youngblood et al., in prep.), and the near-infrared [FeII] line (at 1.644 $\mu$m) profiles of HH 201 appear to be consistent with theoretical predictions of a bow shock \citep{1999MNRAS.307..337T}. We inspected the line profile of the brightest of the detected iron lines, [FeII]$\lambda$8617, and find a double component throughout the head of the bullet and along its tail. In this can be seen in Fig. \ref{specs}, where the continuum-subtracted and Gaussian-fitted spectra for eight pixels are shown, together with the integrated [FeII]$\lambda$8617 intensity map (central panel). Where only one component is seen (not shown in Fig. \ref{specs}), the typically line width is $\sim$ 300 km/s, while for in case of a double-peaked line the FWHM is $\sim$ 30 - 45 km/s for the red component and $\sim$ 20 - 60 km/s for the blue component. The blue component is typically lower in intensity than the red one. Table \ref{specfit} shows the best fit centroid and width values obtained with \textsc{pyspeckit}, using a two component Gaussian fit.
 
\begin{table*}
\begin{center}
\caption{Best fit parameters of the Gaussian fitting to the [FeII]$\lambda$8617 lines shown in Fig. \ref{specs}. Centroids and widths are measured in km/s, values are rounded to the nearest integer.}
\begin{tabular}{lccccc}
\hline
\hline
Position & Coordinates  & Centroid & Width & Centroid & Width \\
 (J2000) & (blue) & (blue) & (red) & (red) \\
\hline
1 & 5 35 11.29 -5 21 52.15 & -108$\pm$12 & 69$\pm$10 & 29$\pm$11 & 43$\pm$9 \\
2 & 5 35 11.25 -5 21 52.49 & -126$\pm$28 & 80$\pm$20 & -25$\pm$10 & 42$\pm$11 \\
3 & 5 35 11.34 -5 21 52.60 & -186$\pm$10 & 42$\pm$8 & -44$\pm$11 & 61$\pm$10 \\
4 & 5 35 11.29 -5 21 52.98 & -46$\pm$12 & 92$\pm$12 & - & - \\
5 & 5 35 11.25 -5 21 51.68 & -138$\pm$11 & 99$\pm$10 & 4$\pm$5 & 32$\pm$8 \\
6 & 5 35 11.38 -5 21 53.90 & -136$\pm$30 & 70$\pm$24 & 0$\pm$10 & 46$\pm$7 \\
7 & 5 35 11.34 -5 21 54.28 & -107$\pm$11 & -30$\pm$8 & -27$\pm$5 & 57$\pm$6 \\
8 & 5 35 11.47 -5 21 54.73 & -168$\pm$12 & 69$\pm$10 & 29$\pm$11 & 43$\pm$9 \\
\end{tabular}
\label{specfit}
\end{center}
\end{table*}

We interpret the detection of two velocity components along the line of sight as a sign of expansion. For the evolutionary scenario of this bullet we therefore suggest that it is currently being disrupted, probably because of it being a high-velocity and high-density object hydrodynamically interacting with the surrounding Orion Nebula, for example by impacting on a high-density region. 

\subsection{Outflows and proplyds}
\begin{figure*}
\mbox{
\subfloat[]{\includegraphics[scale=0.35]{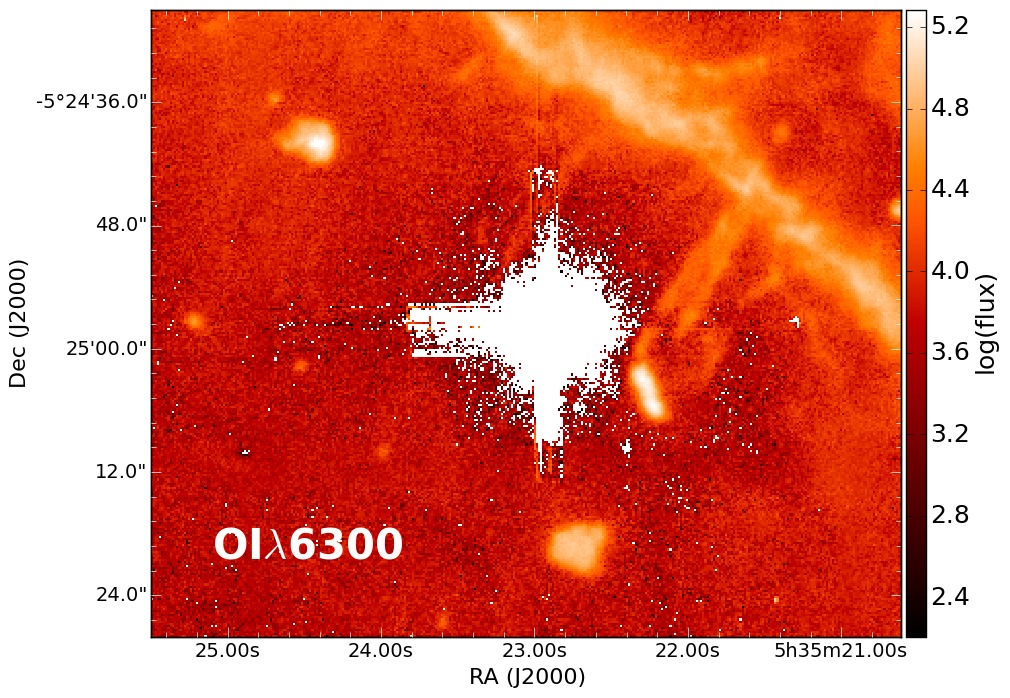}}
\subfloat[]{\includegraphics[scale=0.35]{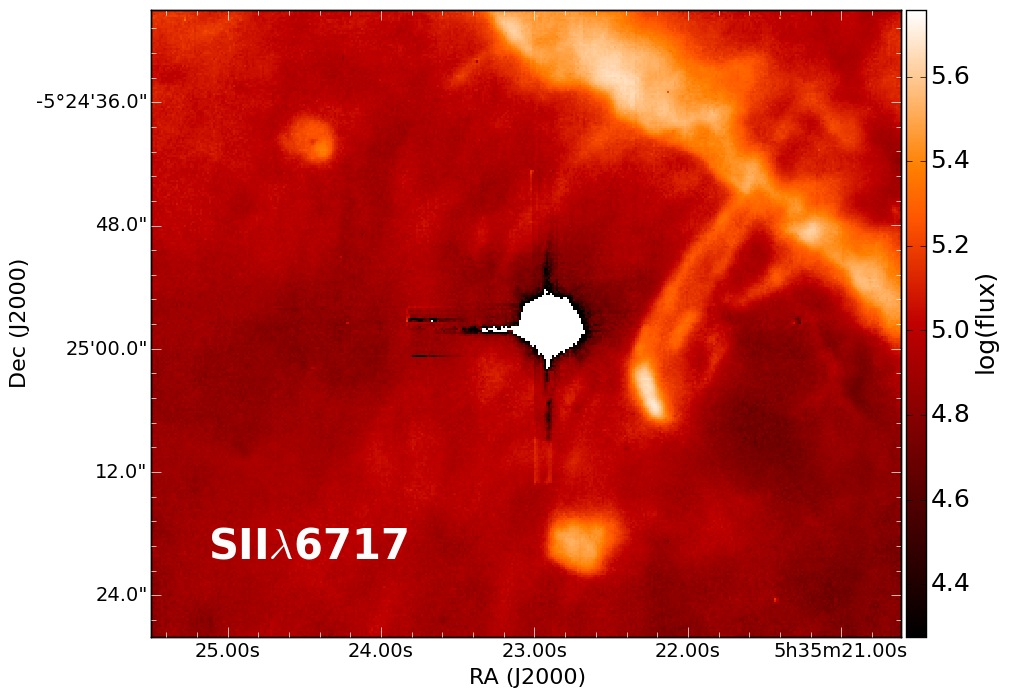}}}
\mbox{
\subfloat[]{\includegraphics[scale=0.35]{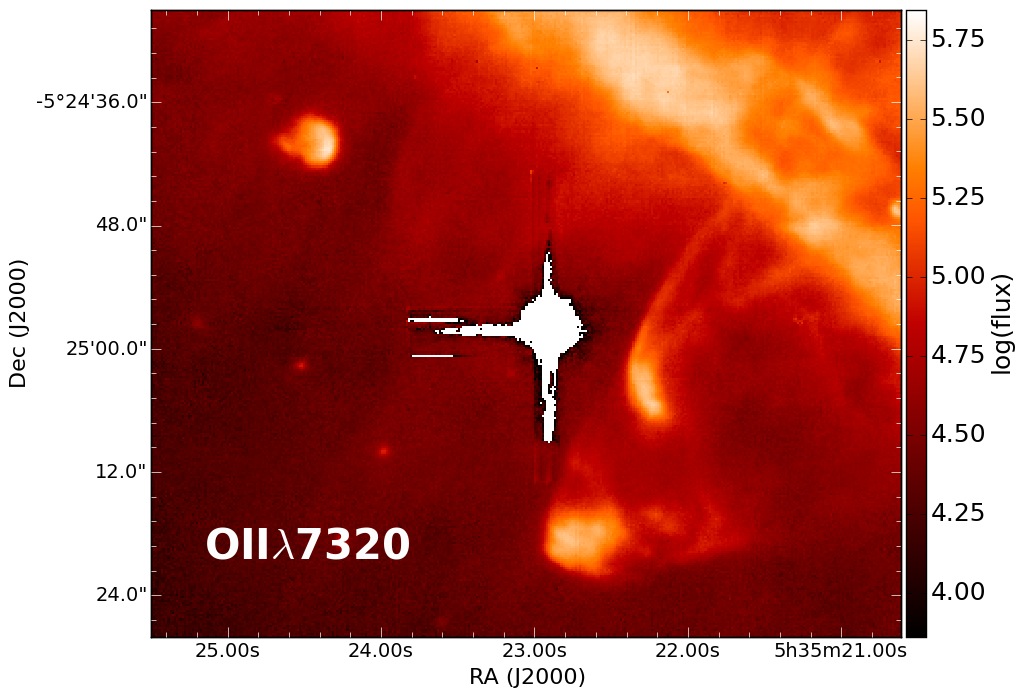}}
\subfloat[]{\includegraphics[scale=0.35]{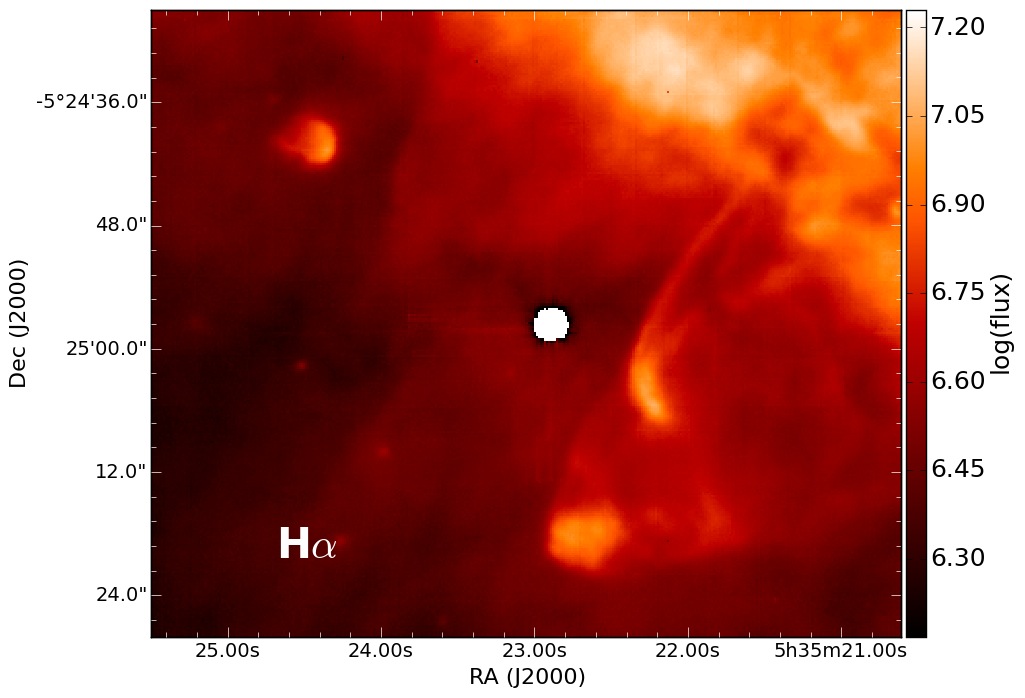}}}
\mbox{
\subfloat[]{\includegraphics[scale=0.35]{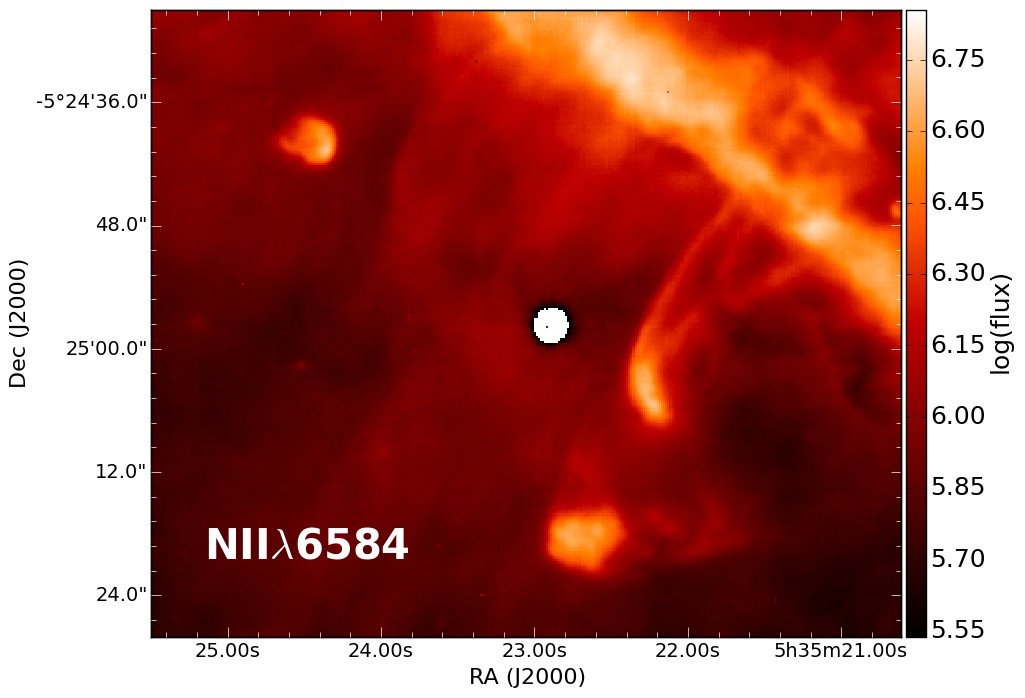}}
\subfloat[]{\includegraphics[scale=0.35]{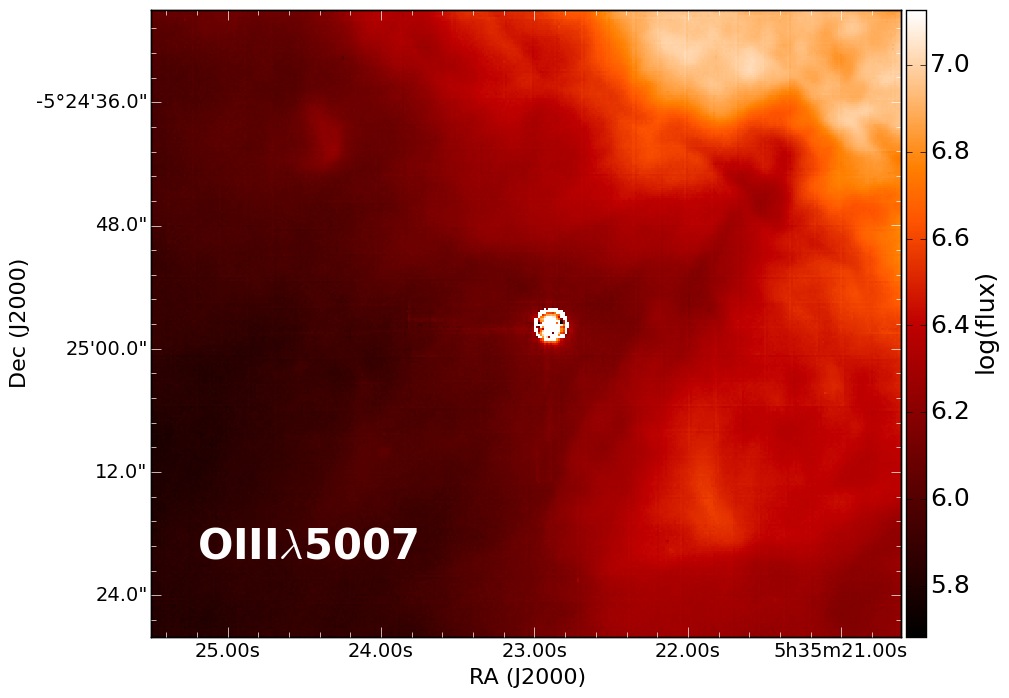}}}
  \caption{Continuum-subtracted integrated intensity maps of HH203 and 204, as well as 244-440  [OI]$\lambda$6300 (a), [SII]$\lambda$6717 (b), [OII]$\lambda$7320 (c), H$\alpha$ (d), [NII]$\lambda$6584 (e) and [OIII]$\lambda$5007 (f). The flux is measured in 10$^{-20}$ erg s$^{-1}$ cm$^{-2}$ pixel$^{-1}$, all maps are linearly scaled to minimum/maximum.}
  \label{jets_int}
\end{figure*}

In terms of the the S$_{23}$ vs. [OII]/[OIII] analysis, the region just south of the Bright Bar containing the HH objects 203 and 204, as well as several of the Orion proplyds including Orion 244-440, is of great interest. The proplyds on this side of the Bar occupy a completely different region in Fig. \ref{SO23}c than the proplyds that lie on the other side of the Bar and that are under the direct influence of the Trapezium stars. This is not surprising, as the physical conditions in the two regions differ in terms of ionising radiation and other feedback mechanisms (like stellar winds and outflows), which lead to the proplyds south of the Bar having higher S$_{23}$ values and displaying lower degrees of ionisation. A 1.2'$\times$1.0' sub-cube was generated so to analyse this region in more detail, integrated intensity maps of [OI], [SII], [OII], H$\alpha$, [NII] and [OIII] are shown in Fig. \ref{jets_int}. 

\begin{figure*}
\mbox{
\subfloat[]{\includegraphics[scale=0.5]{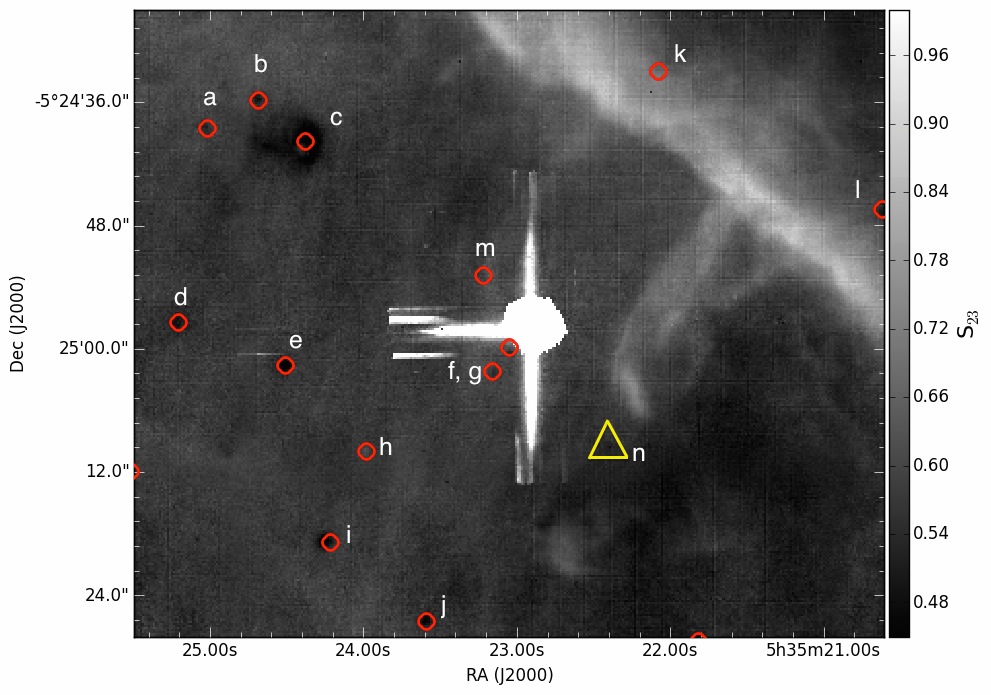}}}
\mbox{
\subfloat[]{\includegraphics[scale=0.5]{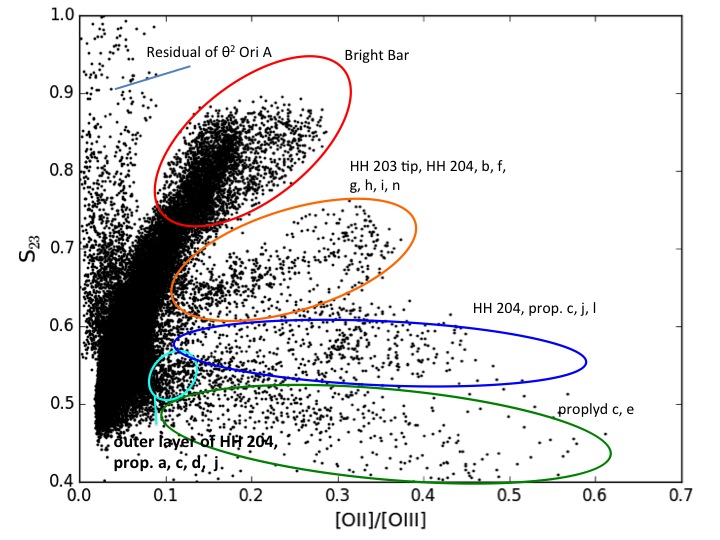}}}
  \caption{S$_{23}$ map of the region containing the Herbig-Haro objects HH203, 204 as well as the proplyd Orion 244-440 (panel a), red circles and letters from $a$ to $m$ indicate the Orion proplyds in this field, the yellow triangle indicated with the letter $n$ marks the position of a candidate proplyd (see text Section 5.2 and Table \ref{proplyds}). Panel (b): S$_{23}$ vs. [OII]/[OIII] parameter space of the same region. Strong residuals of $\theta^{2}$ Ori A from the continuum-subtraction are indicated in panel (b).}
 \label{jetsSO23}
\end{figure*}

Fig. \ref{jetsSO23}b shows the colour-coded S$_{23}$ vs. [OII]/[OIII] parameter space, as well as the corresponding S$_{23}$ map (panel a). The S$_{23}$ vs. [OII]/[OIII] parameter space separates the various objects in this field into different populations: the Bar shows higher S$_{23}$ values than the proplyds and HH objects, while these show a wider range of degrees of ionisation. The clear separation into different populations however is not seen when considering [SII]/H$\beta$ (or [SII]/H$\alpha$) as is shown in Fig. A7, where the same populations are not as clearly separated and overlapping. We interpret this as S$_{23}$ being an indicator of the relative contribution of shocks (and photoionisation) to the excitation of the sulphur atoms, as shocks locally compress matter, reducing the ionisation parameter and therefore enhancing the emission of low- over high-ionisation species (e.g. enhancing [SII] emission with respect to [SIII]). In this scenario, the Bar shows higher S$_{23}$ values because photoionisation from the Trapezium stars produces a larger fraction of [SIII] emission, which is not the case for the objects south of the Bar. 

We cross-matched the identified sources in this field with the \textit{HST/ACS} Atlas of Great Orion Nebula proplyds \citep{2008AJ....136.2136R}, and report their main characteristics in Table \ref{proplyds}, where the types are $i$ =  ionised disk seen in emission, $J$ = jet, $B$ = binary system (from \citealt{2008AJ....136.2136R}), and mj = microjet (from \citealt{2000AJ....119.2919B}). All of the proplyds in this field are associated with ionised disks seen in emission, and in general the objects can be separated into 4 classes, depending on their S$_{23}$ values and degree of ionisation:

\begin{itemize}
\item \textit{S$_{23}$-low and high ionisation (SLHI)}: proplyds $a, c, d, j$ and outer layer of HH 204
\item \textit{S$_{23}$-low and lower ionisation (SLLI)}: proplyds $c$ and $d$
\item \textit{mean S$_{23}$ values and lower ionisation (MSLI)}: HH 204, proplyds $j$, $c$ and $l$
\item \textit{S$_{23}$-high and lower ionisation (SHLI)}: proplyds $b, f,g, h, i$ and $n$, tip of HH 203, HH 204
\end{itemize}

\begin{table*}
\centering
\caption{Orion proplyds. The proplyd Id linking to Fig. \ref{jetsSO23} (column 1), the proplyd number from Ricci et al. (column 2), the coordinates (J2000, column 3), and the type (column 4, i = ionised disk seen in emission, J = jet, B = binary system, mj = microjet, SLLI = S$_{23}$-low and low ionisation, SLHI = S$_{23}$-low and high ionisation, MSLI = medium S$_{23}$ values and low ionisation, SHLI = S$_{23}$-high and low ionisation).}
\begin{tabular}{lccc}
\hline
\hline
Id & Object & Coordinates & Type\footnotemark[1] \\
\hline
a & 250-439 & 5 35 25.02 - 5 24 38.49 & i, SLHI \\
b & 247-436 & 5 35 24.69 -5 24 35.74 & i, J, SHLI, mj \\
c & 244-440 & 5 35 24.38 -5 24 39.74 & i, SLHI + MSLI (out), SLLI (middle), MSLI (disk), mj \\
d & 252-457 & 5 35 25.21 -5 24 57.34 & i, SLHI, mj \\
e & 245-502 & 5 35 24.51 -5 25 01.59 & i, SLLI  \\
f & 231-502 & 5 35 23.16 -5 25 02.19 & i, B, SHLI \\
g & 231-460 & 5 35 23.05 -5 24 59.58 & i, SHLI \\
h & 239-510 & 5 35 23.98 -5 25 09.94 & i, SHLI \\
i & 242-519 & 5 35 24.22 -5 25 18.79 & i, SHLI, mj \\
j & 236-527 & 5 35 23.59 -5 25 26.54 & i, SLHI + MSLI \\
k\footnotemark[2] & 221-433 & 5 35 22.08 -5 24 32,95 & i \\
l & 206-446 & 5 35 20.62 -5 24 46.45 & i, MSLI,  mj \\
m\footnotemark[2] & 232-455 & 5 35 23.22 -5 24 52.79 & i \\
n\footnotemark[3] & 224-510 & 5 35 22.41 -5 25 09.61 & SHLI \\
\hline
\end{tabular}
\label{proplyds}
\\[1.5pt]
\hspace{-9cm}\footnotemark[1]\tiny{From \cite{2008AJ....136.2136R}, this work and \citealt{2000AJ....119.2919B}.}\\
\hspace{-12.15cm}\footnotemark[2]\tiny{Non-detection in this work.}\\
\hspace{-11.85cm}\footnotemark[3]\tiny{New detection from this work.}
\end{table*}

The proplyd 232-455 (indicated with the letter $m$ in Fig. \ref{jetsSO23}a) is a non-detection in our data, as the emission is dominated by the ghost of $\theta^{2}$ Ori A. Not in the  \citeauthor{2008AJ....136.2136R} catalog is the object indicated with the yellow triangle (object $n$) in Fig. \ref{jetsSO23}a, which is found just below HH 203. We tentatively classify this source as a candidate proplyd and assign it the identifier 224-510, as it displays similar S$_{23}$ and [OII]/[OIII] values as the objects in the SHLI category, but was not detected by \citeauthor{2008AJ....136.2136R} and we do not have the spatial resolution of \textit{HST} to better resolve the source and an eventual microjet/outflow. We will now briefly discuss the objects belonging to the four classes\footnote{We will not discuss proplyds 231-502, 231-460, 232-455 and 221-443, as the former three are too close to $\theta^{2}$ Ori A to be properly distinguished, and the latter is confused with the emission coming from the Bright Bar.}.

\subsubsection{Multi-class objects: HH 204 and Orion 244-440}
\begin{figure}
\centering
\includegraphics[scale=0.22]{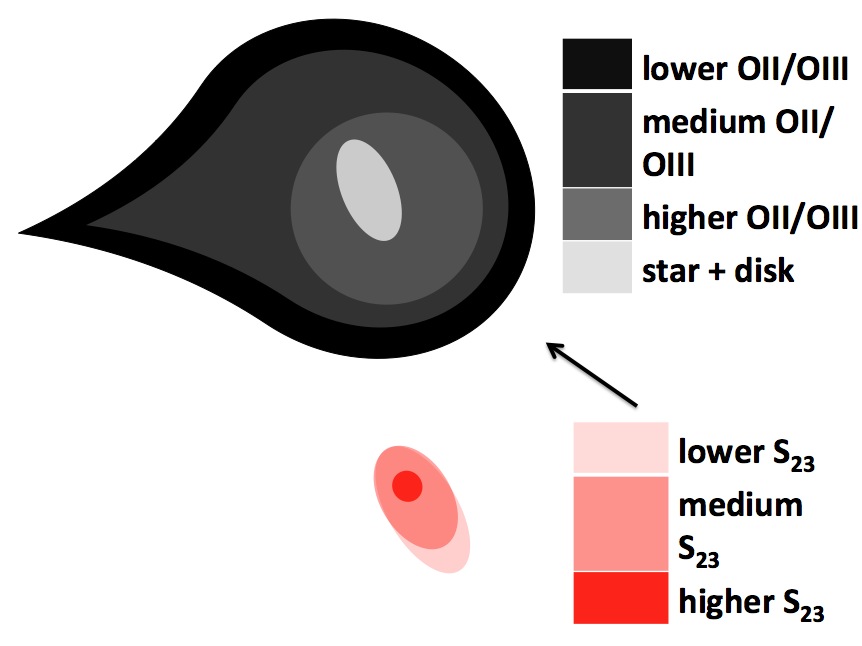}
\caption{Sketch of Orion 244-440: the outer layer displays an [OII]/[OIII] gradient with decreasing ionisation from the outside in (greyscale in upper figure), the star+disk component (lower figure) shows an S$_{23}$ gradient which increases away from the direction of motion. The black arrow indicates the direction of $\theta^{2}$ Ori A.}
\label{ske}
\end{figure}

Both HH 204 and Orion 244-440 seem to have S$_{23}$-high as well as S$_{23}$-low components, but while 244-440 appears to have a layered structure, HH 204 shows a \textit{head-tail} configuration. Orion 244-440 is a giant proplyd containing a YSO, or possibly even a binary system, with a one-sided microjet along the North-East axis \citep{2000AJ....119.2919B}. Fig. \ref{jetsSO23} and Fig. A8 (panels c and d) show that the outer layer of 244-440 appears to be S$_{23}$-low over a wide range of [OII]/[OIII], while the inner part, corresponding to the star+disk component of the proplyd (magenta box marked with the letter \textit{c} in Fig. \ref{jetsSO23}), shows a steep rise in S$_{23}$ values. Complementary to \cite{1999AJ....118.2350H}, we provide a sketch of Orion 244-440 in Fig. \ref{ske}: the outer, S$_{23}$-low, layer of the proplyd, shows a positive [OII]/[OIII] outside-in gradient, confirming the model of \citeauthor{1999AJ....118.2350H} where the ionised density of the outer layers is low, and increases when moving in towards the proplyd; the star+disk component seems to be aligned with the direction towards $\theta^{2}$ Ori A and the direction of motion (indeed, \citealt{1993ApJ...410..696O} find that  some proplyds are oriented toward $\theta^{2}$ Ori A, rather than $\theta^{1}$ Ori C). The star+disk component itself is characterised by a S$_{23}$ gradient that increases in the direction of motion.

\begin{figure}
\centering
\includegraphics[scale=0.3]{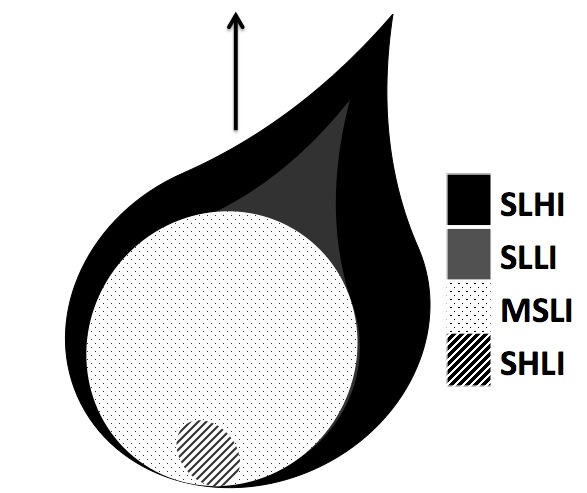}
\caption{Sketch of HH204. The black arrow indicates the direction to $\theta^{2}$ Ori A, while the direction of motion is along the head-tail axis.}
\label{hh204}
\end{figure}

The head of HH 204 (sketched in Fig. \ref{hh204} and highlighted in Fig. A8, panels a and b), on the other hand, displays intermediate S$_{23}$ values, and a S$_{23}$-low tail that shows an ionisation front in the direction of $\theta^{2}$ Ori A. Also, just as 244-440, it covers a broad range of [OII]/[OIII], and because of its layered structure it belongs to more than one of the four classes. 

\cite{2012MNRAS.421.3399N} performed an thorough analysis of HH 204 with integral field data from the Potsdam Multi-Aperture Spectrograph. They find a clearly stratified ionisation structure across HH 204, and discuss the presence of a trapped ionisation front at the head of the bow-shock, due to the a high [OI]/H$\beta$ ratio at that position, on the opposite side of the source of ionisation. In their analysis, the location of the enhanced [OI]/H$\beta$ ratio is also characterised by an enhanced [SII]/H$\beta$ ratio, which is in good agreement with the enhanced S$_{23}$ found in this work at the head of the bow-shock (marked as SHLI in Fig. \ref{hh204}).

\subsubsection{SHLI: HH 203, proplyds 239-510, 242-519, 247-436 and the candidate proplyd 224-510}
All of these objects are classified under SHLI, but they can be further separated into two sub-classes, depending on the presence of a microjet or not. The object HH 203 appears to have a different structure in terms of the S$_{23}$ parameter than HH 204, and only its tip deviates from the general cluster of data points in Fig. \ref{jetsSO23}b that originate from the ambient matter. Its tip classifies as SHLI, as does proplyd 239-510 and the candidate proplyd (identified as $h$ and $n$ respectively in Table \ref{proplyds} and Fig. \ref{jetsSO23}a). Also classified as SHLI are the proplyds 247-436 and 242-519 ($b$ and $i$ respectively), but these two are associated with microjets \citep{2000AJ....119.2919B}: indeed, although wSLLIe are limited by spatial resolution, we seem to be able to distinguish two components in these two proplyds, corresponding to a stellar and an outer (shell-like) component (marked in red and green respectively in Fig. A8, b and Fig. A11a, b).

\subsubsection{SLHI: proplyds 250-439, 252-457 and 236-527}
These three proplyds (identified as $a$, $d$ and $j$ respectively in Table \ref{proplyds}) occupy the S$_{23}$-low but highly ionised region of the scatter plot. Of the three, only 252-457 is associated with a particularly long jet of about 15", more details are discussed in \cite{2000AJ....119.2919B}.

\subsubsection{SLLI and MSLI: proplyds 245-502 and 206-446}
Both of these two sources are single objects in their classes, if one does not count the multi-class objects. Proplyd 245-502 (blue in Fig. A10a, b) is 
Proplyd 206-446 (Fig. A9c, d) is associated with a microjet perpendicular do the disk and faces $\theta^{2}$ Ori A \citep{2000AJ....119.2919B}, and although we are again limited by spatial resolution, the MUSE data seems to hint that this proplyd might have a SLHI component.

\section{Conclusions}
An analysis of the Orion Nebula in terms of ionic and total abundances as well as kinematics of the ionised gas was presented, based on optical integral field observations with the instrument MUSE@VLT. The following points summarise the main results of this paper:
\begin{itemize}
\item The first 5' x 6' ionic and total abundance maps of oxygen, sulphur and nitrogen of the Orion Nebula are presented.
\item The abundance maps are contaminated by a certain degree of structuring that traces the main features of M 42 (the Bright Bar, the Orion S regent, HH 203 and 204, some of the proplyds). Furthermore, the ionic abundances computed from the MUSE data do not agree with literature values. We suggest that a combination of observational limitations (e.g. high level of noise in the electron temperature and [OII] maps, strong dependance of the abundances on the physical parameters) and intrinsic properties (e.g. high-density regions with densities approaching the critical values of the species used to compute the physical parameters) lead to the described discrepancies between previously published data and this work. Within errors, the computed O and S abundances are in agreement with literature values.
\item The computed structure functions (S$_{2}$) prove that this dataset lacks the depth to reproduce previous results, which showed that in the Orion Nebula, the slope of $S_{2}$ corresponds to a broken power law with a steeper index at small scales and a turnover to a shallower index at larger scales. The structure functions computed in this work are much shallower than previous results, as they are highly affected by the large number of pixels with low signal to noise in the velocity maps. We demonstrated how noise can affect the structure function with simulated HII regions, and will discuss this issue in depth in a forthcoming paper. 
\item The only Herbig-Haro object that stands out from the rest is HH 201, the only one of the so called Orion Bullets covered by the MUSE mosaic. This outflow, similarly to the one identified in our analysis about the Pillars of Creation, shows very high S$_{23}$ together with a very high degree of ionisation. From previous studies it seems that this object originates from an explosive event that occurred in the BN-KL star forming region. Based on its cometary shape in the velocity maps and on the double component seen in the [FeII] line, we speculate that this object is currently impacting in a region with high densities and that is is therefore currently being disrupted.  
\item We applied a method developed in our previous publication about the iconic Pillars of Creation in M 16 to the Orion dataset to analyse the outflows and proplyds. In this method, the S$_{23}$ parameter is analysed in terms of the corresponding degree of ionisation (given by the ratio [OII]/[OIII]). We find that the proplyds can be divided into two distinct populations, in correlation with their location. The population of proplyds near the Trapezium cluster have lower S$_{23}$ values and a higher degree of ionisation, while the population of proplyds that lie south of the Bright Bar seem to be more shielded from the intense feedback of the Trapezium stars, as they show higher S$_{23}$ values and lower degree of ionisation. Furthermore, the proplyds and HH objects south of the Bar can be in turn divided into four classes, depending on their S$_{23}$ vs. [OII]/[OIII] values, and we show how they can be clearly distinguished in the S$_{23}$ vs. [OII]/[OIII] parameter space. This demonstrates that this method is very useful to pick out these kind of objects in the S$_{23}$vs. [OII]/[OIII] parameter space.
\item We suggest that the reason for the capability of the S$_{23}$ parameter to distinguish between the Bright Bar, the HH outflows and the proplyds is twofold: this line ratio traces both high degrees of ionisation (given by the higher S$_{23} $values) and the relative contribution of shocks to the excitation mechanism (in the regime of low S$_{23}$ values).
\end{itemize}

\section*{Acknowledgments}
This work is based on MUSE commissioning data obtained with ESO telescopes at the Paranal Observatory under programme 60.A-9100(A). The data reduction and analysis make use of \textsc{aplpy} (http:$//$aplpy.github.io), \textsc{spectral$\_$cube} (spectral-cube.readthedocs.org), \textsc{pyspeckit} (pyspeckit.bitbucket.org, \cite{2011ascl.soft09001G}) and \textsc{glue} (glueviz.org). This research was supported by the Christiane N{\"u}sslein-Volhard Foundation (AFMCL) and the DFG cluster of excellence \textit{Origin and Structures of the Universe} (JED). PMW received funding through BMBF Verbundforschung (project MUSE-AO, grant 05A14BAC). This work was partly supported (LT) by the Italian Ministero dell'Istruzione, Universit{\`a} e Ricerca through the grant \textit{Progetti Premiali 2012 -- iALMA} (CUP C52I13000140001).

\bibliography{orionbib}
\bibliographystyle{mn2e}

\appendix

\section[]{}
\begin{figure*}
\mbox{`
\subfloat[]{\includegraphics[scale=0.35]{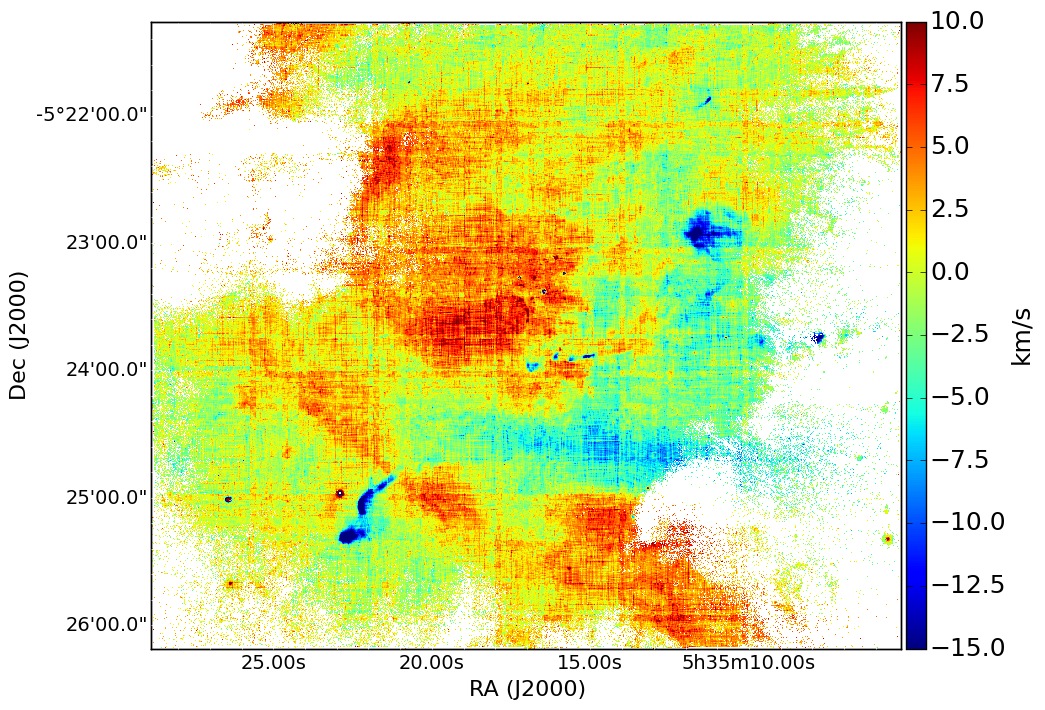}}
\subfloat[]{\includegraphics[scale=0.35]{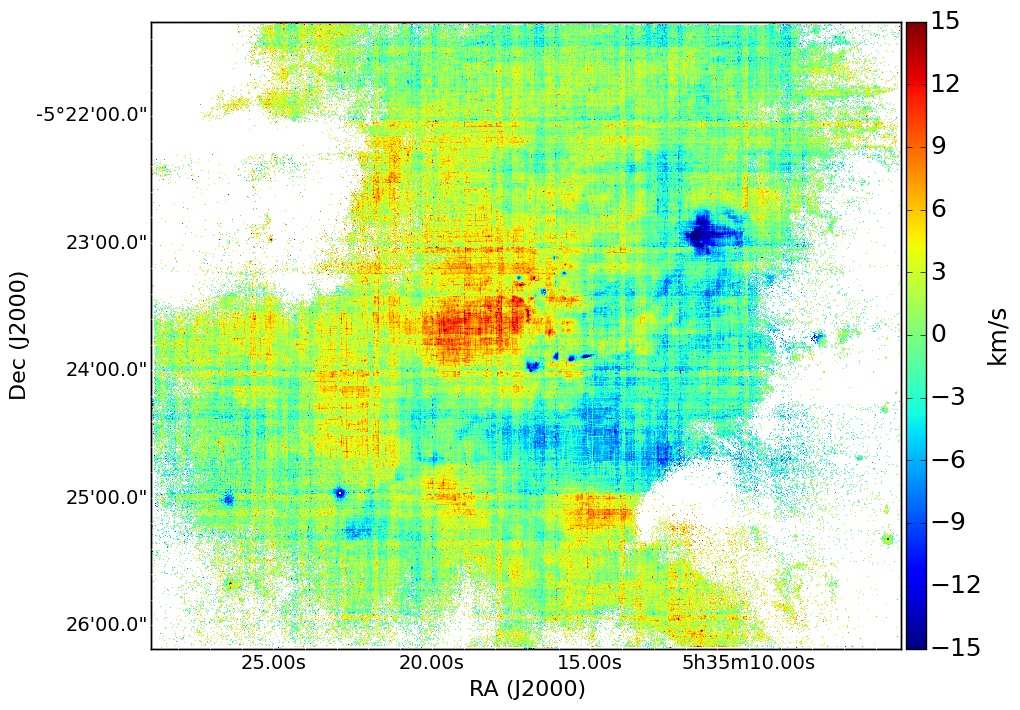}}}
\mbox{
\subfloat[]{\includegraphics[scale=0.35]{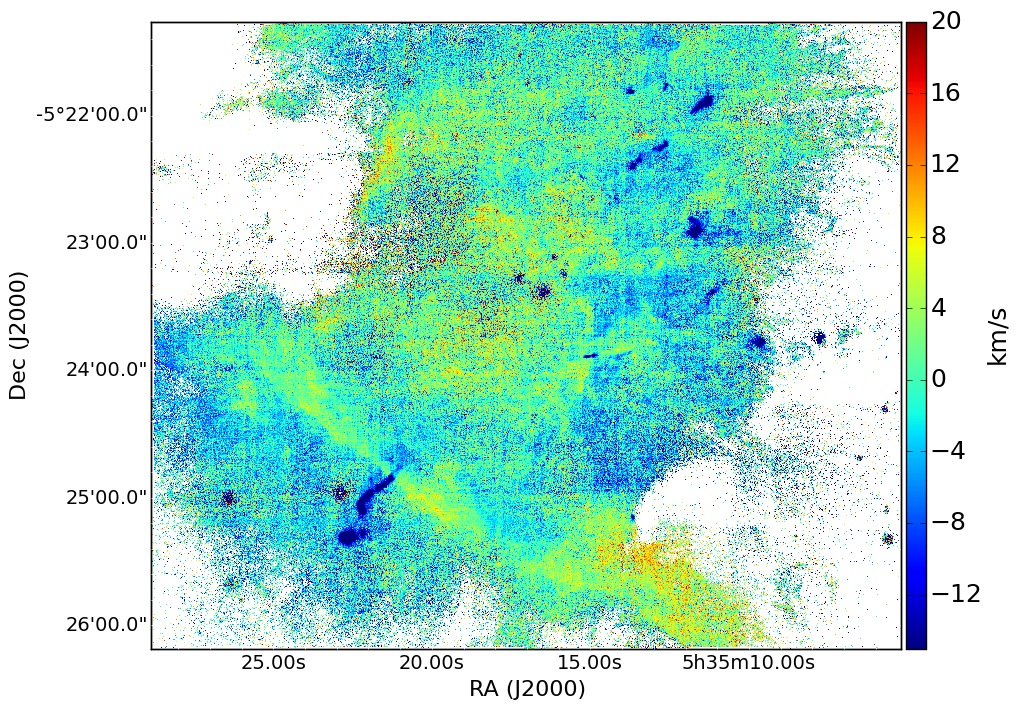}}
\subfloat[]{\includegraphics[scale=0.35]{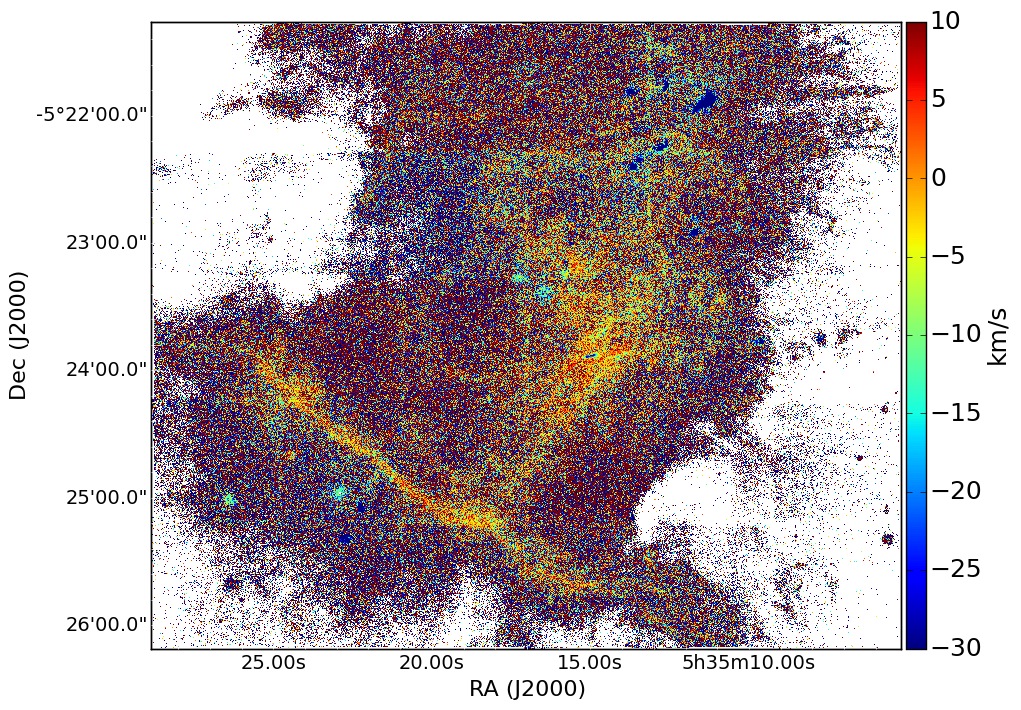}}}
  \caption{Velocity maps of H$\alpha$ (a ), [OIII]$\lambda$5007 (b) and [SII]$\lambda$6731 (c) and [OI]$\lambda$6300 (d) masked based on the mean value of the [OI] line. The indicated velocities correspond to velocities relative to the mean velocity of the surrounding medium (see text Section 5.1).}
  \label{vmaps_masked}
\end{figure*}

\begin{figure*}
\mbox{
\subfloat[]{\includegraphics[scale=0.4]{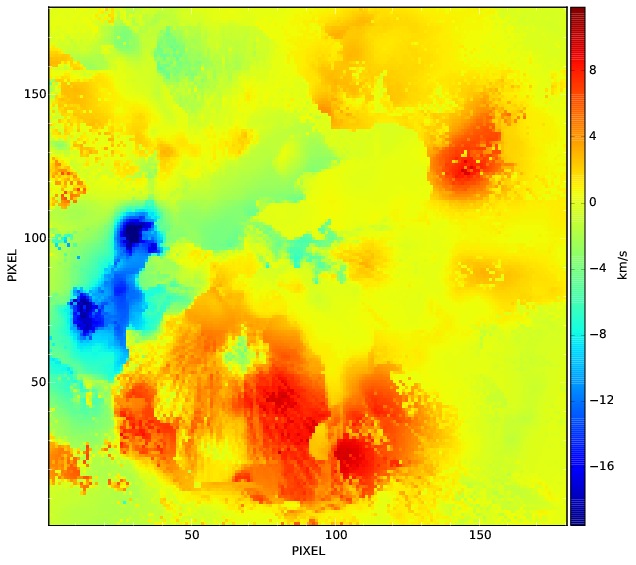}}
\subfloat[]{\includegraphics[scale=0.4]{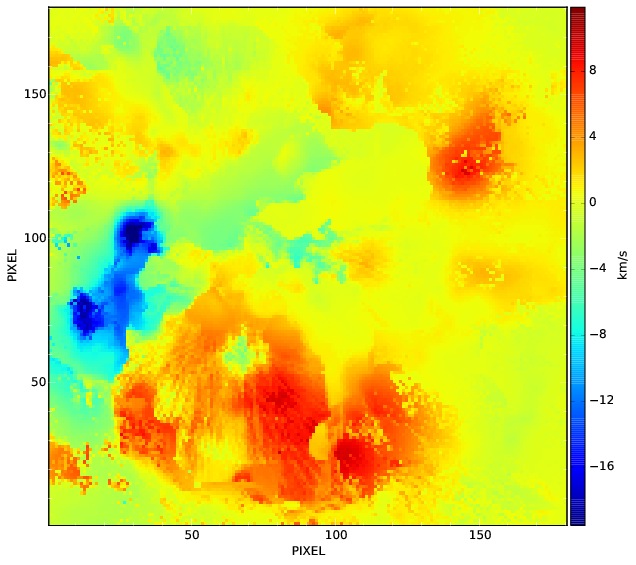}}}
\mbox{
\subfloat[]{\includegraphics[scale=0.4]{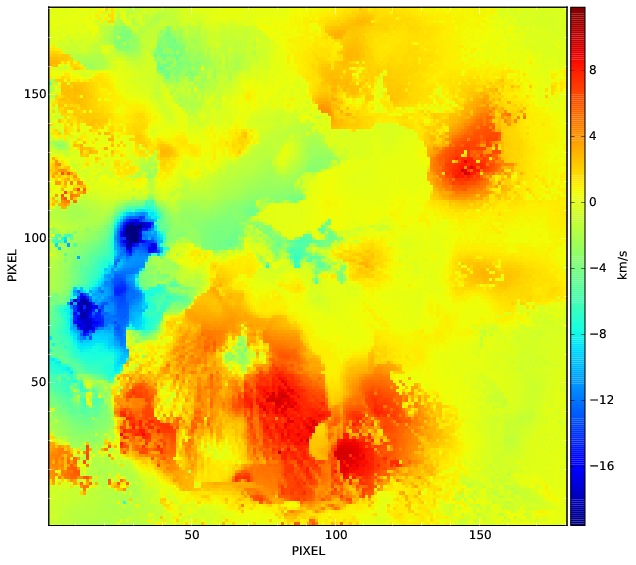}}
\subfloat[]{\includegraphics[scale=0.4]{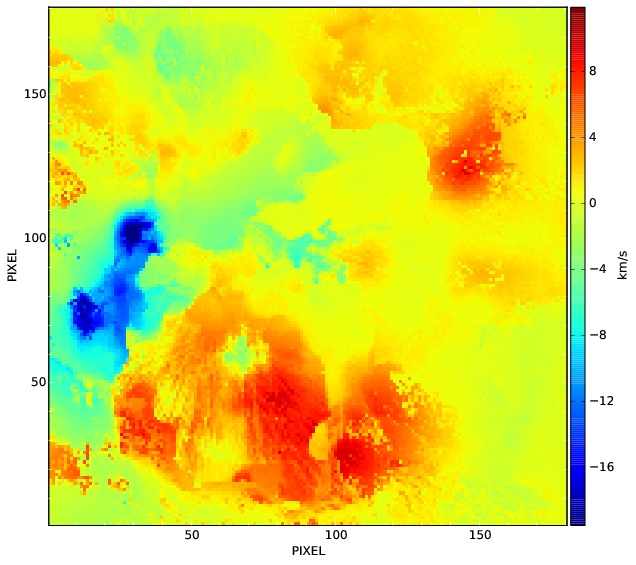}}}
  \caption{[SII] velocity map of our simulated HII region (panel a), post-processed with the ionisation radiative transfer code MOCASSIN. The pixel scale is 0.028 pc. To this we gradually add Gaussian noise by increasing the standard deviation $\sigma$ from 0.01 km/s (panel b) to 10 (panel d in Fig. \ref{sim_maps2}).}
  \label{sim_maps}
\end{figure*}

\begin{figure*}
\mbox{
\subfloat[]{\includegraphics[scale=0.4]{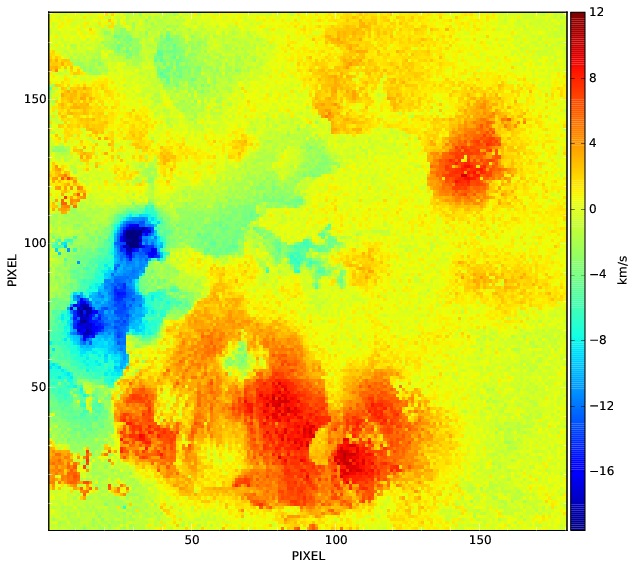}}
\subfloat[]{\includegraphics[scale=0.4]{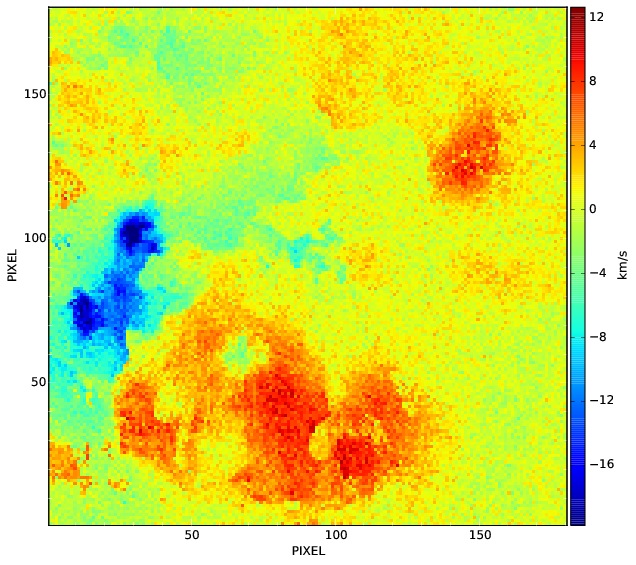}}}
\mbox{
\subfloat[]{\includegraphics[scale=0.4]{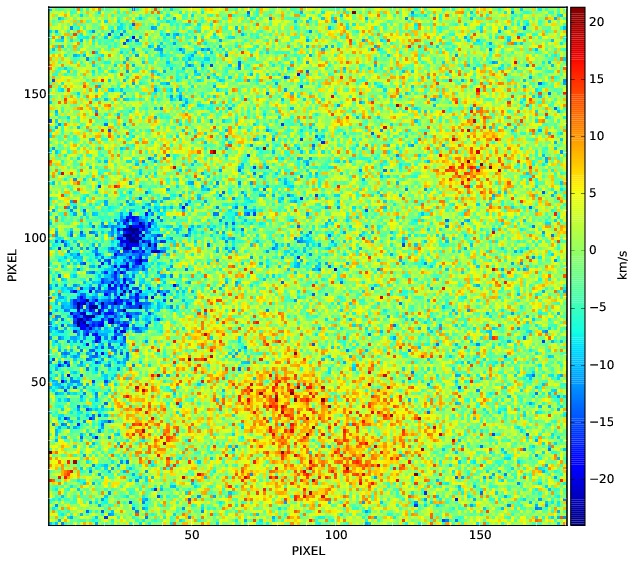}}
\subfloat[]{\includegraphics[scale=0.4]{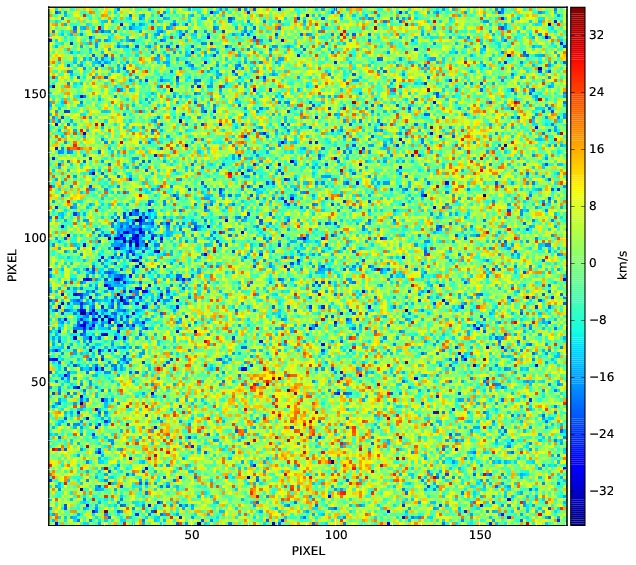}}}
  \caption{See Fig. \ref{sim_maps}.}
  \label{sim_maps2}
\end{figure*}

\begin{figure*}
\mbox{
\subfloat[]{\includegraphics[scale=0.4]{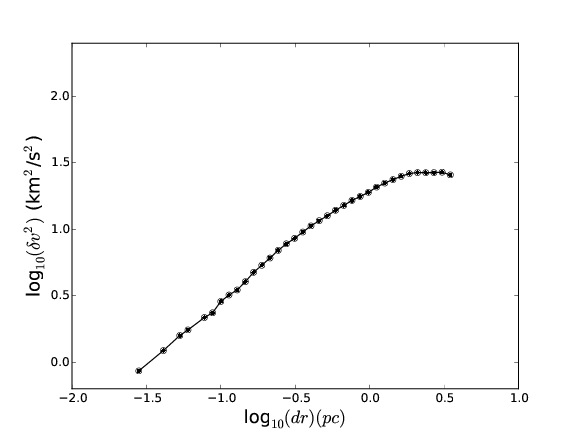}}
\subfloat[]{\includegraphics[scale=0.4]{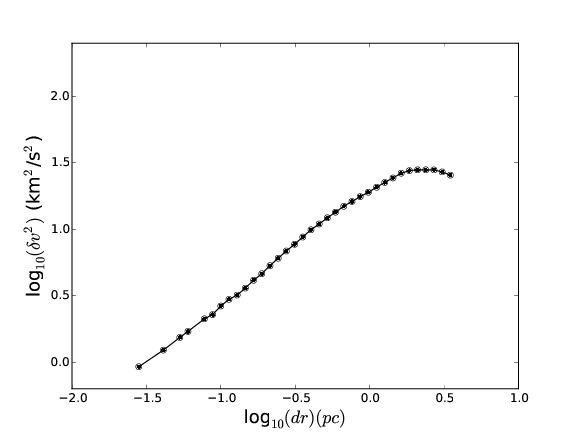}}}
\mbox{
\subfloat[]{\includegraphics[scale=0.4]{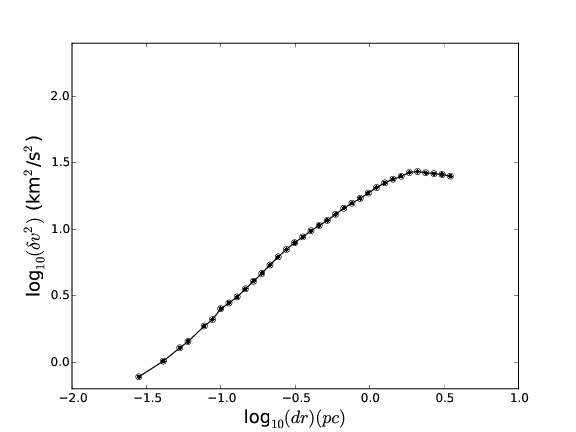}}
\subfloat[]{\includegraphics[scale=0.4]{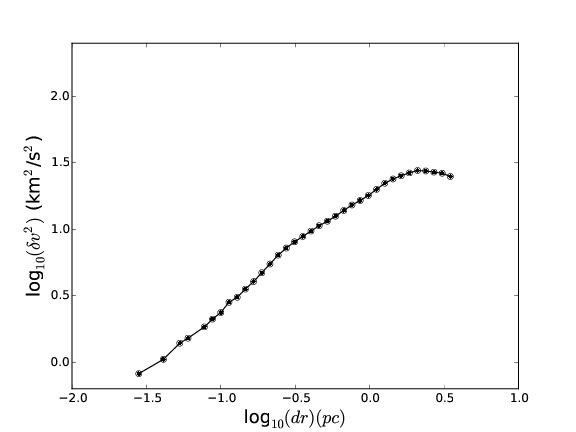}}}
  \caption{Structure function corresponding to the [SII] velocity map of our simulated HII region (panel a). To this we gradually add Gaussian noise by increasing the standard deviation $\sigma$ from 0.01 km/s (panel b) to 10 (panel d in Fig. \ref{sim_strf2}). To show the flattening of the structure function, all plots share the same scaling.}
  \label{sim_strf}
\end{figure*}

\begin{figure*}
\mbox{
\subfloat[]{\includegraphics[scale=0.4]{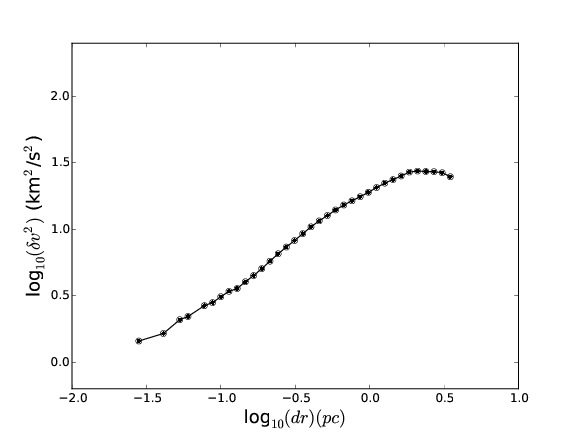}}
\subfloat[]{\includegraphics[scale=0.4]{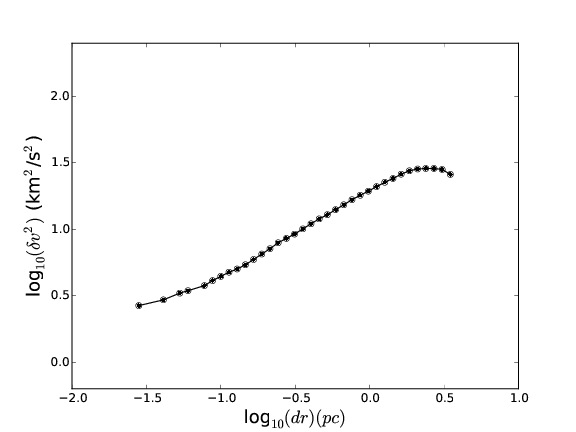}}}
\mbox{
\subfloat[]{\includegraphics[scale=0.4]{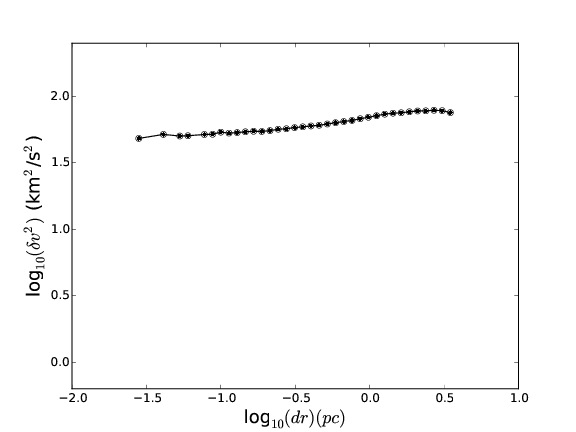}}
\subfloat[]{\includegraphics[scale=0.4]{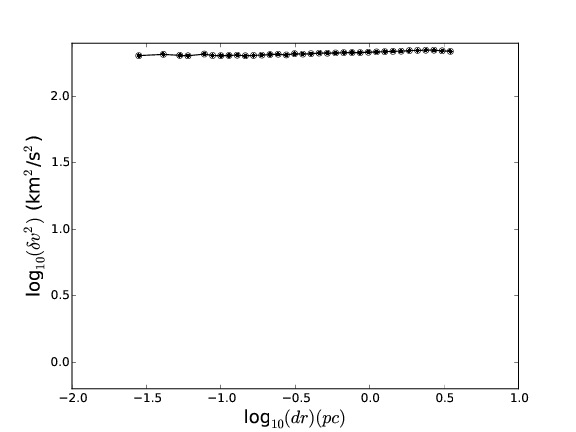}}}
  \caption{See Fig. \ref{sim_strf}.}
  \label{sim_strf2}
\end{figure*}

\begin{figure*}
\centering
\includegraphics[scale=0.7]{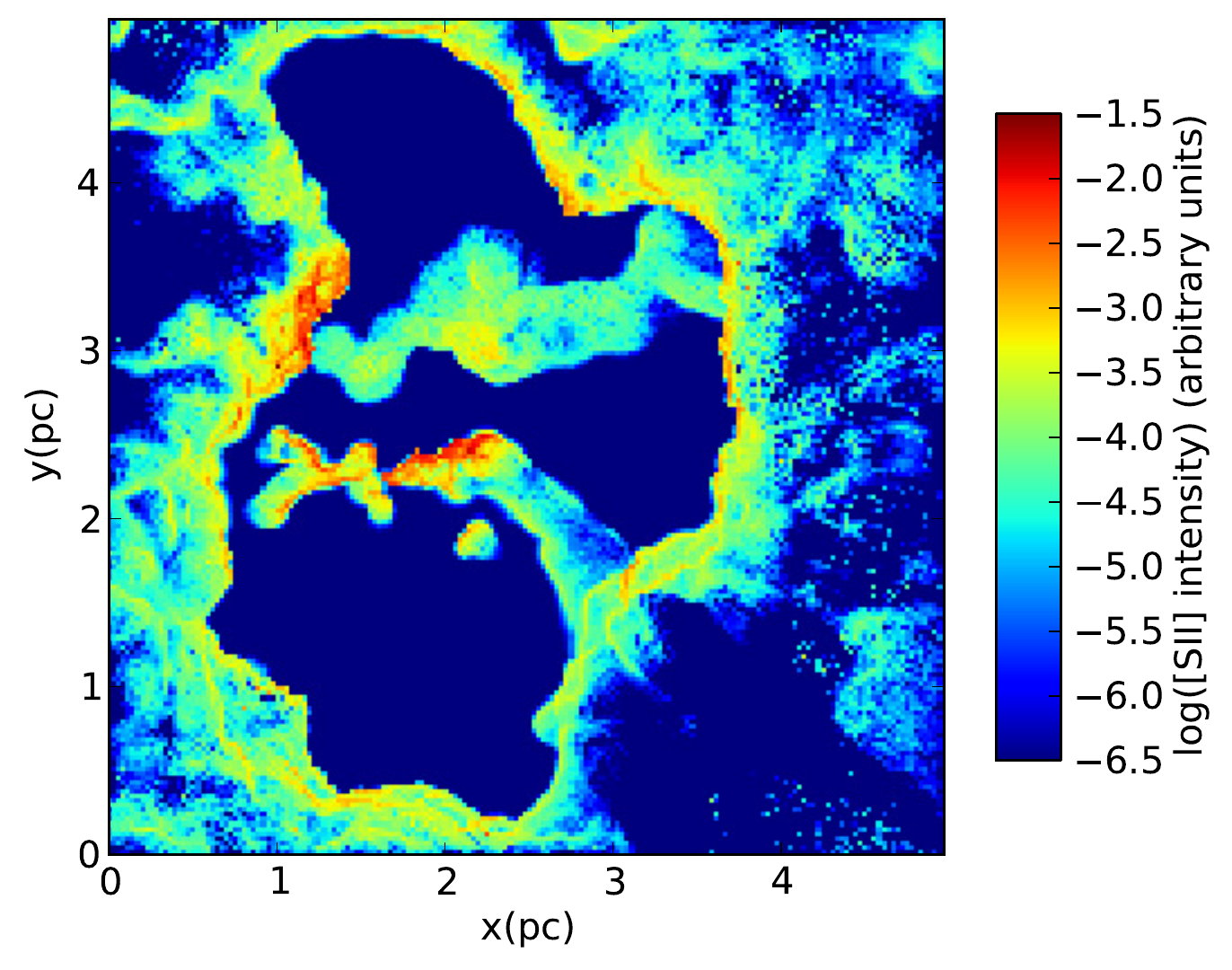}
\caption{Integrated [SII] intensity map of the synthetic star forming molecular cloud, post-processed with \textsc{mocassin}. See text Section 4.}
\label{sim_SII}
\end{figure*}

Fig. \ref {vmaps_masked} shows the velocity maps masked according to the [OI]$\lambda$6300 intensity threshold of $\sim$10$^{-16}$ erg s$^{-1}$ cm$^{-2}$ pixel$^{-1}$ used to compute the second order structure function S$_{2}$. The effect of adding Gaussian noise to the synthetic velocity maps (Fig. \ref{sim_maps} and \ref{sim_maps2}) of a simulated HII region ( Fig. \ref{sim_SII}) is shown in Fig. \ref{sim_strf} and \ref{sim_strf2}, where the structure functions for different values of the standard deviation $\sigma$ are shown. See text Section 5. 

\begin{figure*}
\centering
\includegraphics[scale=0.5]{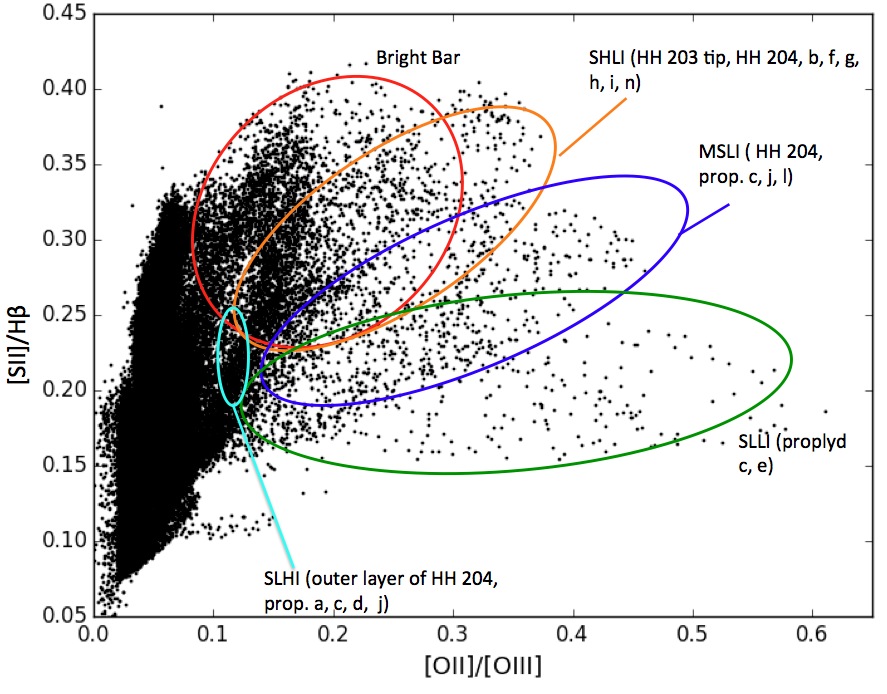}
\caption{[SII]/H$\beta$ vs. [OII]/[OIII] scatter plot, with the same regions as indicated in Fig. \ref{jetsSO23}b. See text Section 5.2.}
\label{SII_Hb}
\end{figure*}

\begin{figure*}
\mbox{
\subfloat[]{\includegraphics[scale=0.35]{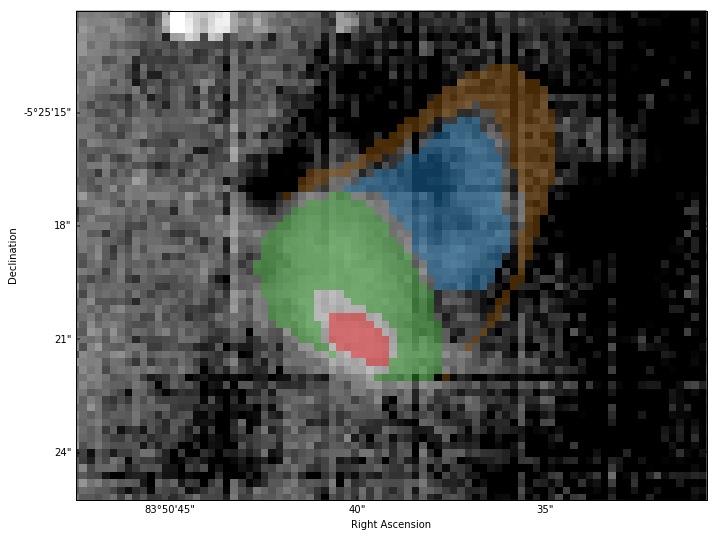}}
\subfloat[]{\includegraphics[scale=0.35]{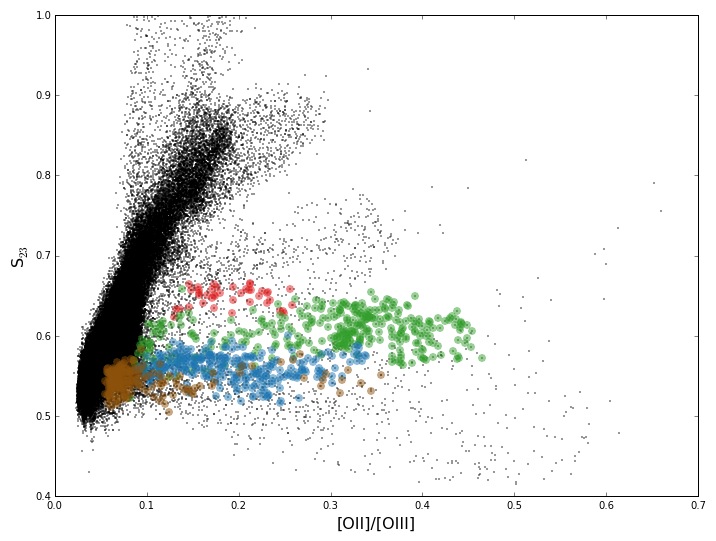}}}
\mbox{
\subfloat[]{\includegraphics[scale=0.35]{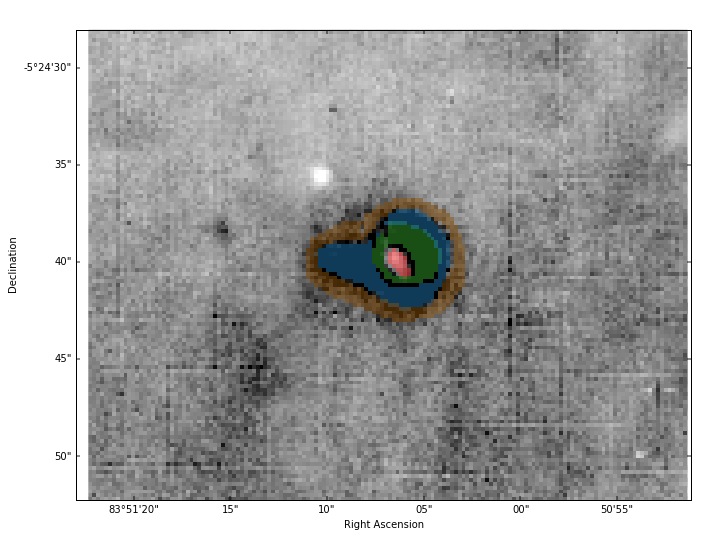}}
\subfloat[]{\includegraphics[scale=0.35]{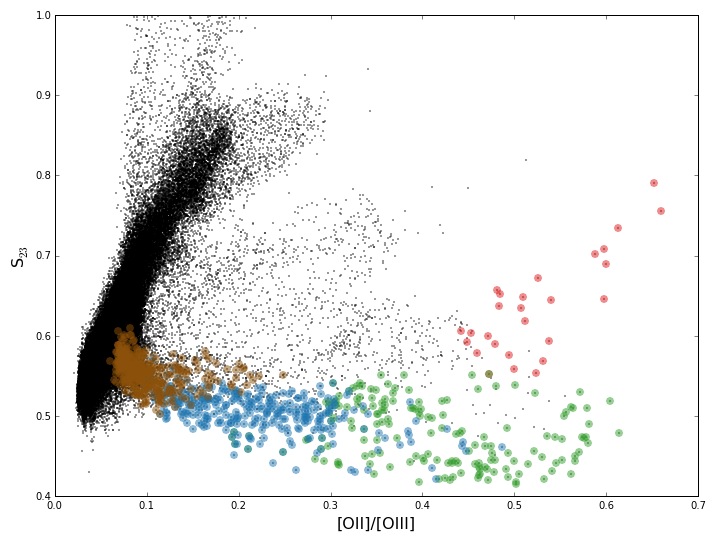}}}
\mbox{
\subfloat[]{\includegraphics[scale=0.35]{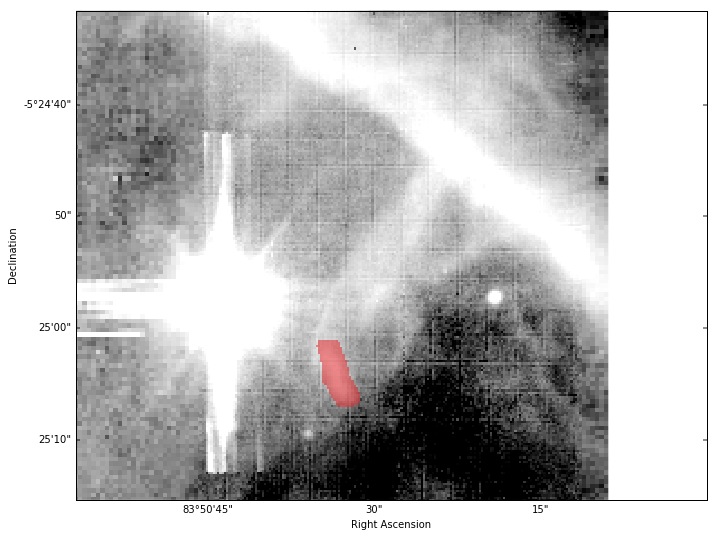}}
\subfloat[]{\includegraphics[scale=0.35]{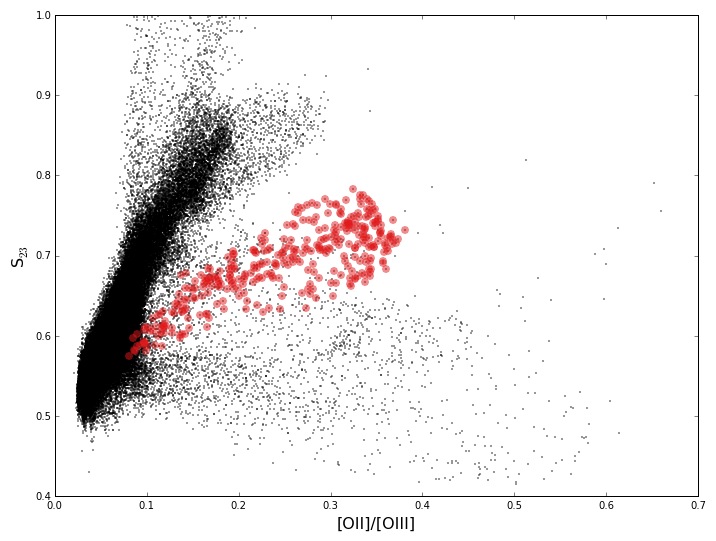}}}
  \caption{S$_{23}$ maps (left panels) and S$_{23}$ vs. [OII]/[OIII] scatter plots (right panels) of HH 204 (panels a and b), proplyd 244-440 (panels c and d) and  HH 203 (panels e and f). Scales and coordinates are the same as in Fig. \ref{jetsSO23}.}
  \label{zoom1}
\end{figure*}

\begin{figure*}
\mbox{
\subfloat[]{\includegraphics[scale=0.35]{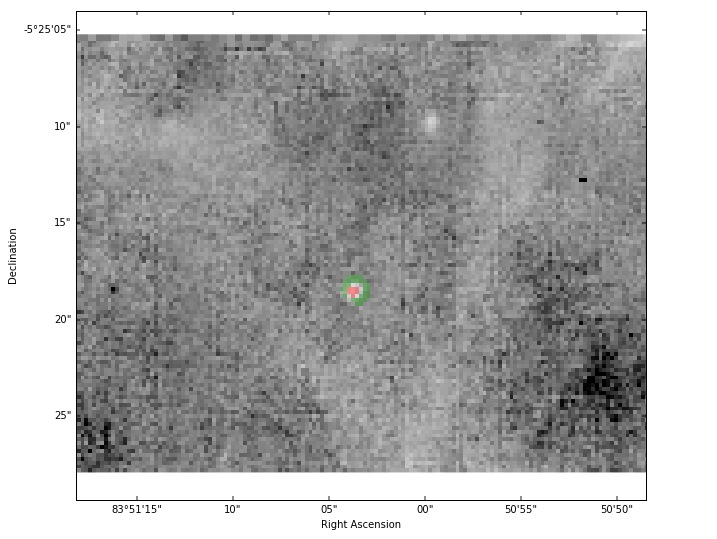}}
\subfloat[]{\includegraphics[scale=0.35]{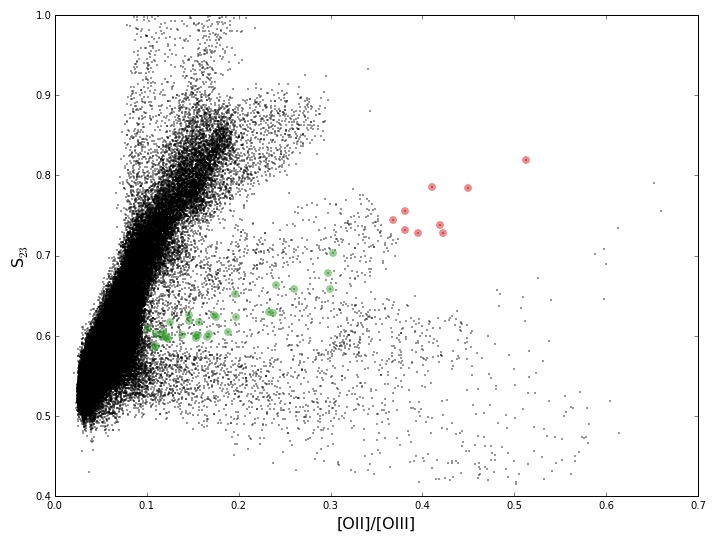}}}
\mbox{
\subfloat[]{\includegraphics[scale=0.34]{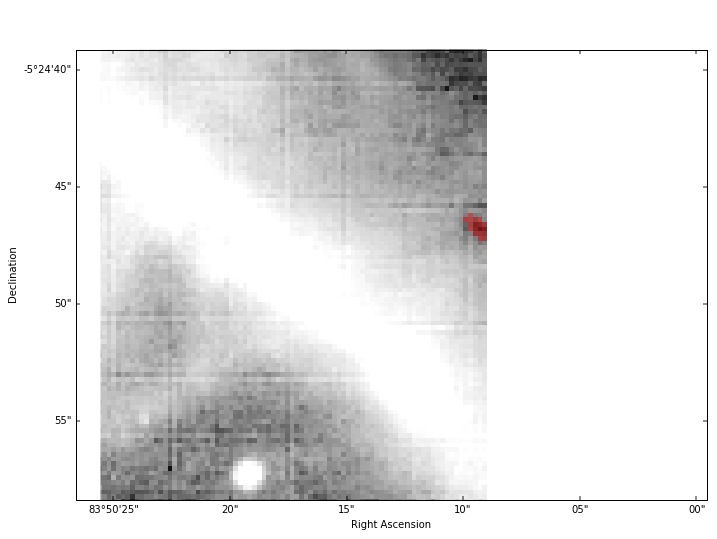}}
\subfloat[]{\includegraphics[scale=0.32]{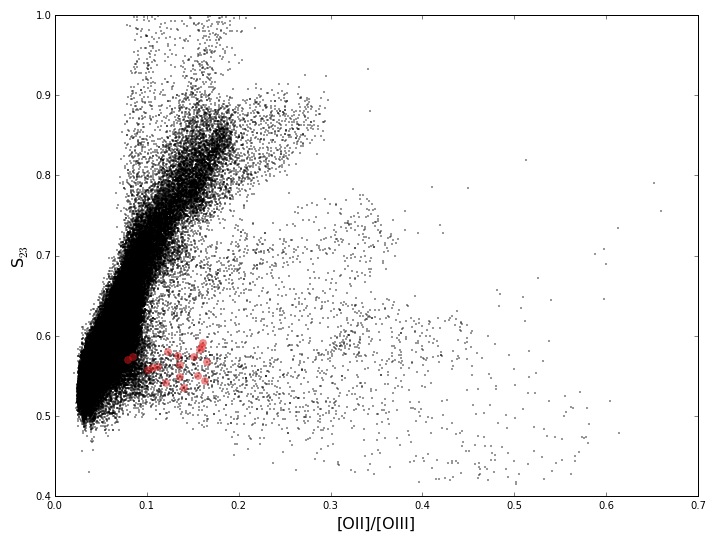}}}
\mbox{
\subfloat[]{\includegraphics[scale=0.35]{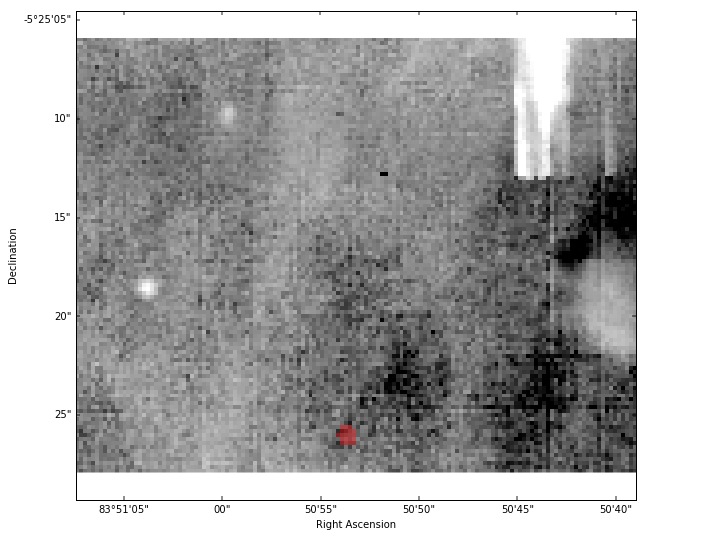}}
\subfloat[]{\includegraphics[scale=0.35]{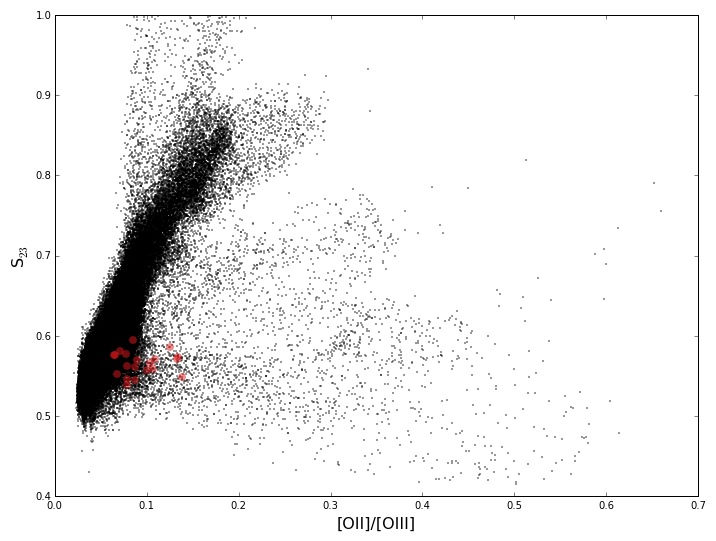}}}
  \caption{Same as Fig. \ref{zoom1}, but for proplyds 242-519 (panels a and b), 206-446 (panels c and d) and 236-527 (panels e and f).}
  \label{zoom2}
\end{figure*}

\begin{figure*}
\mbox{
\subfloat[]{\includegraphics[scale=0.35]{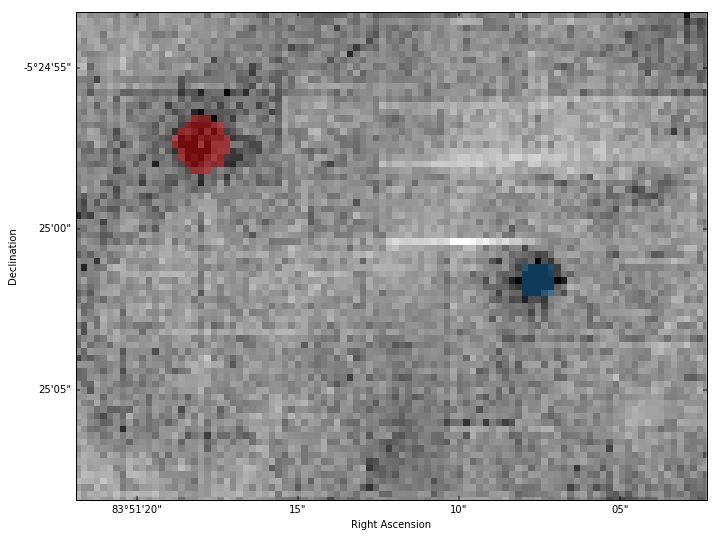}}
\subfloat[]{\includegraphics[scale=0.35]{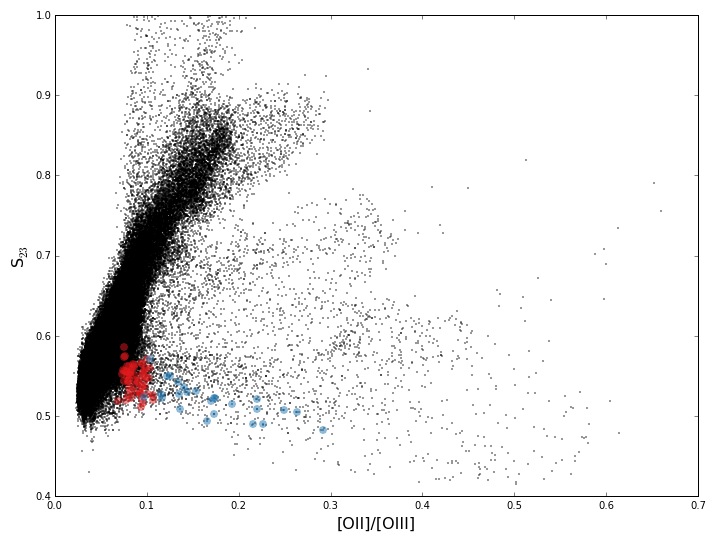}}}
\mbox{
\subfloat[]{\includegraphics[scale=0.35]{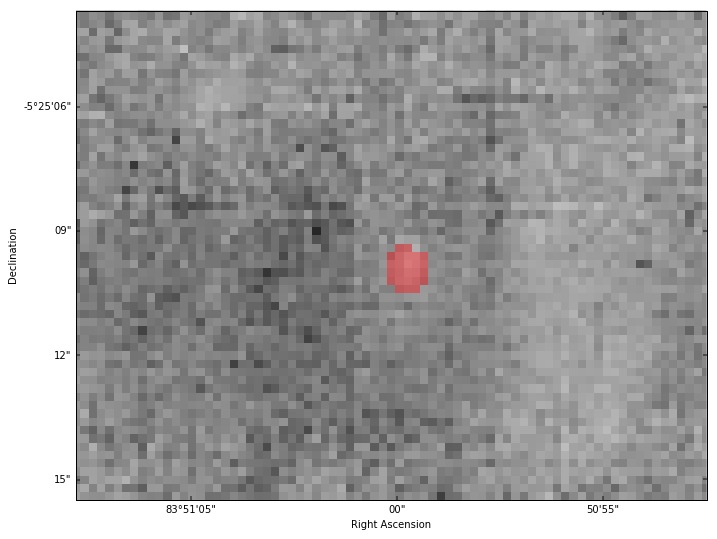}}
\subfloat[]{\includegraphics[scale=0.35]{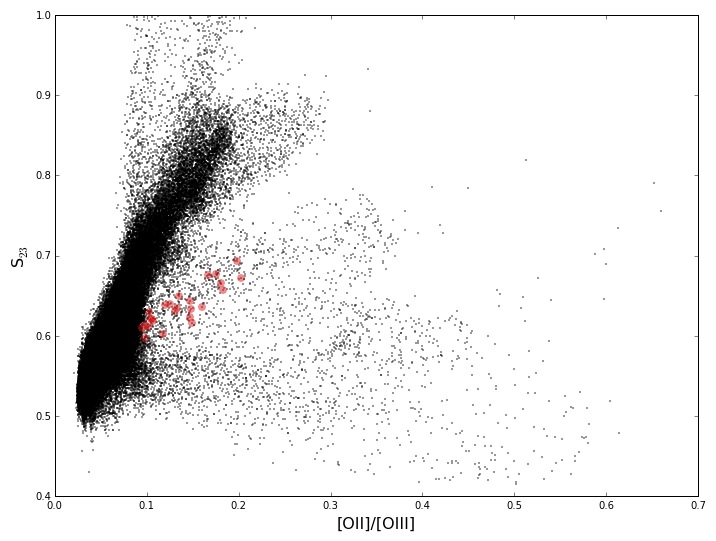}}}
\mbox{
\subfloat[]{\includegraphics[scale=0.35]{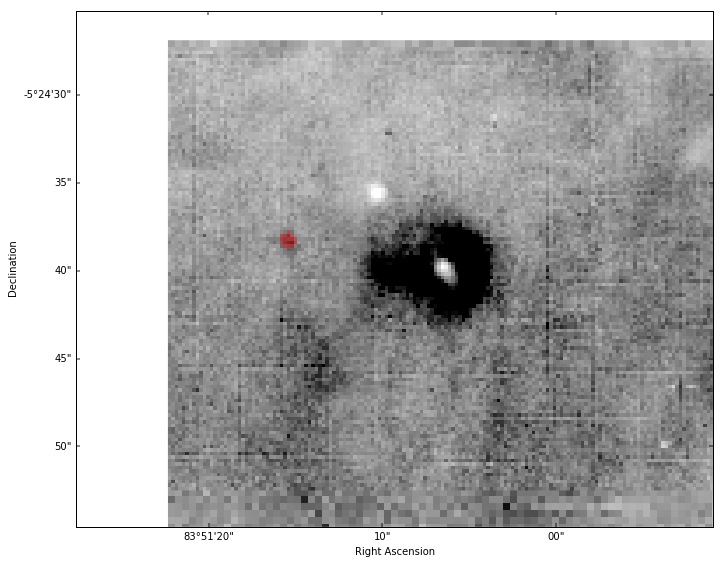}}
\subfloat[]{\includegraphics[scale=0.35]{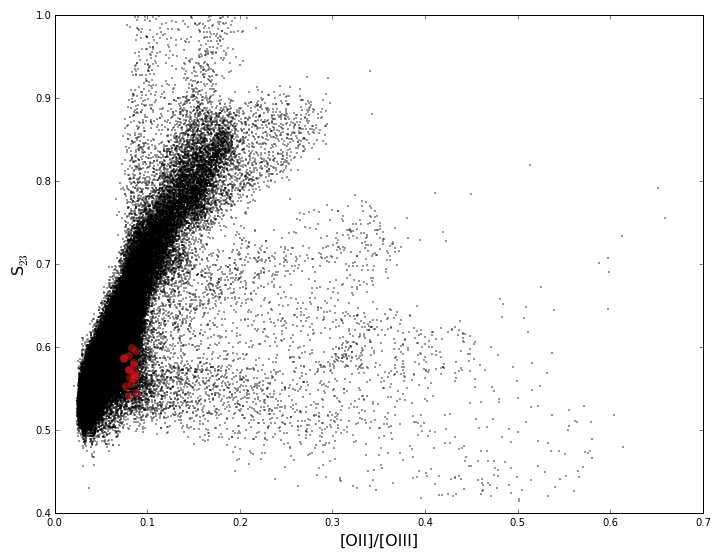}}}
  \caption{Same as Fig. \ref{zoom1}, but for the proplyds 252-457 and 245-502 (red and blue respectively in panels a and b), 239-510 (panels c and d) and 250-439 (panels e and f).}
  \label{zoom3}
\end{figure*}

\begin{figure*}
\mbox{
\subfloat[]{\includegraphics[scale=0.35]{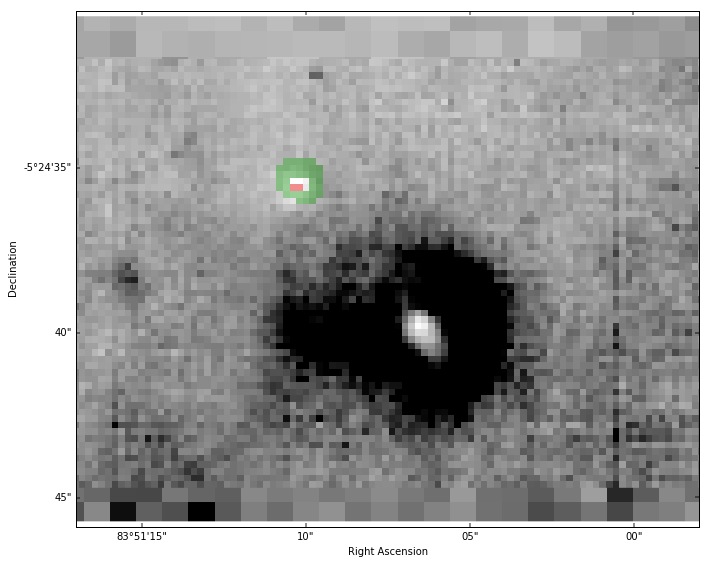}}
\subfloat[]{\includegraphics[scale=0.35]{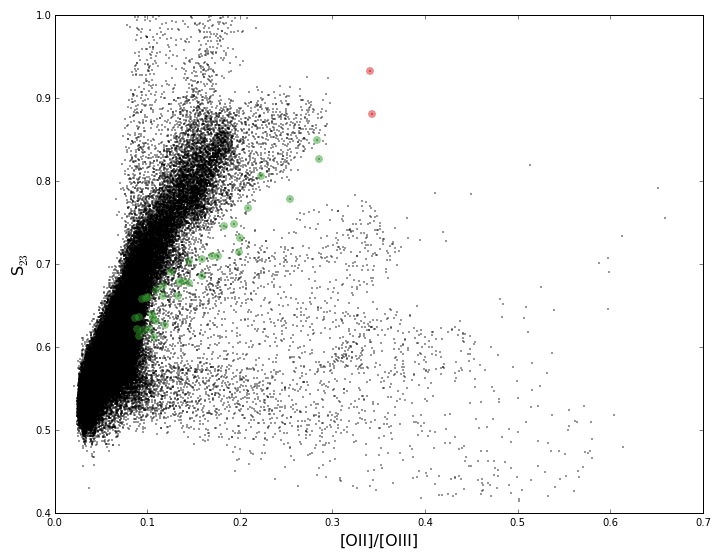}}}
\mbox{
\subfloat[]{\includegraphics[scale=0.35]{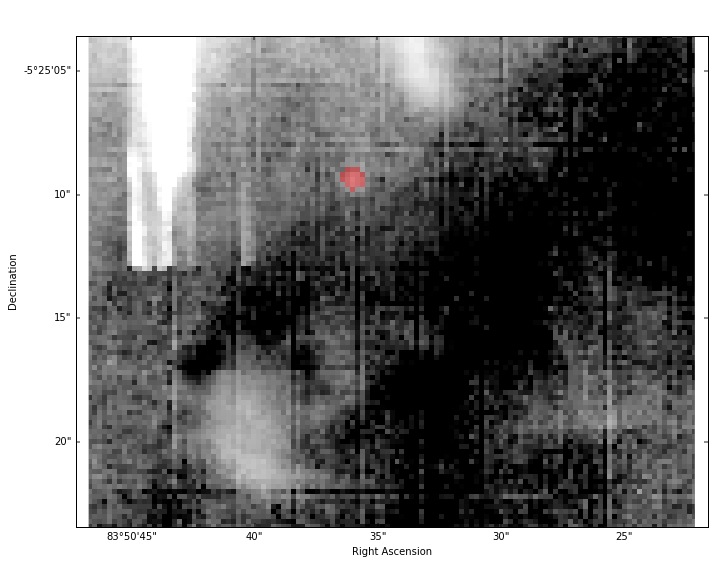}}
\subfloat[]{\includegraphics[scale=0.35]{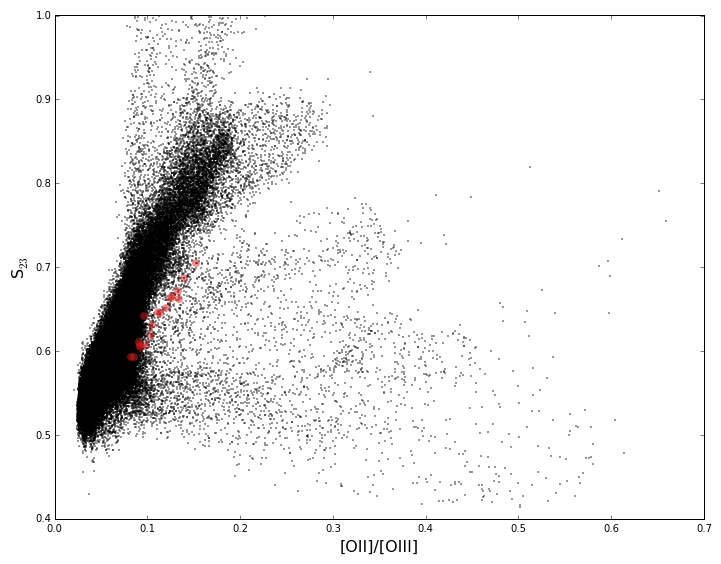}}}
  \caption{Same as Fig. \ref{zoom1}, but for the proplyd 247-436 (panels a and b), the candidate proplyd (224-510, letter $n$ in Fig. \ref{jetsSO23}a, panels c and d) }
  \label{zoom4}
\end{figure*}

Figure \ref{SII_Hb} shows the equivalent of Fig. 18b, but for [SII]/H$\beta$ (see text Section 5.2). Figures from \ref{zoom1} to \ref{zoom4} show the S$_{23}$ maps (left panels) and their corresponding location in the S$_{23}$ vs. [OII]/[OIII] parameter space of the Orion proplyds listed in Table 5 and discussed in Section 5.2. Figures from \ref{zoom1} to \ref{zoom4} are intentionally not continuum-subtracted (a version of continuum-subtracted S$_{23}$ vs [OII]/[OIII] parameter space is shown in Fig. 18) to highlight the stellar components of the proplyds (244-440, 242-519, 247-436).

\bsp

\label{lastpage}

\end{document}